\documentclass[12pt,a4paper]{article}
\counterwithin*{equation}{section}

\usepackage{caption}
\usepackage{cite,mathtools,amsmath,amssymb,graphicx,subfigure}
\usepackage[usenames]{color}
\usepackage[bookmarksopen,colorlinks=true,linkcolor=dark-green,citecolor=dark-red,urlcolor=dark-red,linktocpage=false]{hyperref}
\usepackage{url}
\usepackage{ulem}

\definecolor{dark-blue}{rgb}{0,0,0.6}
\definecolor{dark-green}{rgb}{0,0.4,0}
\definecolor{dark-red}{rgb}{0.6,0,0}

\usepackage[height=23.5cm,width=16.5cm,centering]{geometry}

\def\thefootnote{\fnsymbol{footnote}}
\newcommand{\gsim}{~\mbox{\raisebox{-1.0ex}{$\stackrel{\textstyle >}{\textstyle \sim}$ }}}
\newcommand{\lsim}{~\mbox{\raisebox{-1.0ex}{$\stackrel{\textstyle <}{\textstyle \sim}$ }}}

\newcommand{\beq}{\begin{align}}
\newcommand{\eeq}{\end{align}}
\newcommand{\beqa}{\begin{eqnarray}}
\newcommand{\eeqa}{\end{eqnarray}}
\newcommand{\mpl}{M_{\rm pl}}

\newcommand{\lmk}{\left(}
\newcommand{\rmk}{\right)}

\newcommand{\D}{{\rm d}}

\begin{document}
\begin{titlepage}

\begin{center}
\hfill DESY 20-126 \\
\hfill RESCEU-13/20 \\

\vskip 10mm

{\fontsize{20pt}{0pt} \bf
Occurrence of Tachyonic Preheating
}
\\[3mm]
{\fontsize{20pt}{0pt} \bf
in the Mixed Higgs-$ R^2 $ Model
}

\vskip 10mm

{\large
Minxi He,\footnote{
Email: {\tt he-minxi194"at"g.ecc.u-tokyo.ac.jp}
}$^{1,2}$
Ryusuke Jinno,$^{3}$ Kohei Kamada,$^{2}$ 
} 
\\[2mm]
{\large 
Alexei~A.~Starobinsky,$^{2,4}$ and Jun'ichi Yokoyama$^{1,2,5,6}$
}

\vskip 5mm

$^{1}${\it Department of Physics, Graduate School of Science, }
\\[1mm]
{\it The University of Tokyo, Hongo 7-3-1, Bunkyo-ku, Tokyo 113-0033, Japan}
\\[2mm]
$^{2}${\it Research Center for the Early Universe (RESCEU), Graduate School of Science,}
\\[1mm]
{\it The University of Tokyo, Hongo 7-3-1, Bunkyo-ku, Tokyo 113-0033, Japan}
\\[2mm]
$^{3}${\it Deutsches Elektronen-Synchrotron DESY, Notkestrasse 85, D-22607 Hamburg, Germany}
\\[2mm]
$^{4}${\it L.~D.~Landau Institute for Theoretical Physics, Moscow 119334, Russia}
\\[2mm]
$^{5}${\it Kavli Institute for the Physics and Mathematics of the Universe (Kavli IPMU),}
\\[1mm]
{\it WPI, UTIAS, The University of Tokyo, Kashiwanoha 5-1-5, Chiba 277-8583, Japan}
\\[2mm]
$^{6}${\it Trans-scale Quantum Science Institute,} 
\\[1mm]
{\it The University of Tokyo, Hongo 7-3-1, Bunkyo-ku, Tokyo 113-0033, Japan}

\end{center}
\vskip 4mm

\begin{abstract}
It has recently been suggested that at the post-inflationary stage of the mixed Higgs-$R^2$ model of inflation efficient particle production can arise from the tachyonic instability of the Higgs field. 
It might complete the preheating of the Universe if appropriate conditions are satisfied, especially in the Higgs-like regime. 
In this paper, we study this behavior in more depth, including the conditions for occurrence, analytical estimates for the maximal efficiency, and the necessary degree of fine-tuning among the model parameters to complete preheating by this effect. 
We find that the parameter sets that cause the most efficient tachyonic instabilities obey simple laws in both the Higgs-like regime and the $R^2$-like regime, respectively. We then estimate the efficiency of this instability. 
In particular, even in the deep $R^2$-like regime with a small non-minimal coupling, 
this effect is strong enough to complete preheating although a severe fine-tuning is required among the model parameters.
We also estimate how much fine-tuning is needed to complete preheating by this effect. 
It is shown that the fine-tuning of parameters for the sufficient particle production 
is at least $ < \mathcal{O}(0.1) $ in the deep Higgs-like regime with a large scalaron mass, while it is more severe $\sim {\cal O}(10^{-4})-{\cal O}(10^{-5})$ 
in the $R^2$-like regime with a small non-minimal coupling. 

\end{abstract}

\end{titlepage}

\tableofcontents
\thispagestyle{empty}
\newpage

\renewcommand{\thefootnote}{$\diamondsuit$\arabic{footnote}}
\setcounter{page}{1}
\setcounter{footnote}{0}

\section{Introduction}
\label{Sec-1}

Reheating is an essential ingredient for a successful inflationary universe model (see e.g. \cite{Sato:2015dga} for a review), through which high-energy particles are produced and thermalized in the ``empty" space, converting the Universe to a radiation-dominated one after its quasi-exponential expansion during inflation. Proper analysis of cosmic history during reheating regime is required to express
the pivot scale of curvature perturbation in terms of the number of $e$-folds of inflation, in order to confront the prediction of inflation with observational data~\cite{Liddle:2003as,Martin:2010kz}. For inflationary models with inflaton oscillations after the end of inflation, reheating can occur in two characteristic regimes in general, namely, (1) preheating which usually involves short but rapid and non-perturbative processes transferring a significant fraction  of the energy from inflaton to other matter fields through broad parametric resonance~\cite{Kofman:1994rk,Shtanov:1994ce,Kofman:1997yn,Greene:1997fu} or tachyonic processes~\cite{Felder:2000hj,Felder:2001kt,Kofman:2001rb}, and (2) perturbative reheating where the inflaton field decays perturbatively transferring all its energy to other particles~\cite{Starobinsky:1980te,Abbott:1982hn,Dolgov:1982th}.

This paper studies the preheating process in a two-field inflationary model, the mixed Higgs-$ R^2 $ one, which involves a non-minimally coupled Higgs field and the geometric $ R^2 $ term~\cite{Wang:2017fuy,Ema:2017rqn,He:2018gyf,Gundhi:2018wyz,Enckell:2018uic,Cheong:2019vzl}. Phenomenologically, 
this model considers a combination of two inflationary models that are most favored by observations,
namely, the Higgs inflation with a non-minimal coupling to gravity~\cite{CervantesCota:1995tz,Bezrukov:2007ep,Barvinsky:2008ia}\footnote{``Higgs inflation" in this paper solely means the original Higgs inflationary model which uses a large non-minimal coupling to the scalar curvature $ R $, among many variants of the model exhausted in Ref.~\cite{Kamada:2012se}.}, in which the Standard Model (SM) 
Higgs field discovered at the Large Hadron Collider is identified with the inflaton, and the $R^2$-inflation~\cite{Starobinsky:1980te,Starobinsky:1983zz} which uses a particular type of scalar-tensor gravity and thus has a scalar degree of freedom called scalaron in addition to usual massless tensor degrees of freedom of General Relativity.
Theoretically, the mixed Higgs-$ R^2 $ model can be considered as a UV-extension of the Higgs inflation~\cite{CervantesCota:1995tz,Bezrukov:2007ep,Barvinsky:2008ia} because it is found that the presence of the quadratic curvature term lifts the cutoff scale of the Higgs inflation to the Planck scale $ M_{\rm pl} $~\cite{Ema:2017rqn,Gorbunov:2018llf}. 
It has been pointed out that the pure Higgs inflation becomes strongly coupled at a relatively small energy scale, 
which raises a question on the validity of the scenario~\cite{Burgess:2009ea,Barbon:2009ya,Burgess:2010zq,Hertzberg:2010dc,Barvinsky:2009ii}.
This is also the reason why it is impossible to reconstruct the Higgs potential at inflationary scale 
only from the low-energy scale observables~\cite{Bezrukov:2009db,Bezrukov:2014ipa}. 
Although the cutoff scale is background-dependent and the perturbative unitarity turned out to hold during inflation~\cite{Bezrukov:2010jz}, we cannot give a reliable prediction on reheating because the energy scale of the spike-like behavior in the mass of the Goldstone mode
at the preheating stage in the pure Higgs inflation, 
recently discovered in Refs.~\cite{DeCross:2015uza,JinnoThesis,Ema:2016dny,Sfakianakis:2018lzf}, exceeds the cutoff scale at the reheating stage and the system enters the strong coupling regime.~\footnote{
See Ref.~\cite{Hamada:2020kuy} for the effect of higher dimensional operators on this phenomenon.
}
Thus, it is desirable to have a UV-extension of the Higgs inflation in order to provide further understanding. The extension with the $R^2$ term is the most straightforward way.\footnote{See Refs.~\cite{Giudice:2010ka,Barbon:2015fla,Lee:2018esk} for other  proposals of the UV-extensions of the Higgs inflation.} 
As is pointed out in Refs.~\cite{Salvio:2015kka,Netto:2015cba,Calmet:2016fsr,Liu:2018hno,Ghilencea:2018rqg}, a large quadratic curvature term in the mixed Higgs-$ R^2 $ model can emerge from the renormalization group running.  Also, Refs.~\cite{Ema:2019fdd,Ema:2020zvg} discuss the origin of $R^2$ from the viewpoint of scattering amplitude and the non-linear sigma model.

As shown in the early works, the prediction of the mixed Higgs-$R^2$ model on the power spectrum of primordial scalar (curvature) perturbations and its scale dependence well matches the WMAP and Planck observations~\cite{Akrami:2018odb}, similarly to its two single-field limits, i.e. the Higgs inflation and the $R^2$-inflation. 
The reason for that shown in Ref.~\cite{He:2018gyf} is the attractor behavior during inflation that allows an effective $ R^2 $-inflation description of this model. It has been argued that the Higgs- and $R^2$-inflation are distinguishable by precise observations thanks to the difference in their reheating temperatures~\cite{Bezrukov:2011gp}. 
Thus, identifying the reheating mechanism in the present model is essential to determine the ratio of mixing between its two limiting single-field counterparts and to distinguish it from them using observational data.

Despite the close relation with the Higgs inflation, the preheating mechanisms are surprisingly different in the mixed Higgs-$ R^2 $ model. 
In the pure Higgs inflation, the reheating is driven by post-inflationary oscillation of a scalar field, namely the SM Higgs field. 
In the earlier works, the resonant production of the transverse mode of weak gauge bosons has been identified as the dominant process of reheating~\cite{Bezrukov:2008ut,GarciaBellido:2008ab} (see also Ref.~\cite{Repond:2016sol}). 
Later, as mentioned above, a more efficient process has been pointed out, which involves the longitudinal modes of weak gauge bosons during the first stage of preheating.
The effective mass of the longitudinal modes is different from that of transverse modes and receives large contribution from a Higgs field velocity which causes high spikes in the mass and therefore may complete reheating quickly~\cite{Ema:2016dny}. 
However, the energy scales of these spikes are beyond the cutoff scale of the theory at that epoch, 
that decreases the reliability of its predictions. 
On the other hand, at the reheating epoch in the mixed Higgs-$R^2$ model, the single-field description no longer holds 
and multi-field (the Higgs field and the scalaron) dynamics is essential for the reheating. 
In the previous study by the present authors~\cite{He:2018mgb}, it has been found that 
the violent preheating mechanism in the pure Higgs inflation is largely weakened by the multi-field dynamics. 
The energy density of produced particles has been calculated analytically, and the result has shown the inefficiency of the spike preheating.
This occurs because the spikes get much milder in this model. 
Besides, the cutoff scale issue is perfectly avoided because this scale is raised up to $ M_{\rm pl} $.\footnote{Preheating process in similar models is studied in, e.g. Ref.~\cite{Fu:2019qqe}.}
But the dominant process for the completion of (p)reheating has not been identified in our previous study. 

Soon after the work~\cite{He:2018mgb}, Bezrukov {\it et. al.}~\cite{Bezrukov:2019ylq} showed that, with some special choices of model parameters in the Higgs-like regime (the definition of this regime is given in Sec.~\ref{Sec-2}), the preheating in this model can be completed by tachyonic instability of the (physical) Higgs field and the longitudinal modes of weak gauge bosons right after the first stage mentioned above. 
The shape of the two dimensional potential of the (physical) Higgs field $h$ and the scalaron $\varphi$ 
consists of the two potential valleys in $\varphi>0$ where inflation takes place and the potential hill between them at $h=0$.
If appropriate model parameters are chosen, the system can climb up the potential hill around $h=0$ 
during the oscillations of $\varphi$ around the origin.
As a result of this, dynamics of the background fields and of reheating as a whole becomes chaotic in the sense introduced in Ref.~\cite{Podolsky:2002qv}, see also Ref.~\cite{Jin:2004bf}. Also the Higgs field as well as the longitudinal modes of gauge bosons are tachyonic in this regime. Thus, tachyonic preheating can take place\footnote{Tachyonic preheating was first studied in theories with spontaneous symmetry breaking~\cite{Felder:2000hj,Felder:2001kt,Kofman:2001rb}. 
Tachyonic instability for the spectator field can also be induced by large field space curvature, see, e.g. Ref.~\cite{Iarygina:2020dwe}.}, and it is indeed found to be strong enough to complete preheating at some parameter points.
This possibility is interesting, but it is not quantitatively clear 
which choices of model parameters allow this tachyonic preheating and how much fine-tuning is needed among them.
The purpose of the present paper is to obtain deeper understanding about it. 

In this paper, we analyze this instability of the physical Higgs field with both analytical and numerical methods. 
We first find the conditions on the model parameters for the tachyonic effect to be prominent. 
We then analyze the efficiency of this effect and estimate the energy density of produced particles, so that we can estimate the necessary degree of fine-tuning to complete preheating by the tachyonic instability.
We show that a relatively weak fine-tuning is needed for the completion of preheating in the Higgs-like regime, while in the $R^2$-like regime severe fine-tuning is necessary.
Although we analyze the physical Higgs only, the effect on the longitudinal modes of the weak gauge bosons or the Nambu-Goldstone modes can be evaluated in a similar way. Their contribution to the preheating process is estimated to be comparable to that of the Higgs field (see Sec.~\ref{Sec-5}). 

Before proceeding to the main part, let us clarify some subtleties and limitations of the present study. 
In order to focus on the dynamics of the physical Higgs field, we adopt a simplified model, taking it as a real singlet field. This toy model apparently shows two distinct trajectories along the valleys of the scalar potential in the region with $\varphi>0$. 
This model therefore gives rise to domain walls along the Higgs direction during or after inflation as it stands. However, these domain walls have nothing to do with the static ones which appear in the field theory of a singlet scalar field with a double-well (Higgs-like) potential, they are temporary and disappear in the flat space-time limit when $\varphi\to 0$. Moreover, even during preheating, they disappear for the part of each oscillation of the scalaron $\varphi$ where it becomes negative. Note the realistic SM Higgs field does not yield any domain walls, too.
We do not take into account backreaction from produced particles to the background fields, either. 
We limit our study to the first scalaron oscillation since the tachyonic instability is expected to be the strongest source of particle production during this period, so backreaction can be neglected temporarily. 
Further, we say that {\it the preheating is completed} if the energy density of produced particles becomes comparable to the background energy density.  
Note that we will not investigate how the produced particles become thermalized and how the Universe enters the  radiation-dominated stage finally without late-time scalar field (re)domination. 
As pointed out in Ref.~\cite{Bezrukov:2019ylq}, even if the model parameters do not allow tachyonic preheating at the first scalaron oscillation,
the tachyonic instability can still be effective at subsequent oscillations. 
In such cases, backreaction of produced particles on the background scalar field dynamics can be important and the simple treatment in our present paper is not directly applicable. 
This is beyond the scope of the paper and we leave it for the future study.

The structure of this paper is as follows. In the next section, we briefly review the model and present the parameter values which realize the maximal tachyonic instability in both the Higgs- and $R^2$-like regimes. 
We then derive two simple relations between the model parameters which (approximately) hold in the two regimes respectively. 
In Sec.~\ref{Sec-3}, we analytically estimate the maximal efficiency of the tachyonic instability and see whether the energy transfer from the background fields to Higgs particles is strong enough to deprive the energy of coherent background field oscillations immediately.
Note that, strictly speaking, backreaction should be taken into account properly to determine if all the energy of the coherent field oscillations is transferred to particles. 
In Sec.~\ref{Sec-4}, we study the necessary degree of fine-tuning of the model parameters to complete preheating by this tachyonic effect with both numerical and semi-analytical methods.
We also discuss instability of the longitudinal modes of the gauge fields in Sec.~\ref{Sec-5}. 
Section~\ref{Sec-6} is devoted to the conclusions, discussion on some perspectives of this tachyonic instability, and the future directions of the study of the model.

\section{Mixed Higgs-\texorpdfstring{$ R^2 $}{Lg} model and tachyonic instability}
\label{Sec-2}

\subsection{Brief review of the mixed Higgs-\texorpdfstring{$ R^2 $}{Lg} model}

Let us start with a brief review of the mixed Higgs-$R^2$ model. 
The Lagrangian is originally given in the Jordan frame (the quantities defined there are denoted with a subscript~``J")~\cite{Wang:2017fuy,Ema:2017rqn,He:2018gyf,Gundhi:2018wyz}
\begin{align}
    S_{\mathrm J}
    &= \int\!\! d^4 x \sqrt{-g_{\mathrm J}} {\cal L}_\mathrm{J} \notag \\
    &=
    \int\!\! d^4 x \sqrt{-g_{\mathrm J}} 
    \left[ \left(\frac{M_{\mathrm{pl}}^2}{2} + \xi |{\mathcal H}|^2 \right)R_{\mathrm J} + \frac{M_{\mathrm{pl}}^2}{12 M^2}R_{\mathrm J}^2 
    - g_{\mathrm J}^{\mu\nu} \partial_\mu  {\mathcal H} \partial_\nu {\mathcal H}^\dagger - \lambda |{\mathcal H}|^4 \right] ~,  \label{jordanaction}
\end{align}
where $ R_{\mathrm J} $ is the Ricci scalar, $M_\mathrm{pl}$ is the reduced Planck mass, and $ {\mathcal H} $ is the SM Higgs. There are three model parameters, the scalaron mass $ M $, the non-minimal coupling $ \xi>0 $\footnote{The conformal coupling corresponds to the case $ \xi =-1/6 $. The case with $ \xi <0 $ is studied e.g. in \cite{Pi:2017gih}.}, and the self-coupling of Higgs $ \lambda $. We consider the case where the SM Higgs vacuum is stable and keeps non-critical, $\lambda = {\cal O}(10^{-2})$, up to the inflationary scale, which allows us to take $\lambda$ to be a constant in our analysis~\cite{Bezrukov:2009db,Bezrukov:2014ipa}.
Here we take the sign convention $g_{\mu\nu} = (-,+,+,+)$ for the metric at the flat limit. 
For our primary purpose to present the inflationary background dynamics as well as to see the tachyonic instability in the physical Higgs field
at the post-inflationary epoch, it is sufficient to focus on the real scalar field $h$ with $ {\mathcal H} =(0, h)/\sqrt{2} $, and to neglect the Nambu-Goldstone modes and the SM SU(2)$_L \times$ U(1)$_Y$ gauge fields. 
The contributions from these fields are discussed in Sec.~\ref{Sec-5}. 

By a conformal transformation \cite{Maeda:1987xf,Maeda:1988ab}
\begin{align}
    g_{\mathrm{E}\mu \nu} (x)=e^{\sqrt{\frac{2}{3}}\frac{\varphi(x)}{\mpl}} {g_\mathrm{J}}_{\mu \nu}(x)
    \equiv e^{\alpha\varphi(x)}{g_\mathrm{J}}_{\mu \nu}(x) ~,
\end{align}
where we define the ``scalaron'' field $\varphi$ as
\begin{align}
    \label{def:psi}\sqrt{\frac{2}{3}}\frac{\varphi}{\mpl}\equiv \ln \lmk\frac{2}{\mpl^2}\left|\frac{\partial {\cal L}_\mathrm{J}}{\partial {R_\mathrm{J}}}\right|\rmk ~,
\end{align}
we transform the action into the Einstein frame where our analysis is to be done. The resulting new action is expressed in terms of two scalar fields as
\begin{align}
    \label{einsteinaction}
    S_{\mathrm{E}} 
    &=
    \int d^4 x \sqrt{-g_{\mathrm E}} 
    \left[\frac{M_{\mathrm{pl}}^2}{2} R_{\mathrm{E}} -\frac{1}{2} g_{\mathrm E}^{\mu\nu}\partial_\mu \varphi \partial_\nu \varphi 
    - \frac{1}{2}e^{-\alpha \varphi} g^{\mu\nu}_{\mathrm E} \partial_\mu h \partial_\nu h -U(\varphi, h) \right] ~,
    \\ 
    \label{potential}
    U(\varphi, h) 
    &= 
    \frac{\lambda}{4} e^{-2 \alpha \varphi}h^4  + \frac{3}{4} M_{\mathrm{pl}}^2 M^2 
    \left[1 - \left(1 + \frac{\xi}{M_{\mathrm{pl}}^2} h^2\right)e^{-\alpha \varphi} \right]^2 ~. 
\end{align}
The subscript ``E" used to denote the quantities defined in the Einstein frame will be omitted afterwards for convenience. The physical results are the same in both frames even though the superficial values of some quantities might be different. 
One remarkable issue in this model is that the scalar sector is weakly coupled~\cite{Ema:2017rqn,Gorbunov:2018llf} if 
\begin{equation} \label{weakm}
    M \lesssim \sqrt{\frac{4 \pi}{3}} \frac{M_\mathrm{pl}}{\xi}, 
\end{equation}
and the strong coupling scale comes from the gravity sector, $\Lambda_c \simeq M_\mathrm{pl}$, 
whereas in the pure Higgs inflation the system is strongly coupled at much smaller scales $E > \Lambda \simeq M_\mathrm{pl}/\xi$. 

\begin{figure}[h]
	\centering
	\includegraphics[width=.5\textwidth]{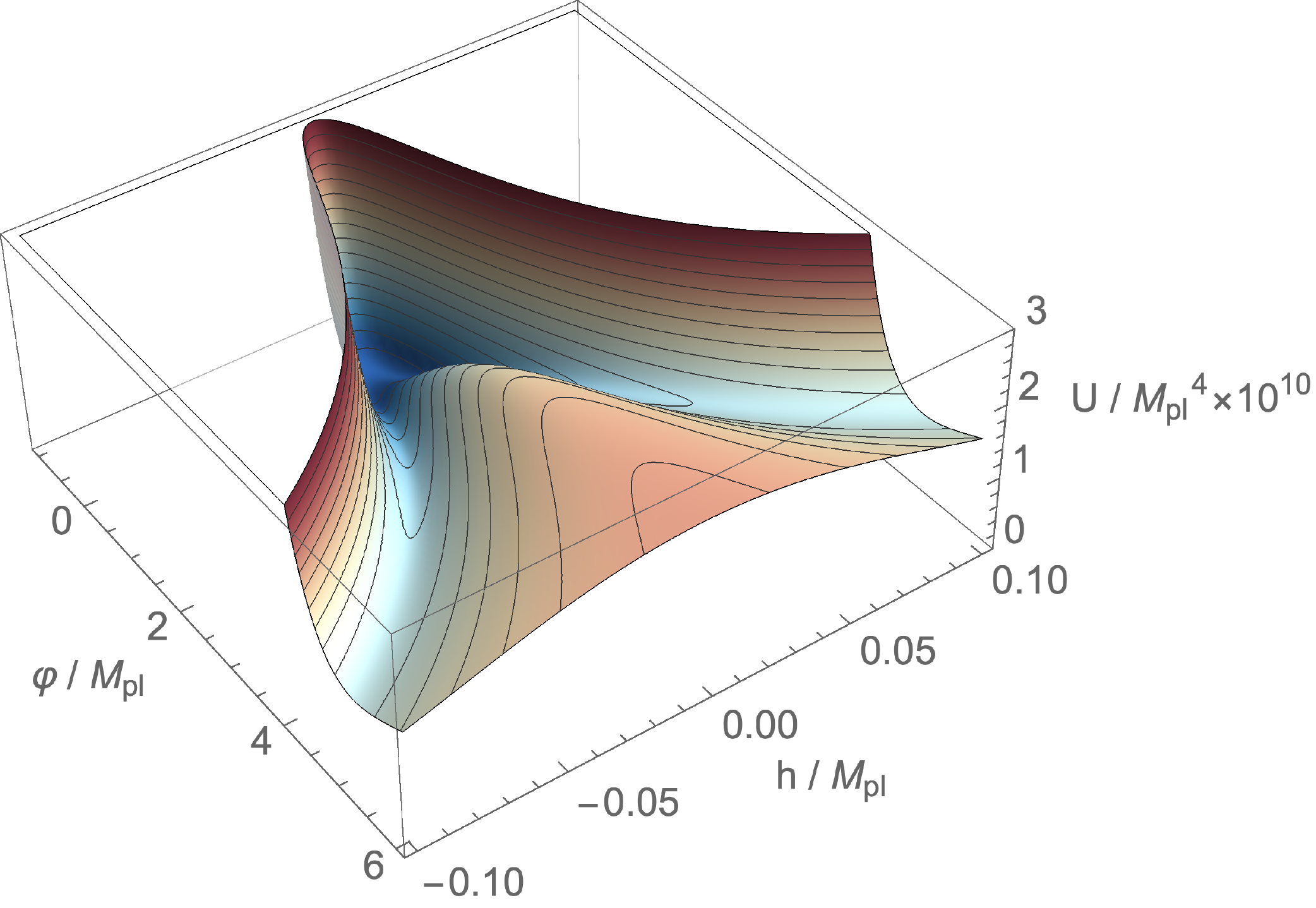}
	\caption{An example for the shape of the potential is shown. The parameters are chosen $\lambda = 0.01$ and  $ \xi=3000 $ with 
	satisfying Eq.~\eqref{eq-relation-M-xi}.}
	\label{fig-potential}
\end{figure}

The inflationary dynamics of this model is driven by two scalar fields with mixed potential and kinetic terms with non-flat field space. 
As is clarified in the previous studies, for sufficiently large quartic coupling $\lambda$,\footnote{For an extremely small quartic coupling $\lambda$, the attractor behavior disappears, and this model can exhibit a multi-field dynamics~\cite{Wang:2017fuy,Gundhi:2018wyz}.} there is an attractor behavior in this model due to the shape of the potential with two ``valleys''.  
With arbitrary initial conditions in large $ \varphi $ regime (which is required for inflation to last long enough), the scalar fields will always fall into one of the valleys in the two-field potential and then drive slow-roll inflation\footnote{The real trajectories do not exactly coincide with the valleys of the potential due to effects from the curvature of the field space and turning of the trajectories. The effects are small, though.}. 
The scalaron and the Higgs field on the valley satisfy the following relation 
\begin{align}\label{eq-valley}
	h^2
	&=
	\frac{e^{\alpha \varphi} - 1}{\displaystyle \frac{\xi}{\mpl^2} + \frac{\lambda}{3\xi M^2}} ~. 
\end{align}
These valleys can also be found by omitting the kinetic term of the Higgs field in the Jordan frame, 
so that the Higgs field can be integrated out by taking $ h^2=\xi R_{\rm J}/\lambda $. 
Figure~\ref{fig-potential} shows the typical shape of the potential. 
We can see the two valleys as well as the hill between them in the scalar potential. 

Putting the valley condition (Eq.~\eqref{eq-valley})  back into the action in the Einstein frame, 
one obtains the effective single-field inflation model whose potential is equivalent to the one of the $R^2$ inflation~\cite{He:2018gyf}, with an effective mass of scalaron given by 
\begin{align}\label{eq-relation-M-xi}
	\tilde{M}^2
	&\equiv 
	\frac{M^2}{\displaystyle 1 + \frac{3\xi^2M^2}{\lambda \mpl^2}} ~. 
\end{align}
The value of ${\tilde M}$
is determined as ${\tilde M} = M_c=\pi\sqrt{24\mathcal{P}_{\mathcal{R}}}/N_{\rm inf} \simeq 1.3 \times 10^{-5} \times (54/N_{\rm inf}) M_\mathrm{pl} $~\cite{Starobinsky:1983zz,JinnoThesis,He:2018gyf} by the observed amplitude of the power spectrum of primordial curvature perturbation, $ \mathcal{P}_{\mathcal{R}} \simeq 2.1\times 10^{-9} $ on the pivot scale $ k_0 = 0.05~ \rm{Mpc}^{-1} $~\cite{Akrami:2018odb}. As a result, we are allowed to define the energy density of inflation as $ U_{\rm inf} \equiv 3 M^2_{\rm pl} \tilde{M}^2 /4 $. Besides, the constraint above (Eq.~\eqref{eq-relation-M-xi}) can be rewritten in a simpler form as \cite{He:2018mgb}
\begin{align}\label{eq-observation-constraint}
    \frac{\xi^2}{\xi_c^2} + \frac{M_c^2}{M^2} = 1 
\end{align}
with the two critical values being defined as
\begin{align}
	\xi_c \simeq 4.4 \times 10^3 \times \sqrt{\frac{\lambda}{0.01}}~, 
	~~~~
	M_c \simeq 1.3 \times 10^{-5} M_\mathrm{pl}. 
\end{align}
Here we have fixed $ \lambda=0.01 $ and the e-fold number when the pivot scale left the horizon to be 54 for definiteness. 
The constraint reduces the number of independent parameters ($M$ and $\xi$) to only one. 
Hereafter, we will not take $M$ and $\xi$ as independent parameters 
and take one of them or $\theta$ defined in Eq.~\eqref{eq-theta-reparametrization} to characterize the model, depending on cases for convenience. 

The parameter space is divided into three regions. 
The first region is the non-perturbative regime 
with $ \xi \gsim  \xi_s  \equiv 1/\sqrt{1/\xi_c^2+3M_c^2/4 \pi M_\mathrm{pl}^2} \simeq 4.4 
\times 10^{3} $, where $\xi_s$ is slightly smaller than $\xi_c$ and is 
determined by the combination of Eqs.~\eqref{weakm} and \eqref{eq-observation-constraint}.  
We cannot give any reliable predictions in this parameter region 
and we will not consider it any further. 
The second is the Higgs-like regime with $ \xi_c/\sqrt{2} \simeq  3.1\times 10^3 $, 
where the former inequality comes from the condition $\xi/\xi_c > M_c/M$. 
The third is  the $ R^2 $-like regime with $ 0 \leq \xi \lsim  \xi_c/\sqrt{2} $. 

In the post-inflationary epoch, 
the effective single field description is no longer viable and the system obeys the equations of motion for the scalaron and the Higgs field as well as the Friedmann equation, 
\begin{align}
\ddot{\varphi}+3H\dot{\varphi}+\frac{\alpha}{2}e^{-\alpha \varphi}\dot{h}^2+\frac{\partial U}{\partial \varphi}
= 0, \label{eq:eom1}
&\\
\ddot{h}+3H\dot{h}- \alpha \dot{\varphi}\dot{h}+e^{\alpha \varphi}\frac{\partial U}{\partial h} 
= 0, \label{eq:eom2}
&\\
{\rm with}~~~~~~~3M^2_\mathrm{pl}H^2=\frac{1}{2}\dot{\varphi}^2+\frac{1}{2}e^{-\alpha \varphi}\dot{h}^2 + U(\varphi,h),  \label{eq:eom3}
&
\end{align}
where we take the Friedmann background with $a$ and $H={\dot a}/a$ being the scale factor and the Hubble parameter, respectively. 
Note that the field values become small, $\alpha|\varphi|\ll 1$, 
and hence in this epoch, it is convenient to work on the simplified potential (see Eq.~\eqref{potential}), 
\begin{align}
    \label{potential-1}
    U(\varphi, h) =  \frac{\lambda}{4} h^4 + \frac{1}{2} M^2
    \left(\varphi - \frac{\xi h^2}{\alpha M_{\mathrm{pl}}^2}\right)^2 ~.
\end{align}
The feature of the potential around the origin can also be seen in Fig.~\ref{fig-potential}. 
For $\varphi>0$, the potential is described by the valleys that are continued from the inflationary trajectory 
and the hill at $h=0$ between them. At the valleys, the Higgs field and the scalaron are related as
\begin{align}
    h^2=h_\mathrm{valley}^2=\frac{3\alpha\xi M^2\varphi}{\displaystyle \lambda + \frac{3\xi^2M^2}{M_{\mathrm{pl}}^2}}. 
\end{align}
The potential along the valley is then given by 
\begin{align}
	U(\varphi,h_\mathrm{valley})=U_\mathrm{valley}(\varphi)=\frac{M^2\varphi^2}{2\left(\displaystyle 1 + \frac{3\xi^2M^2}{\lambda M_{\mathrm{pl}}^2}\right)} ~,
\end{align}
with the effective mass for the valley and the ``isocurvature'' mass (or the Higgs mass) being
\begin{align}\label{eq-scalaron-Higgs-mass-positive-phi-valley}
	 m^2_{\varphi,\mathrm{valley}} \equiv\frac{\partial^2 U_\mathrm{valley}}{\partial \varphi^2} = \frac{M^2}{\displaystyle 1 + \frac{3\xi^2M^2}{\lambda M_{\mathrm{pl}}^2}}~, ~~m_h^2\equiv\frac{\partial^2 U}
	{\partial h^2}=6\alpha\xi M^2\varphi ~.
\end{align}
On the other hand, the potential along the hill is evaluated as  
\begin{equation}
    	U(\varphi,h=0)=U_\mathrm{hill}(\varphi)=\frac{M^2}{2} \varphi^2 ~, 
\end{equation}
and the mass as 
\begin{equation} \label{eq-scalaron-Higgs-mass-positive-phi-hill}
  m^2_{\varphi} = M^2, ~ ~m_h^2 = -3 \alpha \xi M^2 \varphi,  
\end{equation}
where the latter is the source of the tachyonic instability as we will see. 
For $\varphi<0$, there is only one minimum along $h=0$, where the potential is 
\begin{align}
	U(\varphi,h=0)=U_0(\varphi)=\frac{1}{2} M^2\varphi^2,  
\end{align}
which leads to the mass of the scalaron and the Higgs field around the valley
\begin{align}\label{eq-scalaron-Higgs-mass-negative-phi}
    m_{\varphi}^2= M^2~,~~m_h^2=3\alpha\xi M^2|\varphi| ~.
\end{align}
These expressions are useful for the analytic estimate of the dynamics in the preheating stage, 
which will be studied in the subsequent sections. 
Note that in both cases $ m_h^2 \gg m_{\varphi}^2 $ holds unless $ \xi \lsim 1 $. 

\subsection{Parameters with the longest duration of tachyonic instability}
\label{Sec-2-2}

After inflation, the scalaron and the Higgs field oscillate around the global minimum $ (\varphi=0, h=0) $ in the two-dimensional potential and start to reheat the Universe. 
They first go down the potential valley in $\varphi>0$ driving inflation.
They then climb up the valley in $\varphi<0$ with small but rapid oscillations in the Higgs direction, and come back to the global minimum. 
This is a bifurcation point in the Higgs direction that gives the distinctive feature in the evolution of the system, as seen in Fig.~\ref{fig-potential}. 
Without fine-tuning in the model parameters (which we refer to ``usual" cases below), 
they again climb up one of the valleys in $\varphi>0$ with small and rapid oscillation in the Higgs direction, which was pointed out in Ref.~\cite{MinxiThesis,He:2018mgb}.\footnote{Similar behaviors are also found in other models \cite{Fan:2019udt}.} 
One also sees that the scalaron is only slowly oscillating around the origin. 
This hierarchy in the period of oscillations comes from the mass hierarchy in the scalaron and the Higgs field, as explained in the previous subsection. 
Remarkably, as pointed out in Ref.~\cite{Bezrukov:2019ylq}, if the model parameter is chosen carefully, it is possible for the Higgs field to keep $h\simeq 0$ when the scalaron comes back to the region $\varphi>0$, i.e. the inflaton climbs up the hill in $ (\varphi >0, h\simeq 0) $. 
The Higgs field feels a tachyonic instability during this period, which we study in more detail in the subsequent sections. Note also that the Toda-Brumer necessary criterion \cite{TODA1974335,BRUMER1974391} for the appearance of classical global dynamical chaos in the background field evolution $Det\,||\frac{\partial^2 U}{\partial\phi\partial h}||<0$ is fulfilled in this regime.

\begin{figure}[h]
	\centering
	\begin{subfigure}
		\centering
		\includegraphics[width=.48\textwidth]{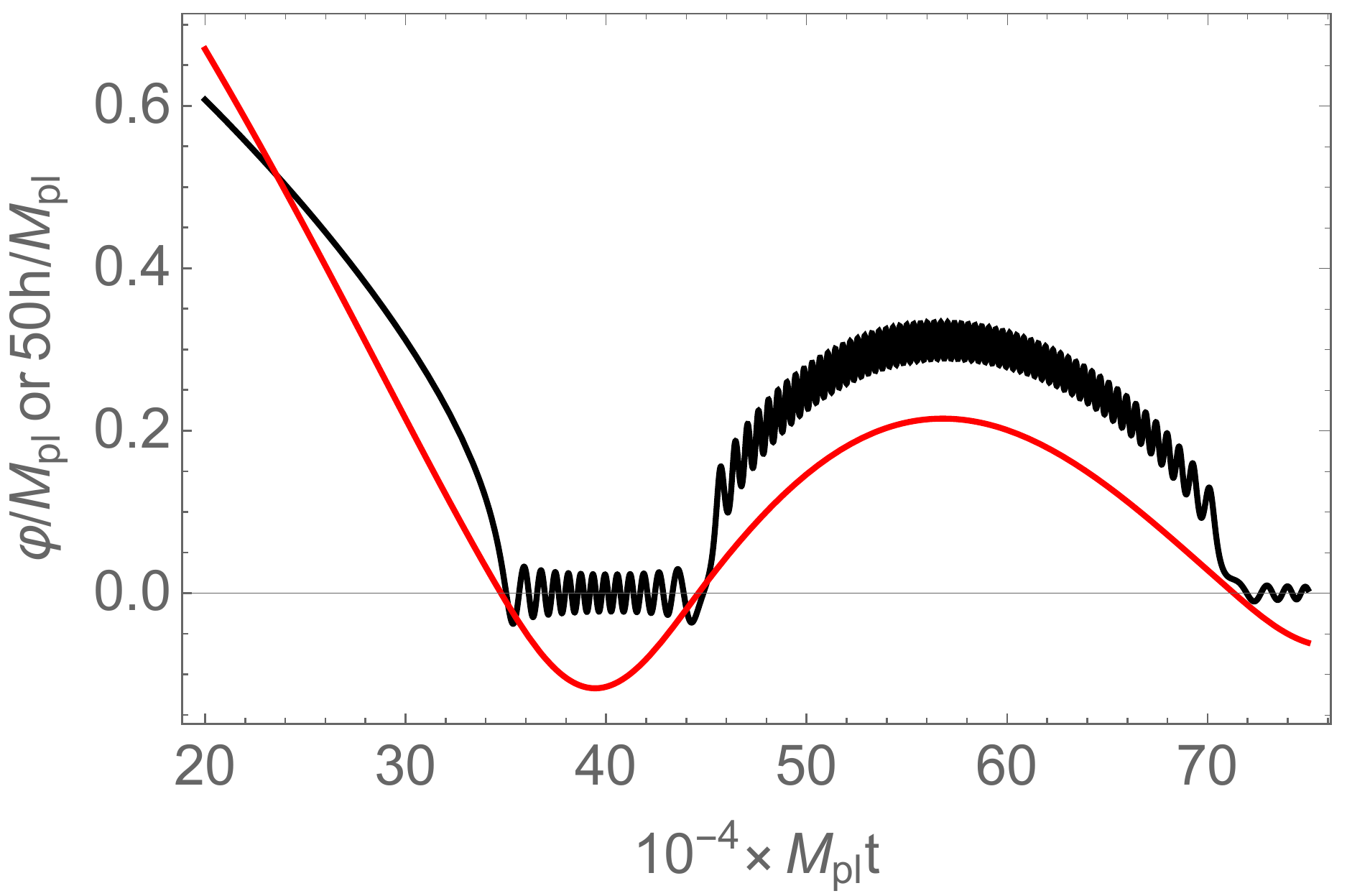}
	\end{subfigure}
	\begin{subfigure}
		\centering
		\includegraphics[width=.48\textwidth]{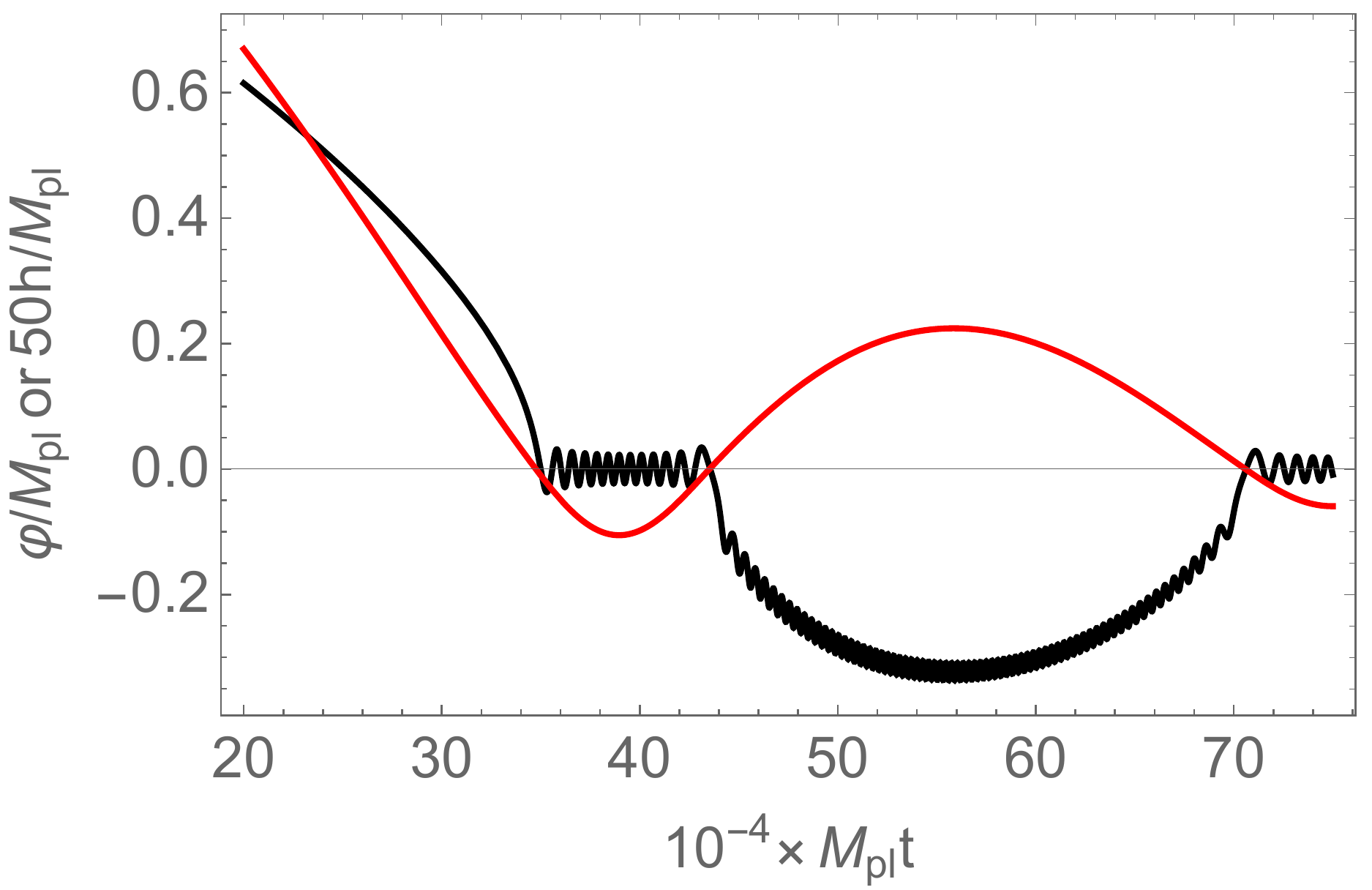}
	\end{subfigure}
	\begin{subfigure}
		\centering
		\includegraphics[width=.48\textwidth]{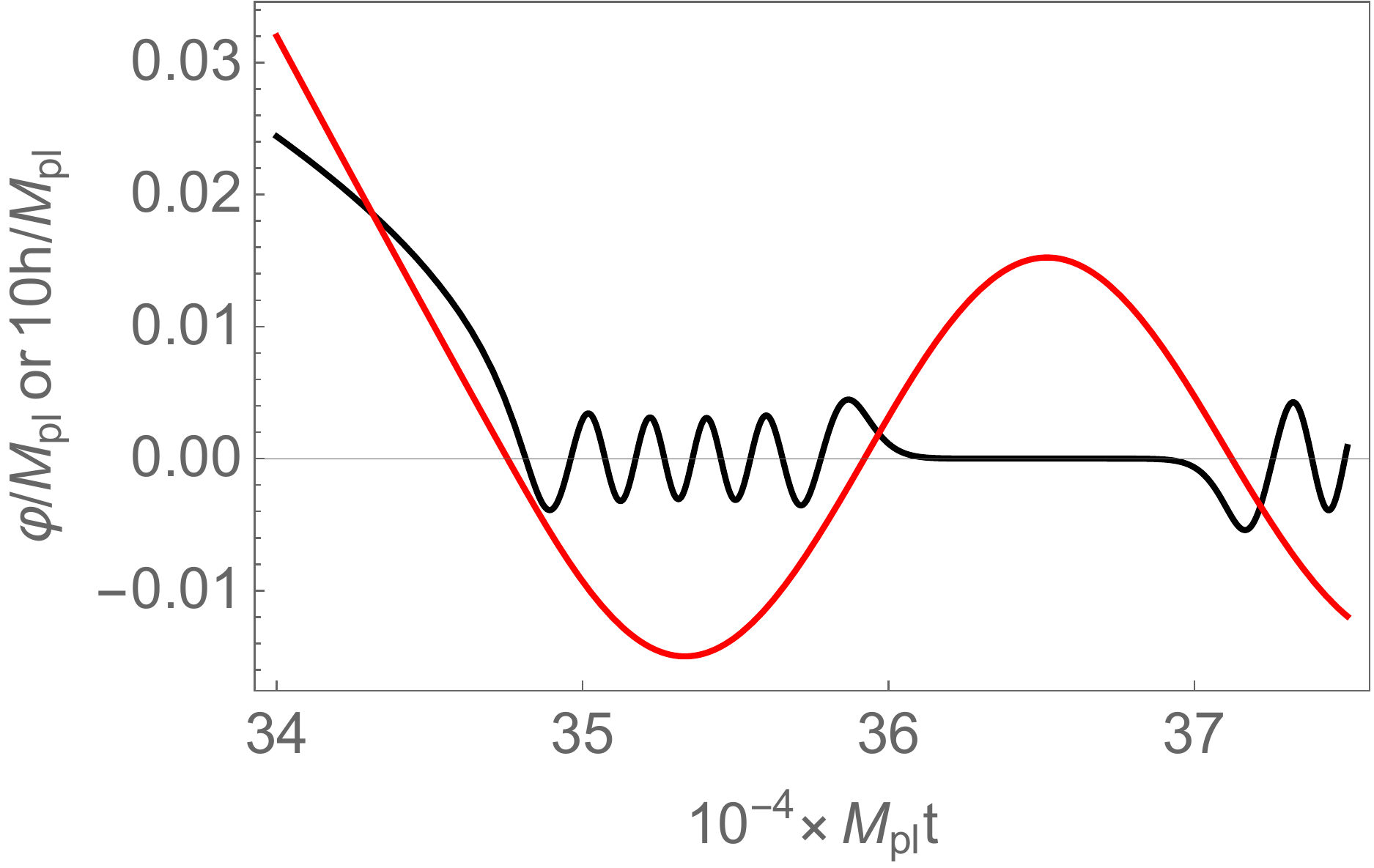}
	\end{subfigure}
	\begin{subfigure}
		\centering
		\includegraphics[width=.48\textwidth]{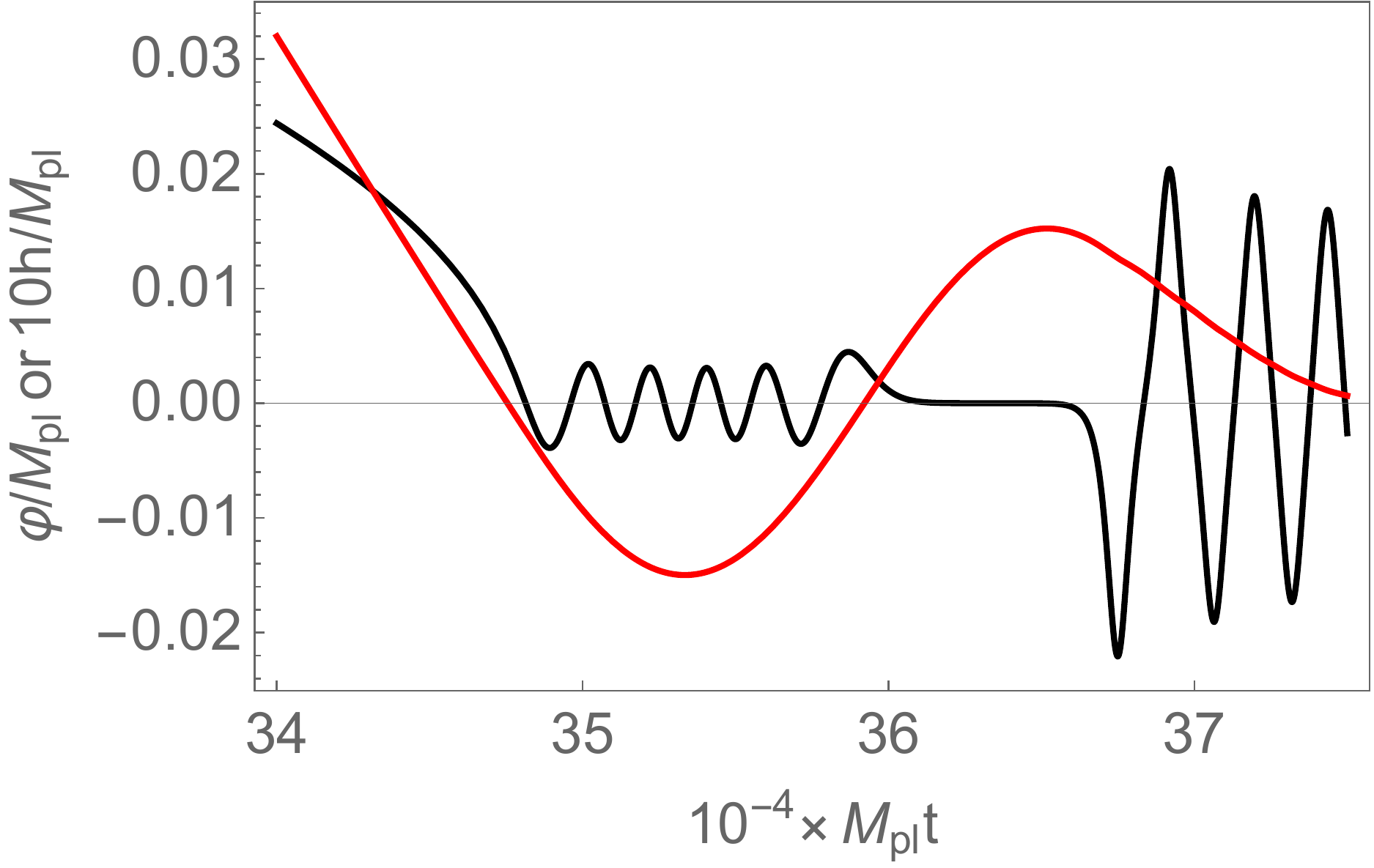}
	\end{subfigure}
	\caption{Evolution of the scalaron and the Higgs field in the post-inflationary epoch. 
	Red and black lines represent the evolution of the scalaron and the Higgs field, respectively. 
	The parameters are chosen as $\lambda=0.01$ and using the observational value of the parameter $\tilde M$ given below Eq.~\eqref{eq-relation-M-xi}. 
	The values of $\xi$ is chosen as follows. 
	Upper left: $ \xi=4000 $. The Higgs enters the valley with positive Higgs field value. Upper right: $ \xi=4100 $. The Higgs enters the valley with negative Higgs field value. Lower left: $ \xi\simeq \xi_{N=10} $ where $\xi_N $ is defined later in this section. Higgs stays on the ``hill" during the whole period of $ \varphi >0 $. Lower right: $ \xi=\xi_{N=10} (1-\epsilon) $ where $\epsilon \sim \mathcal{O}(10^{-12})$. Higgs exits the tachyonic regime halfway. 
	Note that in order to obtain such fine-tuned evolution numerically, we need 16-digit or more precision of $\xi$, 
	but the values themselves do not have any meanings since other parameters that are determined by observations 
	do not have such a high precision. }
	\label{fig-Higgs-oscillation}
\end{figure}

\begin{figure}[h]
	\centering
	\begin{subfigure}
		\centering
		\includegraphics[width=.45\textwidth]{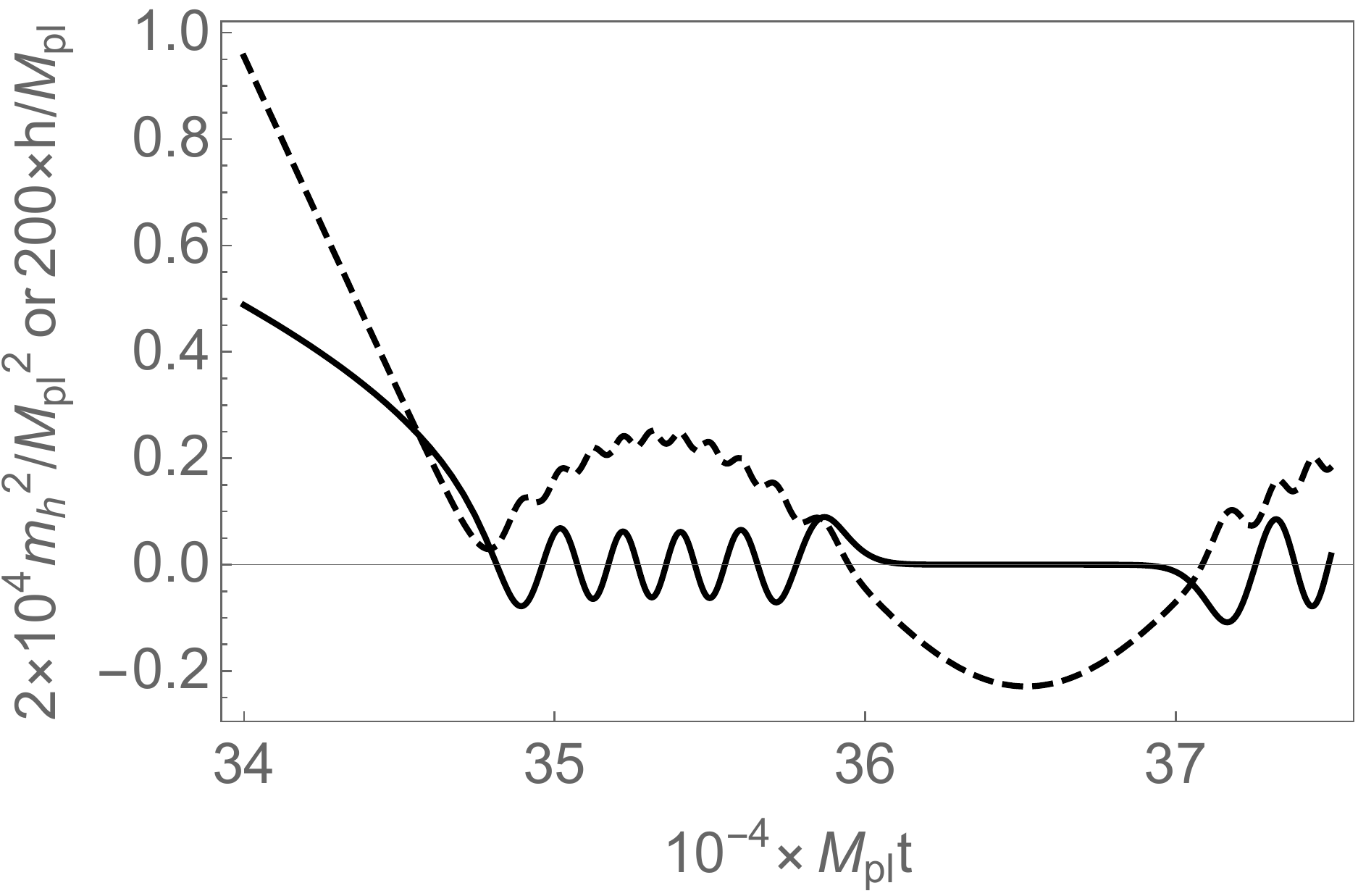}
	\end{subfigure}
	\begin{subfigure}
		\centering
		\includegraphics[width=.45\textwidth]{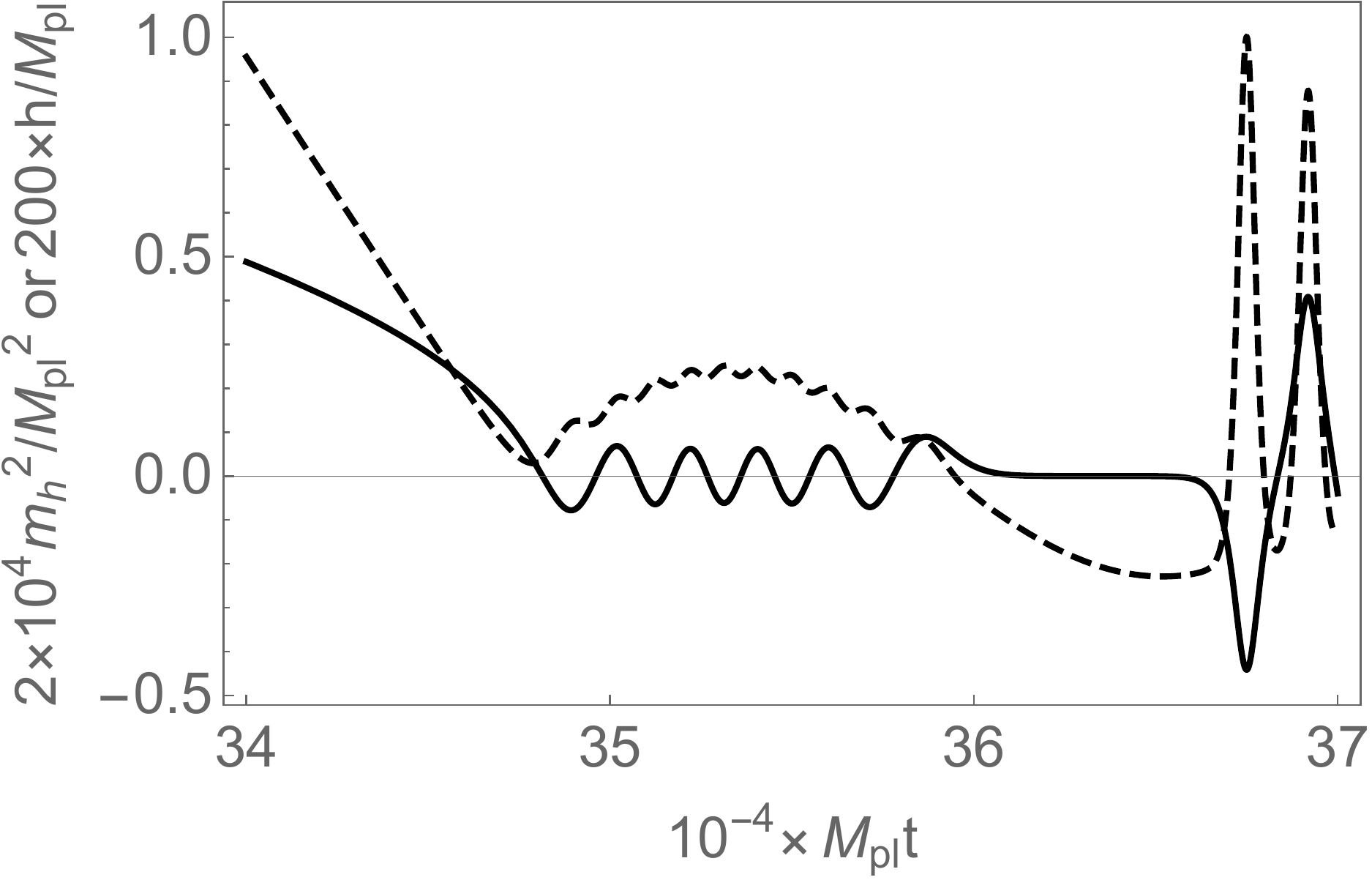}
	\end{subfigure}
	\caption{Time evolution of the  mass squared of the Higgs field along with the Higgs field evolution. 
	The dashed and solid lines represent the evolution of $ m^2_h $ for Higgs field defined as $ m^2_h \equiv \partial^2U/\partial h^2 $ and the value of the Higgs field, respectively. The parameters are the same to the lower panels of Fig.~\ref{fig-Higgs-oscillation}.
	The left panel shows the case where the Higgs field stays on the ``hill" during the whole period of $ \varphi >0 $ 
	and the right panels shows the case where the Higgs field exits the tachyonic regime halfway.}
	\label{fig-Higgs-oscillation-mass}
\end{figure}

Figures~\ref{fig-Higgs-oscillation} show four examples of the field evolution obtained numerically.
Here we solved the full homogeneous equations of motion and the Friedmann equation from the action \eqref{einsteinaction} and \eqref{potential} without any small field approximations. 
The upper panels show a realization of the ``usual'' cases. 
They clearly show that even when the inflation takes place with $h>0$, the Higgs field can develop both positive and negative field values at the first oscillation of the scalaron during reheating. 
The lower panels show the hill-climbing case with a fine-tuning. 
In the most fine-tuned case, it is possible to climb down to the origin $h=0$ without falling into the potential valleys during the period when $\varphi>0$. 
In a less fine-tuned case, due to the tachyonic instability of the Higgs field, 
the system falls into one of the potential valleys and the Higgs field oscillates with a relatively large amplitude. In Fig.~\ref{fig-Higgs-oscillation-mass}, we show the evolution of the squared mass of the Higgs field in the cases of the lower panels in Fig.~\ref{fig-Higgs-oscillation}. They explicitly show that the Higgs field is tachyonic during the hill-climbing epoch. 

Let us now investigate the condition for the hill-climbing in more detail. 
The idea can be understood simply as follows. 
During the first scalaron oscillation in the regime $\varphi<0$, the Higgs field evolution can be characterized by the number of its oscillations around the point $h=0$ and by its final phase when the scalaron comes back to the origin, which we denote $t=t_{\mathrm{enter},0}$. 
Once we change the parameter $\xi$ or $M$ continuously, the phase also changes continuously. 
We have exact hill-climbing when the phase is close to a multiple of $\pi$.
The parameter is therefore related to the number of oscillations of the Higgs field during the period when $\varphi<0$.

The frequency (or mass) of the Higgs oscillation and the time duration of the scalaron being in the regime $ \varphi <0 $, 
which is closely related to the width of spikes in the mass of the longitudinal mode of the gauge bosons~\cite{He:2018mgb}, can be estimated with Eq.~\eqref{eq-scalaron-Higgs-mass-negative-phi} as  
\begin{align}
    \Delta t_{\rm osc} \simeq \frac{1}{2} \frac{2\pi}{M} =\frac{\pi}{M} ~,~~ m^2_h & = 3\alpha\xi M^2 | \varphi (t)|. 
\end{align}
Since the kinetic energy vanishes at the highest point at $\varphi_1(<0)$ after the first zero-crossing, the field value is evaluated as~\cite{He:2018mgb}
\begin{align}\label{eq-energy-at-phi1}
    U_0(\varphi_1)= C^2_1 U_{\rm inf} 
\end{align}
which leads to
\begin{align}\label{eq-varphi1-xi}
    \alpha | \varphi_1 | =C_1 \frac{M_c}{M} ~.
\end{align}
Here $C_1$ is a numerical constant of order of the unity
that parameterizes the energy loss after inflation, and we found $C_1 \simeq 0.25$ numerically~\cite{He:2018mgb}\footnote{$C_1$ here corresponds to $C_m$ in Ref.~\cite{He:2018mgb}. We change the subscript so that the notation in this paper looks more consistent.}. 
Strictly speaking, the mass of the Higgs field during this period is time-dependent, 
but here we introduce another numerical constant of order of the unity, $C_{m_h}$, 
to estimate the averaged Higgs mass during the part of its oscillation when $\phi(t)<0$ as 
\begin{align}
    \overline{m_h} = M(3\xi\alpha)^{1/2} \left< |\phi(t)|^{1/2} \right> = C_{m_h}M(3\xi\alpha|\phi_1|)^{1/2}~.
\end{align}
With this averaged Higgs mass and using Eq.~\eqref{eq-varphi1-xi}, we evaluate the accumulated phase of the Higgs field oscillation during this period as 
$\overline{m_h} \Delta t_{\rm osc} = \pi C_{m_h} \sqrt{3 C_1\xi M_c/M}$. 
Then we can conjecture that the exact hill-climbing of the scalaron after first oscillation happens when 
\begin{align}
    N\pi-\Delta \phi    =& \pi C_{m_h} \sqrt{3 C_1\xi \frac{M_c}{M}} ~, 
\end{align}
is satisfied, 
where $N$ is an integer that represents of the number of half-oscillations of the Higgs field 
and $\Delta \phi < \pi$ is a small phase shift needed for the exact hill-climbing.  
Using Eq.~\eqref{eq-relation-M-xi}, we can write the above equation as 
\begin{align}
    N =& \frac{\Delta \phi}{\pi}  + C_{m_h} \sqrt{3C_1 \xi} \left( 1-\frac{3}{\lambda} \frac{M_c^2}{M^2_{\rm pl}} \xi^2 \right)^{1/4} \label{eq-exact-N-xi}\\
    =& \frac{\Delta \phi}{\pi}  +(3\lambda)^{1/4} C_{m_h} \sqrt{C_1 \frac{M_{\rm pl}}{M}} \left( 1-\frac{M_c^2}{M^2} \right)^{1/4} \label{eq-exact-N-M}
\end{align}
in terms of $ \xi $ and $ M $, respectively. We can solve $ \xi $ in terms of $ N $ as 
\begin{align}
    \xi_N = \sqrt{\frac{\lambda}{6}} \frac{M_{\rm pl}}{M_c} \left[ 1 \pm \sqrt{1- \frac{4}{3\lambda C_{m_h}^4 C_1^2} \frac{M_c^2}{M_{\rm pl}^2} \left( N- \frac{\Delta\phi}{\pi} \right)^4} \right]^{1/2}
\end{align}
where ``$+$" corresponds to the Higgs-like regime while ``$ - $" the $R^2$-like regime. 
Introducing a new parameter $\theta$ as
\begin{equation} \label{eq-theta-reparametrization}
    \cos \theta \equiv \frac{\xi}{\xi_c}, \quad \sin \theta \equiv \frac{M_c}{M}, \quad \left(0<\theta<\frac{\pi}{2}\right)
\end{equation}
so that it satisfies Eq.~\eqref{eq-observation-constraint}, we can rewrite Eq.~\eqref{eq-exact-N-xi} as
\begin{equation}
    N = \frac{\Delta \phi}{\pi}  +C_{m_h} \sqrt{\frac{3 C_1 \xi_c}{2} \sin 2\theta} . \label{eq-exact-N-theta}
\end{equation}
Here $0<\theta<\pi/4$ covers the Higgs-like regime (including the non-perturbative regime) whereas $\pi/4<\theta<\pi/2$ covers the $R^2$-like regime. 
With this parameterization, we can treat two regimes on the same footing. 
From Eq.~\eqref{eq-exact-N-theta} we can estimate the maximal number of the Higgs half-oscillations in the $\varphi<0$ period as
\begin{equation}
    N_\mathrm{max} \simeq C_{m_h} \sqrt{\frac{3 C_1 \xi_c}{2}} \simeq 40 C_{m_h}, 
\end{equation}
in either the Higgs-like and $R^2$-like regime, that also corresponds to the number of the exact hill-climbing cases in each regime. 

\begin{figure}[h]
	\centering
	\includegraphics[width=.5\textwidth]{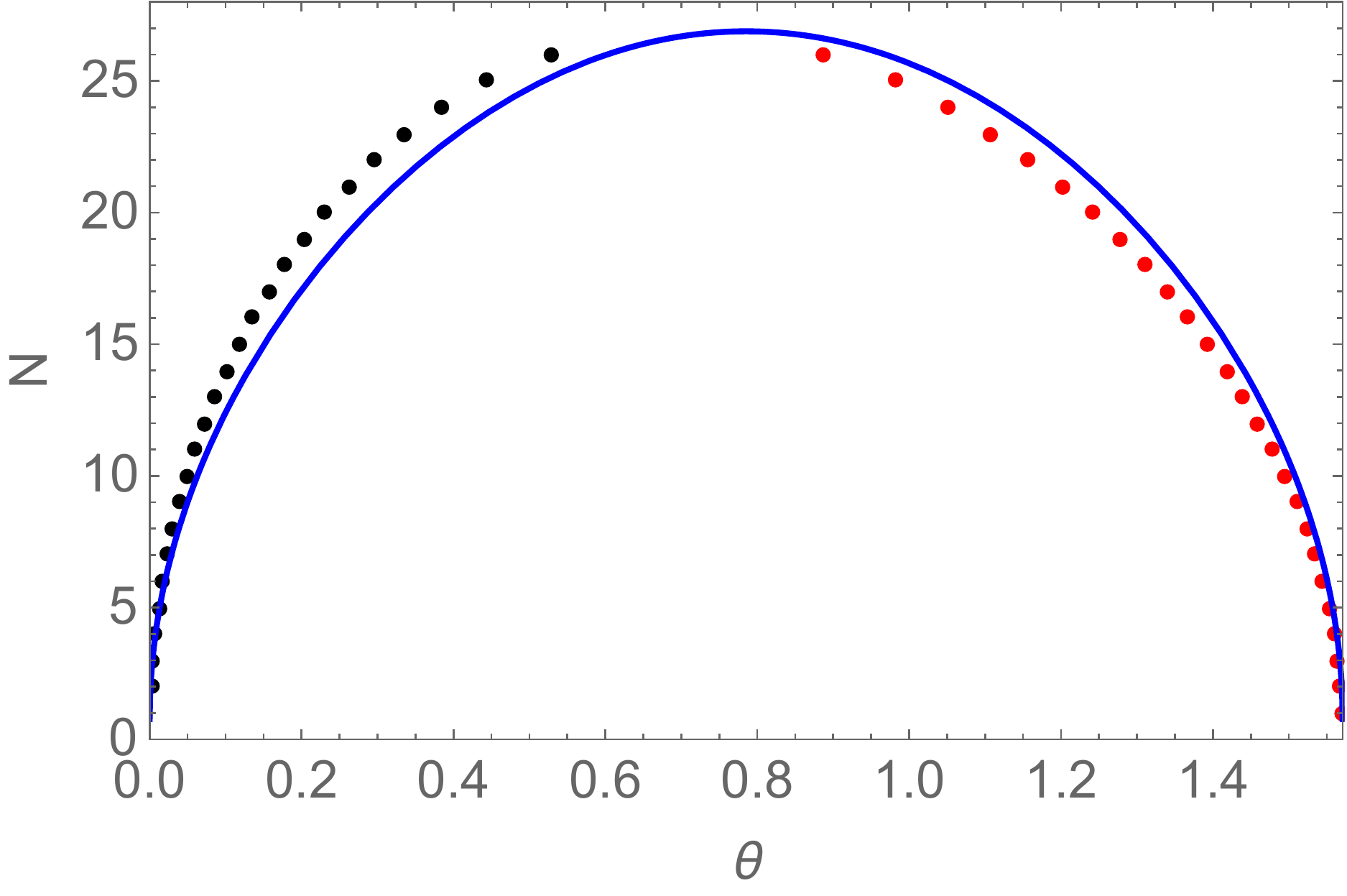}
	\caption{The value, $N$,  of the number of the Higgs field half-oscillations during the $\varphi<0$ period for each exact hill-climbing case. The black dots are for Branch 1 and the red for Branch 2. All the parameters $\theta$ that realize the exact hill-climbing have their corresponding $N$.  The blue line represents the relation Eq.~\eqref{eq-exact-N-theta} with $C_{m_h}=0.64$ and $\Delta \phi = 2.4$. }
	\label{fig-N-theta}
\end{figure}

With the help of these relations, we numerically find all the parameters that realize the exact hill-climbing.\footnote{
In the Higgs-like regime $ M \gg M_c $, from the relation \eqref{eq-exact-N-M} one obtains
\begin{align}\label{eq-square-law-B1M}
    \frac{M_N}{M_{N+1}} \approx \frac{(N+1)^2}{N^2},
\end{align}
while in the $ R^2 $-like regime $ \xi \ll \xi_c $ the relation \eqref{eq-exact-N-xi} leads to 
\begin{align}\label{eq-square-law-B2xi}
    \frac{\xi_N}{\xi_{N+1}} \approx  \frac{N^2}{(N+1)^2} .
\end{align}
Once we find a parameter value that realizes the exact hill-climbing, we can use these simple relations to find easily the next parameter value that realizes the exact hill-climbing. 
} 
Figure~\ref{fig-N-theta} shows the parameter $\theta$ and the corresponding number of Higgs half-oscillations during the $\varphi<0$ period for the exact hill-climbing. 
We can see that Eq.~\eqref{eq-exact-N-theta} gives a qualitatively good explanation on the appearance of the hill-climbing behavior. 
We find that $N_\mathrm{max}=26$, which is explained by taking $C_{m_h}\simeq 0.64$.

Note that would we neglect the Universe expansion and coupling between the scalaron and the Higgs field, so that scalaron oscillations become harmonic and those of the Higgs field quasi-harmonic with the adiabatically changing frequency $m_h(t)\propto |\phi(t)|^{1/2}$, we would obtain
$C_{m_h}=\pi^{-1}\int_0^{\pi}\sqrt{\sin x}\,dx= 4\pi^{-1/2}\,\Gamma(3/4)/\Gamma(1/4)\approx 0.7628$ - not a bad approximation. However, the $\theta$ parameters for the exact hill-climbing in the Figure~\ref{fig-N-theta} are calculated with much greater precision.

In Fig.~\ref{fig-N-theta}, we also see that the behavior of parameters for the exact hill-climbing are classified into two branches corresponding to the Higgs-like and $R^2$-like regimes. 
We call the former Branch 1 and the latter Branch 2. 
Hereafter we label the parameters that realize the exact hill-climbing with the corresponding number of the Higgs field half-oscillation during $\varphi<0$ as $\theta_N^i$ (or $M_N^i$, $\xi_N^i$) with $i=1,2$ being the label for each branch. 
It is also convenient to describe the system with $ M_N^1$ (or $ (M^1_N)^{-1} $) for  Branch 1 and $\xi_N^2$ for Branch 2. 
In Branch 1, as $(M^1_N)^{-1} $ increases, $N$ goes from 1 to 26, 
and in Branch 2, as $\xi_N^2$ increases, $N$ goes from 1 to 26. 
In the former case, the small number of Higgs oscillations for a smaller $(M^1_N)^{-1}$
is due to the small $ \Delta t_{\rm osc} $. 
In the latter case, the small number of Higgs oscillations for a smaller $\xi_N^2$  results from the small $ m_h $ even with large $ \Delta t_{\rm osc} $. 
Note that $N <8$ in Branch 1 corresponds to the non-perturbative regime and we do not consider it any further. 
Now we have found all the parameter values that realize the exact hill-climbing, which leads to the full tachyonic instability of the Higgs after the second zero-crossing of $ \varphi $. In the next section, we estimate the efficiency of such instability and see whether it can complete preheating or not.

\section{Maximal efficiency of the tachyonic instability}
\label{Sec-3}

In this section, we calculate the efficiency of the tachyonic instability in the case of exact hill-climbing
to study the possibility of completing preheating by this phenomenon. 
We here choose the parameters so that exact hill-climbing occurs, that is, the scalaron climbs up the potential hill and comes down to the origin keeping $h\simeq 0$.
The caveat is that we do not take backreaction into account, and hence our results do not reflect the exact evolution of the system but just give an estimate on whether or not the energy density of the produced particles is comparable to that of the background.
In particular, we do not see whether the energy density of the homogeneous field oscillation completely disappears or whether the produced particles are thermalized. 
We simply assume that, once the energy density of the produced particles becomes comparable to that of the whole system, backreaction is significant enough to drain the remaining oscillation energy into particles and the system approaches to the thermal state.

\subsection{Equation of motion for the Higgs perturbation}

We investigate the evolution of fluctuations around 
the homogeneous background of the scalaron and the Higgs field. 
The linearized equations of motion for the fluctuation of scalaron $ \delta \varphi_k $ and Higgs $ \delta h_k $ in the Fourier space are given by 
\begin{align}
    \ddot{\delta \varphi}_k +3H \dot{\delta \varphi}_k +\left( k^2_p+\frac{\partial^2U}{\partial \varphi^2}- \frac{\alpha^2}{2} e^{-\alpha\varphi}\dot{h}^2 \right)\delta\varphi_k&=-\alpha e^{-\alpha \varphi}\dot{h} \dot{\delta h}_k -\frac{\partial^2 U}{\partial \varphi \partial h} \delta h_k \label{eq-exact-eom-delta-scalaron},  \\
    \ddot{\delta h}_k +\left(3H- \alpha \dot{\varphi}\right)\dot{\delta h}_k +\left( k^2_p +e^{\alpha\varphi}\frac{\partial^2U}{\partial h^2} \right)\delta h_k &=\alpha \dot{h} \dot{\delta \varphi}_k -e^{\alpha \varphi} \left( \alpha \frac{\partial U}{\partial h}+ \frac{\partial^2 U}{\partial h\partial \varphi} \right)\delta \varphi_k  \label{eq-exact-eom-delta-higgs},
\end{align}
where 
$ k_p\equiv k/a $, and the dot denotes the derivative with respect to the physical time $t$. 
Here we adopt the spatially-flat gauge for the metric. 
In the following we focus on the physical Higgs field (and scalaron), and do not consider the phase direction of the Higgs field 
or the gauge fields. We discuss them in Sec.~\ref{Sec-5}. 
Redefining 
new fields $ \tilde{\delta \varphi}_k \equiv a^{3/2} \delta \varphi_k $ and $ \tilde{\delta h}_k \equiv a^{3/2} \delta h_k $, we remove the Hubble friction terms 
\begin{align}
    \ddot{\tilde{\delta \varphi}}_k +&\left( k^2_p +\frac{\partial^2 U}{\partial \varphi^2} -\frac{\alpha^2}{2} e^{-\alpha\varphi} \dot{h}^2 -\frac{9}{4}H^2 -\frac{3}{2} \dot{H} \right) \tilde{\delta \varphi}_k \notag \\
    & = -\alpha e^{-\alpha \varphi} \dot{h} \dot{\tilde{\delta h}}_k +\left( \frac{3\alpha}{2} H e^{-\alpha\varphi} \dot{h} -\frac{\partial^2U}{\partial\varphi \partial h} \right) \tilde{\delta h}_k \label{eq-exact-eom-delta-scalaron-redefine}, \\
    \ddot{\tilde{\delta h}}_k +&\left( k^2_p +e^{\alpha \varphi} \frac{\partial^2 U}{\partial h^2} +\frac{3\alpha}{2} H\dot{\varphi} -\frac{9}{4}H^2 -\frac{3}{2} \dot{H} \right) \tilde{\delta h}_k \nonumber \\ &=\alpha \left( \dot{\varphi} \dot{\tilde{\delta h}}_k+ \dot{h} \dot{\tilde{\delta \varphi}}_k \right) 
    -\left( \frac{3\alpha}{2} H\dot{h} + \alpha e^{\alpha \varphi} \frac{\partial U}{\partial h} +e^{\alpha \varphi} \frac{\partial^2U}{\partial h\partial \varphi} \right) \tilde{\delta \varphi}_k \label{eq-exact-eom-delta-higgs-redefine},
\end{align}
where the remaining friction terms and the coupling between the two fluctuations are placed on the right hand side. 
These terms come from the non-canonical kinetic term and off-diagonal elements in the mass matrix. 

Although we can in principle solve the full mode equations~\eqref{eq-exact-eom-delta-scalaron-redefine} and \eqref{eq-exact-eom-delta-higgs-redefine} numerically,
it costs too much to investigate the whole parameter space.
We here instead evaluate the parameter dependence of the efficiency of the tachyonic preheating analytically.
Note that, since $\partial^2 U/\partial \varphi^2 \simeq  M^2 > H^2, |{\dot H}|, {\dot h}^2/M_\mathrm{pl}^2$, the scalaron field does not experience tachyonic instability. 
Moreover, ${\dot h} = \partial U/\partial h = \partial^2 U/\partial \varphi \partial h = 0$ holds in the exact hill-climbing case, and hence the mixing between the scalaron and Higgs fluctuations vanishes. 
Therefore, the scalaron fluctuations are never amplified during this period. 
For this reason, we hereafter focus only on the Higgs fluctuation $ \tilde{\delta h}_k $. 

The mode equation for the Higgs fluctuations~\eqref{eq-exact-eom-delta-higgs-redefine} 
can be further simplified. 
Since the mixing between Higgs and scalaron fluctuations vanishes as mentioned above, we can omit the terms involving the scalaron fluctuations. 
Also, after inflation, the background value of $ \varphi /M_{\rm pl} $ (and $ h/M_{\rm pl} $) are at most $ \mathcal{O}(10^{-1}) $. Consequently, we can take $\exp(\alpha \varphi) \simeq 1$ and the non-canonical part of the Higgs kinetic term in Eq.~\eqref{einsteinaction} does not play an important role.
We also see that the tachyonic mass directly coming from the potential dominates over the other ``mass terms", and hence the latter are neglected in the following arguments. 
Taking the inequalities $|\partial^2 U/\partial h^2| \gg M^2 > {\tilde M}^2 \gtrsim  H^2, |{\dot H}|$ into account,
the Hubble induced terms are negligibly small compared to the tachyonic mass of the Higgs field. 
Moreover, from the Friedmann equation
\begin{align}
	3M^2_{\rm pl} H^2 =\frac{1}{2} \dot{\varphi}^2 +\frac{1}{2} e^{-\alpha \varphi} \dot{h}^2 + U(\varphi, h) ~,
\end{align}
we obtain $ H\dot{\varphi} < M_{\rm pl}H^2 $. 
In particular, since the energy density of the system is  
mostly stored in the potential in the late hill-climbing period (when particle production is the most efficient), we find ${\dot \varphi}^2 \ll U(\varphi,h) \simeq 3 H^2 M_\mathrm{pl}^2$. 
Thus, the mass term is well approximated as $k_p^2 + \partial^2 U/\partial h^2 = \omega_{h,k}^2$. 
We also see that the friction term is smaller than $M$, $|\alpha {\dot \varphi}| < H < M$. 
Since the time scale of our interest is $M^{-1}$, the friction terms can never be important.  
In summary, we conclude that the mode equation for the Higgs fluctuations are simplified into the one with a time-dependent tachyonic mass as 
\begin{align}
	\ddot{\tilde{\delta h}}_k + \omega^2_{h,k}  \tilde{\delta h}_k \approx 0 ~, \quad \omega^2_{h,k} \equiv k^2_p +m^2_{h} = k_p^2 + \frac{\partial ^2 U}{\partial h^2},  \quad m^2_{h}  = -3 \alpha \xi M^2 \varphi(t),  \label{eq-approx-eom-delta-higgs-redefine}
\end{align}
We investigate tachyonic particle production with this equation in the following. 

\subsection{Tachyonic Higgs mass for the exact hill-climbing}

Next, we examine the duration of the exact hill-climbing period and the time evolution of the Higgs mass squared during this period analytically. 
In the exact hill-climbing case, the background field evolution in $\varphi>0$ can be characterized as follows. 
First, the scalaron crosses zero from the region $\varphi<0$ at $t=t_{\mathrm{enter},0}$. 
Then the scalaron climbs up the potential hill in $\varphi>0$ with the Higgs field $h \simeq 0$. 
Once the scalaron reaches $\varphi = \varphi_2$, it reverses its direction and starts to go down the potential hill, still keeping $h\simeq 0$. 
Finally, the scalaron crosses zero again at $t=t_{\mathrm{exit},0}$. 
We evaluate the duration of this period $\Delta t \equiv t_{\mathrm{exit},0} - t_{\mathrm{enter},0}$ and the tachyonic mass of the Higgs field during this.
The amount of particle production is estimated in the next subsection using the result presented here.

The scalaron field value at the highest point can be evaluated in the same way as $\varphi_1(<0)$. 
Since the kinetic energy vanishes at the highest point $\varphi_2(>0)$, we can write 
\begin{align}
	U_0(\varphi_2) &= C^2_2 U_0(\varphi_1) = C_2^2 C_1^2 U_\mathrm{inf}, 
\end{align}
where a new numerical parameter $ C_2 $ is introduced to take into account the energy loss during hill-climbing. 
Then we obtain
\begin{align} \label{eq-scalaron-highest-value}
    \alpha\varphi_2= C_2 C_1 \frac{M_c}{M_N^i} ~.
\end{align}
The evolution of the scalaron is governed by the potential $ U(\varphi, h=0) \simeq (M_N^i)^2 \varphi^2/2$. 
Thus, the duration of the hill-climbing can be evaluated by half period 
of the oscillation around the origin, 
$ \Delta t\simeq \pi/M_N^i $, which is the same as $ \Delta t_{\rm osc} $ in the previous section. 
The time evolution of the scalaron during the exact hill-climbing is approximated as
\begin{equation} \label{eq-scalaron-exact-hill-climbing}
    \varphi(t) \simeq  \varphi_2 \sin \left(\pi \frac{t-t_{\mathrm{enter},0}}{\Delta t} \right) = \frac{C_2 C_1}{\alpha} \frac{M_c}{M_N^i} \sin \left(\pi \frac{t-t_{\mathrm{enter},0}}{\Delta t} \right). 
\end{equation}
Combining this with Eq.~\eqref{eq-scalaron-Higgs-mass-positive-phi-hill}, we obtain the tachyonic Higgs mass squared as 
\begin{equation} \label{eq-higgs-tachyonic-mass-squared-phi2}
    m_h^2 \simeq -m_{h, \mathrm{max}}^2 \sin \left(\pi \frac{t-t_{\mathrm{enter},0}}{\Delta t} \right), 
\end{equation}
where $m_{h, \mathrm{max}} $ is the absolute value of the Higgs mass at $\varphi_2$, 
\begin{align}
    m^2_{h,\mathrm{max}}= \left|\frac{\partial^2 U}{\partial h^2} \right\rvert_{\varphi=\varphi_2, h=0} =& 3C_2 C_1  \xi_N^i M_N^i M_c  = 3 C_2 C_1 \xi_c M_c^2 \cot \theta_N^i \label{eq-exact-tachyonic-mass} \\
    \simeq & \left\{ \begin{array}{ll}
    \sqrt{3\lambda} C_2 C_1 M_{\rm pl} M_N^1 & \text{for Branch 1}, \\[0.3cm]
    3 C_2 C_1 M_c^2\xi_N^2 & \text{for Branch 2}. 
    \end{array} \right.   \label{eq-approx-tachyonic-mass-large-xi}
\end{align}
We see that $m_{h,\mathrm{max}}^2 / M_N^{i2} = (3/2) C_2 C_1 \xi_c \sin 2 \theta_N^i$. 
From Eq.~\eqref{eq-exact-N-theta}, we conclude that the absolute value of the Higgs mass at $\varphi_2$ is larger than (or at least comparable to) the scalaron mass for the exact hill-climbing case. 
This is the source of efficient particle production through the tachyonic instability.

\begin{figure}[h]
	\centering
	\includegraphics[width=.6\textwidth]{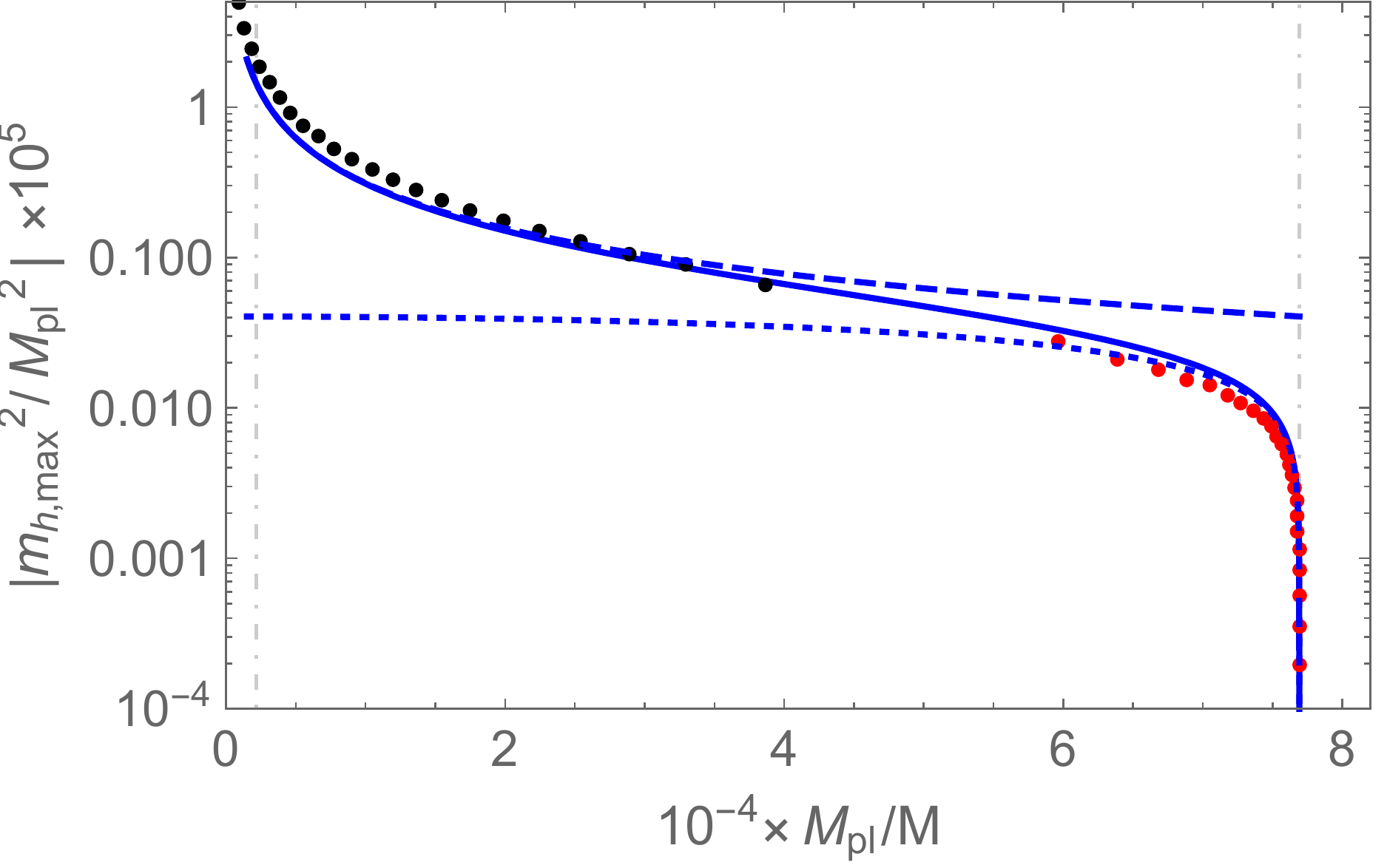}
	\caption{Absolute value of the Higgs mass squared at the highest point $\varphi=\varphi_2$ for the exact hill-climbing case. 
	The black dots are numerical results for each exact hill-climbing case $ M = M_N^1 $ and the red for $ M = M_N^2 $. 
	The solid line shows Eq.~\eqref{eq-exact-tachyonic-mass} with $ C_1=0.25 $ and $ C_2=0.72 $. 
	The dashed and dotted line show the asymptotic formula for Branch 1 and 2 in Eq.~\eqref{eq-approx-tachyonic-mass-large-xi}, respectively. 
	The left dot-dashed boundary is the unitary bound while the right is $ \xi=0 $. }
	\label{fig-tachyonic-mass-comparison}
\end{figure}

By solving the full background equations of motion numerically, as far as the numerical precision allows, 
we estimated the scalaron field value at the highest point $\varphi_2$ for each exact hill-climbing case. 
We confirmed that our analytic estimate of the highest point of the scalaron \eqref{eq-scalaron-highest-value} as well as its time evolution \eqref{eq-scalaron-exact-hill-climbing} works well. 
Figure~\ref{fig-tachyonic-mass-comparison} shows that the absolute value of the Higgs mass squared at $\varphi_2$ obtained from numerical calculation is well fit by taking $C_2 \simeq 0.72$ in Eq.~\eqref{eq-exact-tachyonic-mass}. Note that this value is even closer to the rough estimate $C_{m_h}\approx 0.7628$ obtained in Sec.~\ref{Sec-2-2} neglecting the Universe expansion and coupling between the scalaron and the Higgs field than to the numerical value of $C_{m_h}$ itself.

\subsection{Particle production through the tachyonic instability}

We now investigate the production of the Higgs particles through the tachyonic instability during exact hill-climbing. 
Since the tachyonic mass squared changes from 0 to $- m_{h,\mathrm{max}}^2 = -3 C_2 C_1 \xi_c M_c^2 \cot \theta_N^i$, 
modes with the wave number $k/a<m_{h,\mathrm{max}}$ feels the instability, and the number of particles for such modes is exponentially amplified. 
Assuming that particle production is not efficient when $\omega^2_{h,k}$ is positive,\footnote{
\label{footnote:tachyonic_vs_parametric}
Of course, particle production from parametric resonance can happen even when $\omega^2_{h,k}$ is positive while the Higgs field is oscillating.
However, since it takes $\gtrsim {\mathcal O}(100)$ oscillations for parametric resonance to amplify the particles as suggested from the pure Higgs case~\cite{Bezrukov:2008ut,GarciaBellido:2008ab} (note that these studies do not include the spike contribution as mentioned in Sec.~\ref{Sec-1}, though), we expect that the tachyonic instability is stronger if it takes place.
Note that the purpose of the present study is to identify if the tachyonic instability alone can complete preheating. 
Thus, we do not take other particle production channels into account.
} 
we can estimate the comoving 
occupation number of Higgs particles $n_k$ with $k$ being the comoving momentum, at the time when the scalaron comes back to the origin as (see Eq.~\eqref{eq-tachyonic-nk-approx})
\begin{equation}\label{eq-analytic-occupation-number}
    n_k (t_{\mathrm{exit},0}) \simeq \left|\exp(\Omega_k)-\exp(-\Omega_k)/4\right|^2 \simeq \exp(2 \Omega_k), \quad \Omega_k \equiv \int_{t_\mathrm{enter}(k)}^{t_\mathrm{exit}(k)} dt |\omega_{h,k}(t)|, 
\end{equation}
where $t_\mathrm{enter}(k)$ and $t_\mathrm{exit}(k)$ are the time when $\omega_{k,h}$ crosses zero and the system enters and exits the tachyonic regime, respectively. 
This estimate is consistent with the expectation from the simple WKB approximation, $n_k \simeq \exp(2 \Omega_k)$. 
See App.~\ref{appendix-A} for a detailed derivation of this formula with the violation of adiabaticity taken into account. 

The comoving energy density of the produced particles at $t=t_{\mathrm{exit},0}$ is given by 
\begin{align}\label{eq-tachyonic-rho-definition}
	\rho_{\delta h} (t_{\mathrm{exit},0}) &=\int \frac{d^3k}{(2\pi)^3 } \omega_{h,k}(t_{\mathrm{exit},0}) n_k(t_{\mathrm{exit},0}) = \int \frac{d^3k}{(2\pi)^3} \frac{k}{a(t_{\mathrm{exit},0})} \left|\exp(\Omega_k)-\exp(-\Omega_k)/4\right|^2.
\end{align}
As a rough estimate, we focus on the typical mode $k_t/a(t_{\mathrm{exit},0}) \simeq m_{h,\mathrm{max}}/2$. Then we can approximate as $\Omega_{k_t} \sim m_{h,\mathrm{max}} \Delta t$ (by omitting the time dependence of $m_h^2$ and offset $k^2$ in $\omega_{h,k}^2$) and $\rho_{\delta h} \sim m_{h,\mathrm{max}}^4 \exp(2 \Omega_{k_t})/32 \pi^2 \sim m_{h,\mathrm{max}}^4 \exp(2 m_{h,\mathrm{max}} \Delta t)/32 \pi^2 $. Substituting for $\theta_N^i$ using $m_{h,\mathrm{max}} \Delta t \simeq 2 \pi N$ and Eqs.~\eqref{eq-exact-N-theta} and \eqref{eq-exact-tachyonic-mass}, we have
\begin{align} \label{eq-rho-delta-h-analytic}
    \rho_{\delta h} (t_{\mathrm{exit},0}) &\sim \frac{9  \xi_c^2 M_c^4 \cot^2 \theta_N^i}{32\pi^2} \exp\left(\pi \sqrt{6  \xi_c \sin 2 \theta_N^i} \right) \notag \\[0.1cm]
    & \sim  \frac{\exp\left(2 \pi  N \right)}{32\pi^2} \times \left\{\begin{array}{ll} 9\lambda^2  (M_\mathrm{pl}/N)^4 & \text{for Branch 1}\\[0.2cm] (N {\tilde M})^4 & \text{for Branch 2} \end{array}\right. , 
\end{align}
where in the second approximation we omitted $\Delta \phi$ and took $C_{m_h}=C_1=C_2=1$ for simplicity. 
We see that the exponential amplification is larger for larger $N$ for both branches but the resultant particle production is more effective for Branch 1 due to larger $m_{h,\mathrm{max}}$. 
This is consistent with our naive expectation that the tachyonic preheating is more effective in the Higgs-like regime.
We also see that the tachyonic amplification of Higgs fluctuation can happen even in relatively deep $R^2$-like regime. 
This is because the scalaron mass is smaller for the $R^2$-like regime, which leads to a longer duration of exact hill-climbing and efficient particle production against the smaller tachyonic Higgs mass.

\begin{figure}[h]
	\centering
	\includegraphics[width=.6\textwidth]{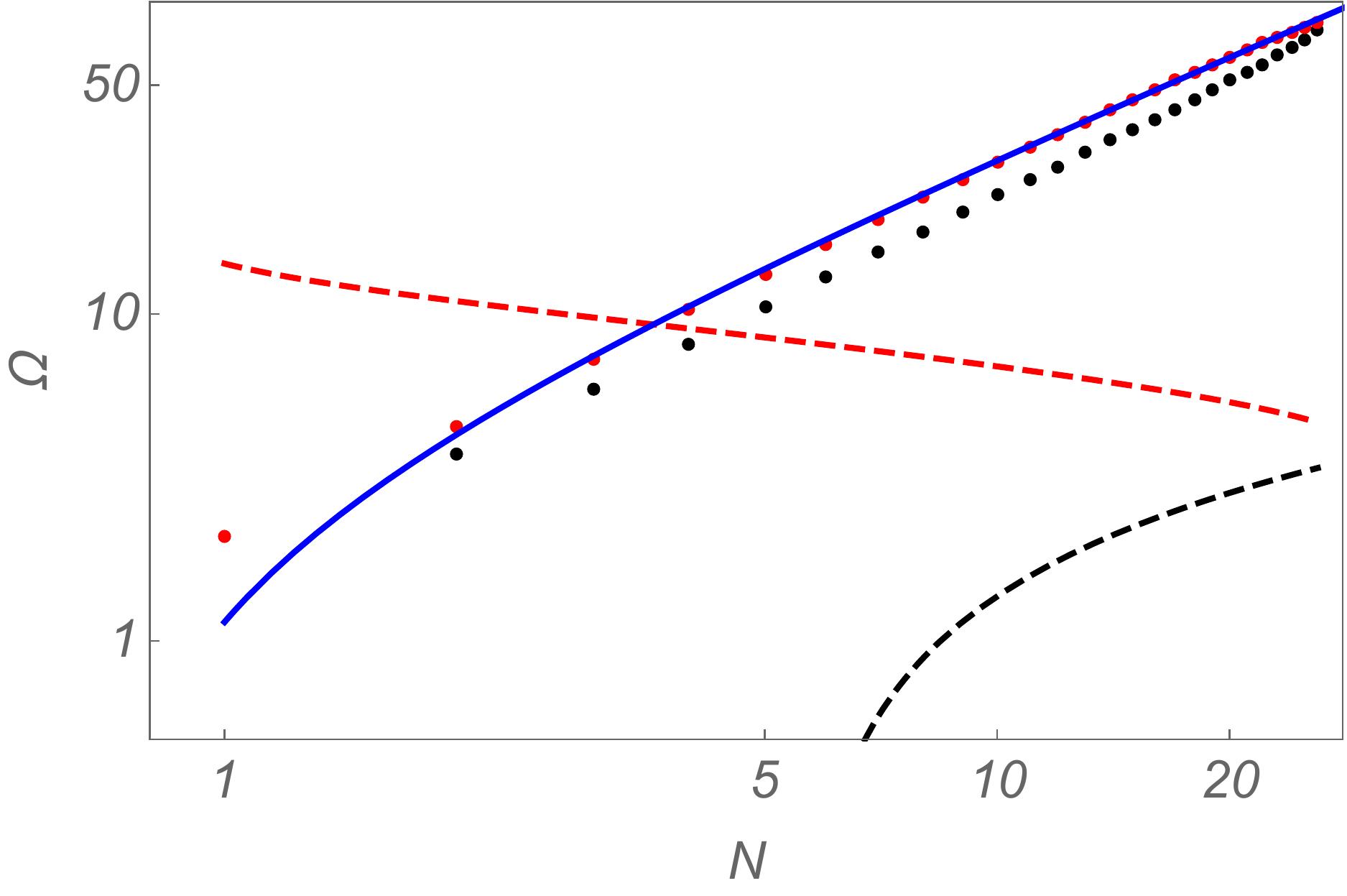}
	\caption{Exponential amplification factor $\Omega$ for each exact hill-climbing case.
	The numerical results (the black dots for Branch 1 and the red for Branch 2) are fitted well with $\Omega \simeq \pi N -2$ (blue line), which is consistent with Eq.~\eqref{eq-rho-delta-h-analytic}. 
	The black and red dashed lines represent $\ln (16 \pi^2 \rho_\mathrm{tot}/m_{h,\mathrm{max}}^4)$	
	for Branch~1  and Branch~2, respectively. 
	Tachyonic instability is effective enough to complete preheating for any $N$ in Branch~1, whereas it is not effective for $N\leq 4$ in Branch~2.}
	\label{fig-omegak}
\end{figure}

In order to see if the tachyonic preheating is strong enough to complete preheating, we compare the energy density of the produced particles with the background energy density, 
\begin{equation}
\rho_\mathrm{tot} \simeq C_2^2 C_1^2 U_\mathrm{inf} = \frac{3 C_2^2 C_1^2}{4} {\tilde M}^2 M_\mathrm{pl}^2.  
\end{equation}
If the former, calculated without taking backreaction into account, is larger than $\rho_\mathrm{tot}/2$, the tachyonic preheating is efficient enough and the backreaction cannot be neglected any further. 
We regard this as the condition for the completion of preheating. 
Figure~\ref{fig-omegak} shows $\Omega\equiv \ln (32 \pi^2 \rho_{\delta h}/m_{h,\mathrm{max}}^4)/2$, which corresponds to our analytic estimate on the effective amplification factor $\Omega_{k_t}$, 
together with $ \ln (16 \pi^2 \rho_\mathrm{tot}/m_{h,\mathrm{max}}^4)/2$ for 
each exact hill-climbing case as functions of $N$ for both branches. 
The latter represents the exponential growth factor needed for preheating to be completed.
Here we use the value of $\varphi_2$ numerically obtained (see Fig.~\ref{fig-tachyonic-mass-comparison}) to determine the coefficient in $m_h^2 (t)= -3 \alpha \xi_N M_N^2 \varphi_2 \sin (M (t-t_{\mathrm{enter},0})) $  
and evaluate $\rho_{\delta h}$ by
integrating Eqs.~\eqref{eq-analytic-occupation-number} and~\eqref{eq-tachyonic-rho-definition}. 
We see that the values of $\Omega$ in Branch 1 are very close to those in Branch 2 with corresponding $N$, as predicted in Eq.~\eqref{eq-rho-delta-h-analytic}, 
and they are fit well with $\Omega \simeq \pi N -2$. 
By comparing $\Omega$ and $\ln (16 \pi^2 \rho_{\mathrm{tot}}/m_{h,\mathrm{max}}^4)/2$ we can determine if the preheating is completed. 
In the Higgs-like regime (Branch 1), the tachyonic instability always completes preheating for the exact hill-climbing case. 
This is because the characteristic energy density $m_{h,\mathrm{max}}^4/32 \pi^2$ without the exponential amplification itself is already as large as the background energy density. On the other hand, due to small $m_{h,\mathrm{max}}^4/32 \pi^2$ in the $R^2$-like regime (Branch 2),  preheating is not completed by the tachyonic instability even at the exact hill-climbing case for $N\leq 4$ (see App.~\ref{appendix-B} for more accurate calculation). 
However, thanks to the long enough $\Delta t$ at sufficient large $N$, 
preheating is completed by the tachyonic instability for the exact hill-climbing case. 
Note that $N=4$ looks on the edge of the completion of preheating, but the analysis in Fig.~\ref{fig-omegak} is relatively qualitative and should not be taken at face value. 
Indeed, as long as our numerical precision allows, we do not find any parameter space in which tachyonic instability completes preheating around $\theta_4^2$. 
As we see in Sec.~\ref{Sec-4}, we conclude that at least we need a 
fine-tuning much more severe than $\mathcal{O}(10^{-5})$ around this value of $N$. 

\begin{figure}[h]
    \centering
    \includegraphics[width=.6\textwidth]{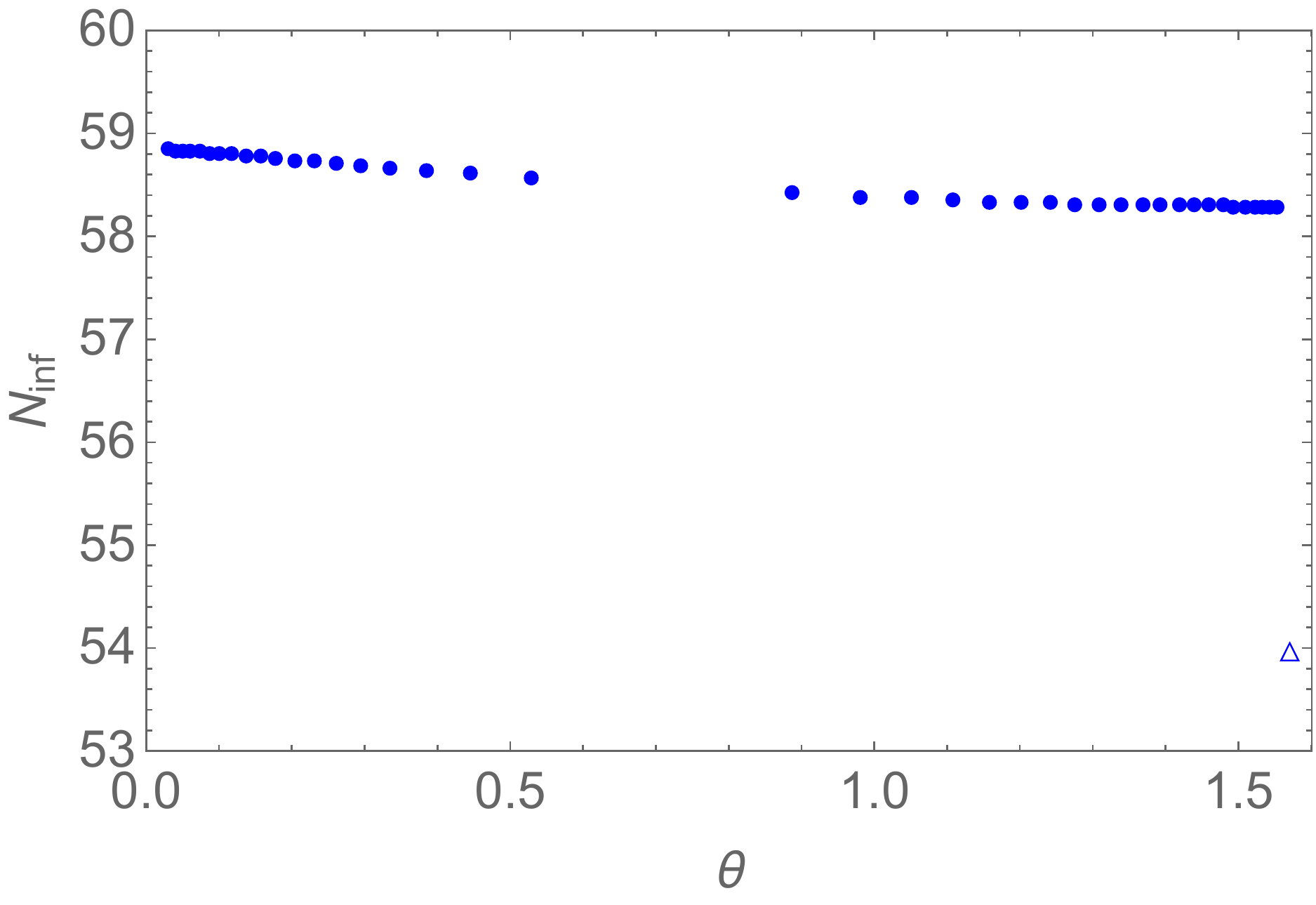}
    \caption{The number of e-folds of inflation with pivot scale $ k=0.002~{\rm Mpc}^{-1} $ is shown for the exact hill-climbing parameters that are out of strong coupling regime and can complete preheating solely by tachyonic instability. The triangle represents the prediction of the $ R^2 $ inflation.}
    \label{fig-preheating-efold}
\end{figure}

To conclude this section, we roughly estimate the duration of preheating in the exact hill-climbing case except for $ N\leq 4 $ in Branch 2 and $ N\geq 7 $ in Branch 1. We here assume that the radiation-dominated epoch starts right after the completion of preheating and the scalar field oscillation will not dominate the universe again. As mentioned above, the tachyonic effect occurs within one scalaron oscillation after the end of inflation. Then from Eq.~\eqref{eq-scalaron-Higgs-mass-positive-phi-valley} and \eqref{eq-scalaron-Higgs-mass-negative-phi}, we have an upper bound of duration for each critical case\footnote{As will be shown in the next section, preheating can be completed by tachyonic instability even for a duration shorter than one scalaron oscillation.}
\begin{align}
	\Delta t_{\rm pre} \simeq \pi \left( \frac{2}{M}+ \frac{1}{2M_c} \right) ~.
\end{align}
In such a short period, the Hubble parameter can be approximately constant which one can take $ H \sim C_1 C_2 M_c/2 $. As a result, we estimate the number of e-folds for tachyonic effect to complete preheating 
\begin{align}
	\Delta N_{\rm pre} \simeq H \Delta t_{\rm pre} \sim C_1 C_2 \pi \left( \frac{1}{4}+\frac{M_c}{M} \right) 
\end{align}
where $ 0.03 \lsim M_c/M \leq 1 $ for $ 0 \leq \xi \leq \xi_s $. Therefore, we have $ 0.2 \lsim \Delta N_{\rm pre} \lsim 0.7 $
which can be regarded as almost instantaneous in a cosmological sense but varies around 0.5 with respect to the change of model parameters. On the other hand, through the relation 
\begin{align}
    k = a_k H_k =  \frac{a_k}{a_{\rm e}} \frac{a_{\rm e}}{a_{\rm pre}}\frac{a_{\rm pre}}{a_0} a_0 H_k
\end{align}
where we denote the scale factor at present, the end of preheating (here we assume that the onset of radiation-dominated epoch is the end of preheating), and the end of inflation as $ a_0 $, $ a_{\rm pre} $, and $ a_{\rm e} $, respectively, one can calculate the number of e-folds of inflation as
\begin{align}
    N_{\rm inf}(k) \equiv \ln \left( \frac{a_{\rm e}}{a_k} \right) =& \ln \left( \frac{a_0 M_{\rm pl}}{k} \right) +\ln \left( \frac{a_{\rm e}}{a_{\rm pre}}\right) + \ln \left( \frac{a_{\rm pre}}{a_0} \right) +\ln \left( \frac{H_k}{M_{\rm pl}} \right) \nonumber \\
    =& \ln \left( \frac{a_0 M_{\rm pl}}{k} \right) -\Delta N_{\rm pre}(\theta) + \ln \left[ \frac{ T_0}{ T_{\rm pre}} \left( \frac{g_0}{g_{\rm pre}} \right)^{1/3} \right] +\ln \left( \frac{M_c}{2M_{\rm pl}} \right) 
\end{align}
where we have assumed that thermalization after preheating is realized within one Hubble time and the total entropy is conserved between the end of thermalization and today, and have approximated the inflation scale $ H_k \simeq M_c/2 $ because it is effective $ R^2 $-inflation. Parameters $ g_0 = 43/11 $ and $ g_{\rm pre} = 106.75 $ are the effective number of relativistic species at present and the end of preheating, while $ T_0 \simeq 2.7 ~{\rm K} $ and $ T_{\rm pre} $ are the corresponding temperatures. The temperature at the end of preheating can be estimated by  
\begin{align}
    \frac{g_{\rm pre} \pi^2}{30} T^4_{\rm pre} \approx \frac{3}{4} C_1^2 C_2^2 M^2_{\rm pl} M_c^2
\end{align}
which gives $ T_{\rm pre} \approx 1.7 \times 10^{28} ~{\rm K} $.
We choose the pivot scale to be $ k/a_0 =0.002 ~{\rm Mpc}^{-1} $\footnote{The $ A_s $ and $ n_s -1 $ were presented in Ref.~\cite{Akrami:2018odb} for $ k =0.05~{\rm Mpc}^{-1} $, so $ N_{0.05}=N_{\rm inf} -3.2 $ where $ N_{\rm inf} $ is given in Eq.~\eqref{eq-e-fold-inflation} for $ k=0.002~{\rm Mpc}^{-1} $.}. As a result, we have 
\begin{align}\label{eq-e-fold-inflation}
    N_{\rm inf} \simeq 59 -\Delta N_{\rm pre}(\theta)
\end{align}
which is shown in Fig.~\ref{fig-preheating-efold}.
Compared with the prediction of the $R^2$-inflation, $ N_\mathrm{inf} \simeq 54 $~\cite{Bezrukov:2011gp}\footnote{The difference in $N_{\rm inf}$ between the Jordan and the Einstein frames exists but it is of the order of the next order correction to the slow-roll approximation which we do not take into account. Therefore, we will also neglect the difference between the two frames.}, we can distinguish them once we will have an accuracy to distinguish $\delta N \sim 5$ experimentally. On the other hand, if we would like to distinguish each parameter for the exact hill climbing we need an accuracy at least up to $ \Delta N \sim {\cal O}(0.1) $. 
Note that this argument is based on the assumption that the Universe becomes radiation-dominated right after the completion of preheating, which does not apply if the rescattering and backreaction prevent the system from entering the radiation domination instantaneously.

\section{Necessary degree of fine-tuning for the tachyonic instability}
\label{Sec-4}

In the previous section, we studied exact hill-climbing cases, in which the model parameters are chosen so that fully efficient tachyonic instability is realized.
However, once they deviate from these values, the Higgs field does not exactly go along the hill but falls down to the valley in the middle of going uphill (or downhill), and as a result the tachyonic particle production terminates. 
Then the question is how much deviation from the exact hill-climbing is allowed for the tachyonic instability to be still sufficiently effective to complete preheating. 
In other words, we study how much fine-tuning among the model parameters is necessary to produce the Higgs fluctuations whose energy density is comparable to the background. 
From Fig.~\ref{fig-omegak}, comparing the data points (amplification factor for the exact hill-climbing) and the solid line (necessary amplification factor to complete preheating), we expect that weaker fine-tuning is required for smaller $N$ in Branch 1 while the necessary fine-tuning is more severe for smaller $N$ in Branch 2.
This is natural in the sense that the tachyonic effect is weaker for $R^2$-like limit because it is well-known that there is no such effect in the $R^2$ inflation. 
In this section, we study quantitatively the required amount of fine-tuning for the completion of preheating.

We first note the following simplifications on the scalar field dynamics. 
Since the falling down from the hilltop to the valley is driven by the tachyonic mass squared of ${\cal O} (m_{h,\mathrm{max}}^2)$, 
the time scale of this dynamics is much smaller than the whole dynamics of the hill-climbing $\Delta t \sim M^{-1}$. 
Therefore we approximate the time evolution as
\begin{equation}
    \varphi(t)\simeq \varphi_2 \sin (M(t-t_{\mathrm{enter},0})), \quad h(t) \simeq 0, \quad \text{for} \quad t_{\mathrm{enter},0}<t<t_\mathrm{drop},
\end{equation} 
where $t_\mathrm{drop}$ is the time when the Higgs field falls down to the potential valley.
The scalar fields oscillate around the potential valley after $t = t_{\rm drop}$. 
However, since the tachyonic instability is typically stronger than the parametric resonance (see footnote~\ref{footnote:tachyonic_vs_parametric}) and lasts sufficiently long, we simply neglect particle production during this epoch. 
we expect that the true amount of the particle production is not much different from our following estimate. 

\begin{figure}[h]
	\centering
	\includegraphics[width=.6\textwidth]{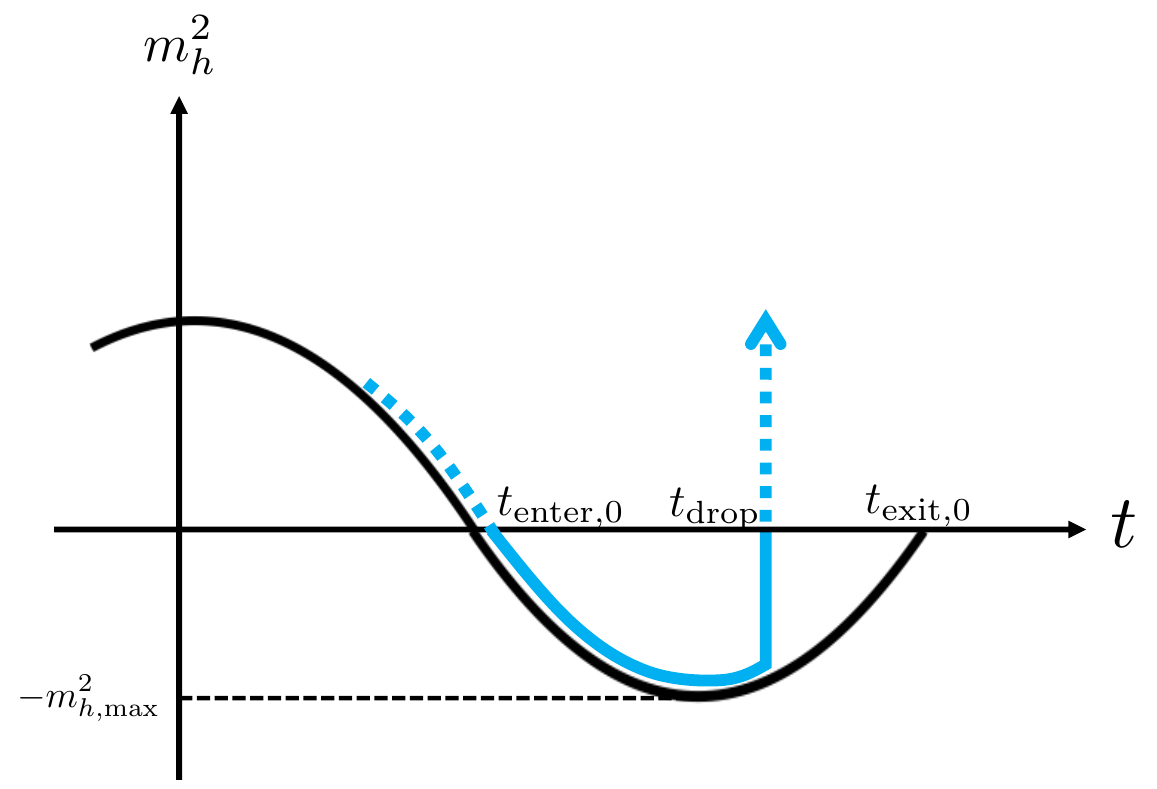}
	\caption{Schematic picture of the time evolution of the mass squared of the Higgs fluctuation.
	The light blue curve represents the evolution of the mass squared of the Higgs fluctuation, while that of the exact hill-climbing case is shown in the black solid curve for comparison. 
	The black dotted curve is the mass squared along the potential valley. 
	The tachyonic mass for the Higgs fluctuation almost follows the exact hill-climbing case until $t=t_\mathrm{drop}$ and then gets shut off almost instantly. 
	Note that in reality the Higgs mass during $t>t_\mathrm{drop}$ shows oscillating features as seen from Ref.~\cite{Bezrukov:2019ylq}. }
	\label{fig-Higgs-mass-non-exact-hill-climbing}
\end{figure}

Practically we adopt the following procedure. 
We obtain the evolution of the scalaron $\varphi(t)$ and the Higgs field $h(t)$ by solving the full background equations of motion \eqref{eq:eom1},~\eqref{eq:eom2}, and~\eqref{eq:eom3} numerically. 
Then we evaluate the mass for the Higgs fluctuation as 
\begin{equation} \label{eq-Higgs-mass-non-exact-hill-climbing}
    m_h^2 (t) = e^{\alpha \varphi(t)} \frac{\partial^2 U}{\partial h^2}(\varphi(t),h(t)). 
\end{equation}
Here we recovered the factor $e^{\alpha \varphi(t)}$ just for slight improvement. 
With this treatment, we find that the tachyonic mass for the Higgs fluctuation almost follows the case of the exact hill-climbing until $t_\mathrm{drop}$, and then gets shut off almost instantly. 
See Fig.~\ref{fig-Higgs-mass-non-exact-hill-climbing} for a rough sketch of the time evolution.
Omitting the other contributions in the mode equation \eqref{eq-exact-eom-delta-higgs-redefine} as well as particle production after falling down to the valley, we evaluate the occupation number of the Higgs fluctuation at late times as
\begin{equation} \label{eq-nk-non-exact}
    n_k = |\exp(\Omega_k)-\exp(-\Omega_k)/4|^2, \quad \Omega_k \equiv \int_{t_\mathrm{enter}(k)}^{t_\mathrm{drop}(k)} \sqrt{m^2_h(t)-\frac{k^2}{a^2}}. 
\end{equation}
Here in the numerical calculation we defined the $k$-dependent drop-off time, 
$t_\mathrm{drop}(k)$, as the time when $\omega_k^2(t)$ crosses zero.  
With these simplifications we evaluate the comoving energy density of the produced Higgs fluctuation when its mass becomes sufficiently small as follows\footnote{For a practical purpose we take $k=m_{h,\mathrm{max}}/100$ as the lower limit of the integration in the numerical calculation.}
\begin{equation} \label{eq-tachyonic-production-estimate}
    \rho_{\delta h} = \int_0^{m_{h,\mathrm{max}}} \frac{d^3 k}{(2\pi)^3}k n_k.  
\end{equation}
Here we neglected the cosmic expansion and took $a=1$. 
See App.~\ref{appendix-C} for analytical estimation of $\rho_{\delta h}$ and discussion about $t_{\rm drop}$.

By imposing a conservative criterion (see Eq.~\eqref{eq-energy-at-phi1}), 
\begin{align} \label{eq-criteria-non-exact}
	\rho_{\delta h} \gsim U_0(\varphi_1)/2 =C^2_1 U_{\rm inf}/2,  
\end{align}
together with Eq.~\eqref{eq-tachyonic-production-estimate}, 
we can identify the parameter range around each exact hill-climbing case, parameterized by $\theta_N^i$, that gives successful amount of particle production. 
Let us define the upper and lower bound of the parameter $\theta$ around $\theta_N^i$ for successful preheating as $\theta_{N+}^i$ and $\theta_{N-}^i$, respectively, and 
also define $\Delta \theta_{\mathrm{eff}, N}^i \equiv \theta_{N+}^i- \theta_{N-}^i$.

\begin{figure}[htbp]
	\centering
    \includegraphics[width=.8\textwidth]{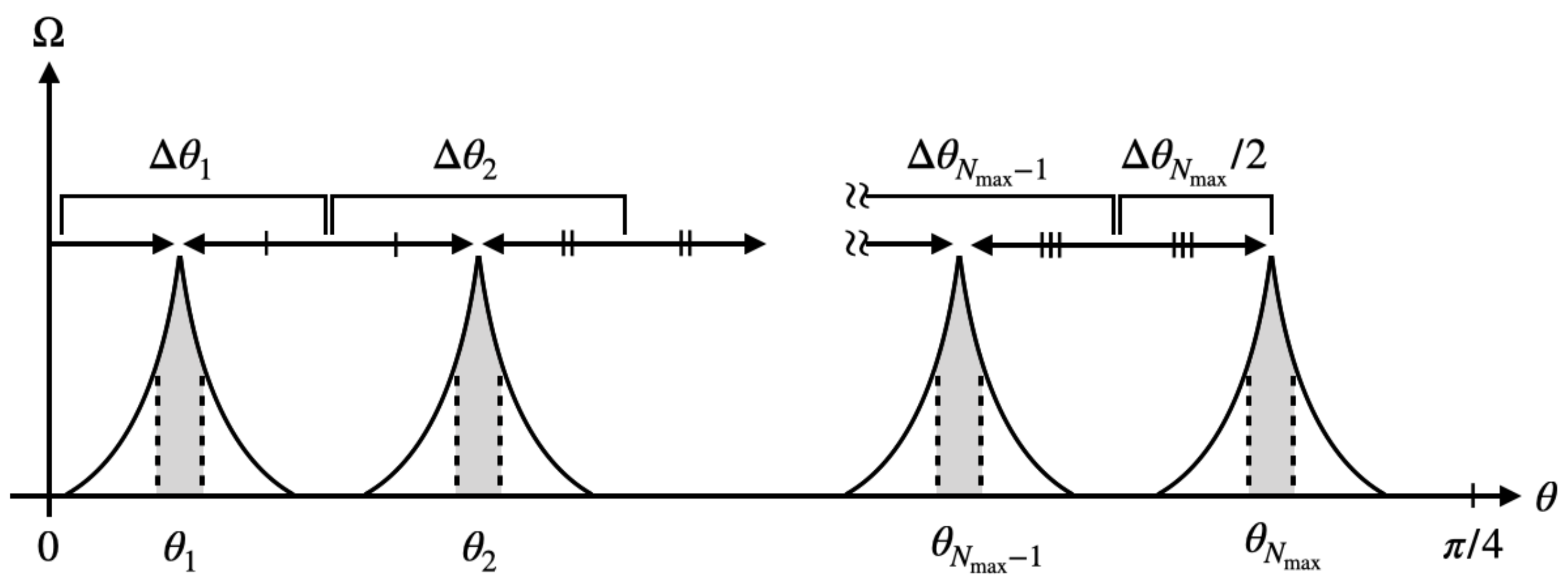}
	\caption{Definition of the effective width of each $ \theta_N $, $ \Delta\theta_N $ is shown. 
	The gray bands are the parameter region where the preheating successfully finishes in one stroke.}
	\label{fig-schematic}
\end{figure}

In order to express the degree of fine-tuning quantitatively, we further define $ \Delta\theta_N^i $ that describes the typical distance between two neighboring values of $\theta_N^i$ as
\begin{align}\label{eq-definition-Delta-theta-N}
    \Delta \theta_N^i &\equiv \frac{\theta_{N+1}^i-\theta_{N-1}^i}{2} \quad \text{for} \quad N=2,3, \cdots N_\mathrm{max}-1 , \notag \\
    \Delta \theta_1^1 &\equiv \frac{\theta_1^1+\theta_2^1}{2}, \quad \Delta \theta_1^2 \equiv \frac{\pi}{2}-  \frac{\theta_1^2+\theta_2^2}{2}, \quad \Delta \theta_{N_\mathrm{max}}^i = 
    \left| \theta_{N_\mathrm{max}}^i-\theta_{N_\mathrm{max}-1}^i\right|,  
\end{align}
with $N_\mathrm{max}=26$. 
See Fig.~\ref{fig-schematic} for a schematic picture of the definition. 
Now we define the degree of fine-tuning for the $N$-th exact hill-climbing case in Branch $i$ as $ \Delta \theta_{\mathrm{eff}, N}^i/\Delta\theta_N^i $. 

\begin{figure}[h]
	\centering
    \includegraphics[width=.6\textwidth]{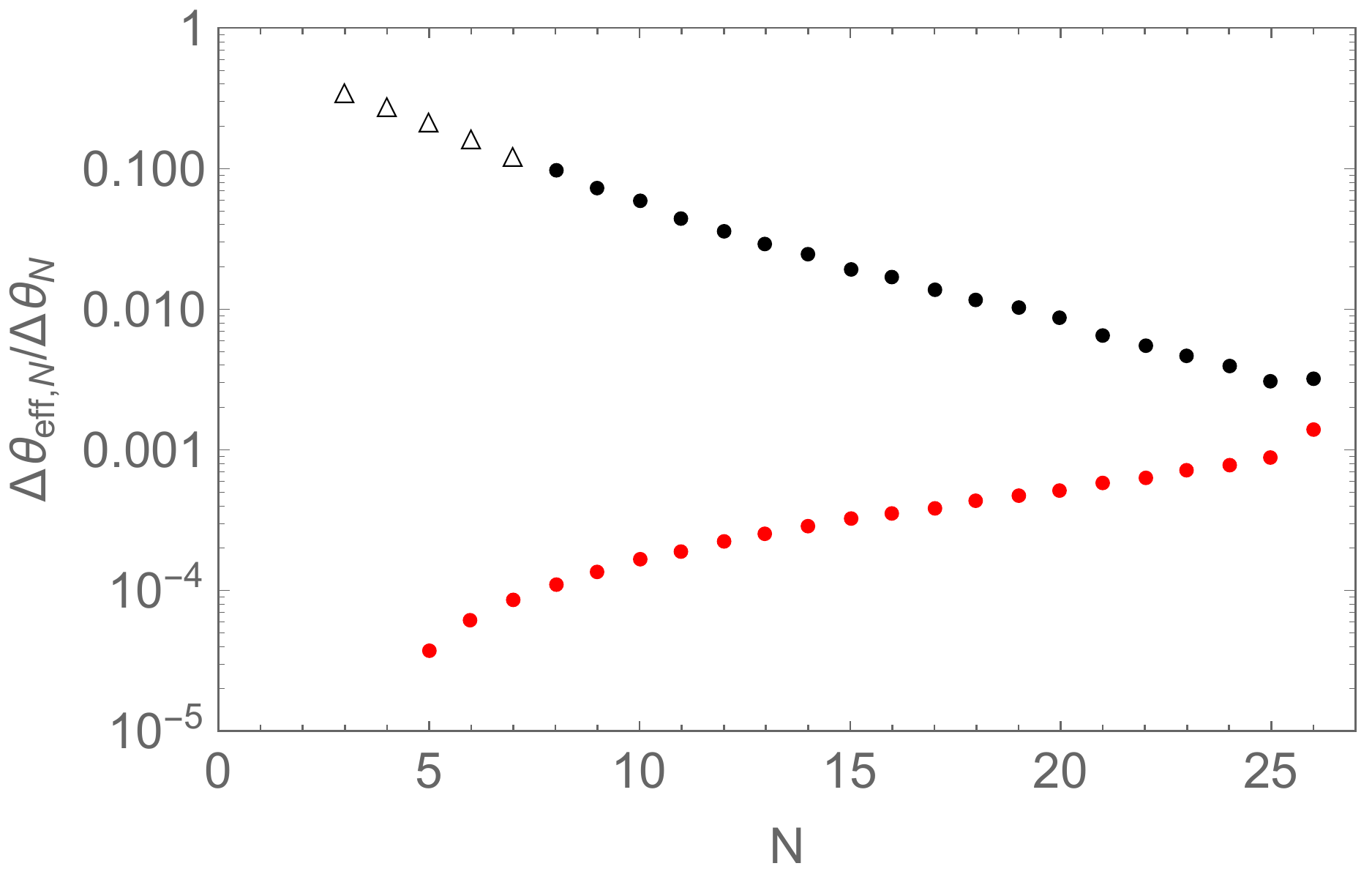}
	\caption{Necessary degree of fine-tuning $ \Delta \theta_{\mathrm{eff}, N}^i/\Delta\theta_N^i $ for 
	the successful tachyonic preheating for Branch 1 (black) and 2 (red). 
    The five black empty triangles are for the cases beyond the unitarity bound. 
	In Branch 2, the data points for $ N \leq 4 $ do not exist because the tachyonic preheating is not efficient even for the exact hill-climbing trajectories. The sudden lift for $N=26$ comes from the definition of $\Delta\theta^i_{N_{\mathrm{max}}}$ in Eq.~\eqref{eq-definition-Delta-theta-N}: 
	the interval $\Delta \theta ^i_N$ increases as $N$ increases, and hence $\Delta\theta^i_{N_{\mathrm{max}}} =\left| \theta^i_{N_{\mathrm{max}}} -\theta^i_{N_{\mathrm{max}}-1} \right|$ underestimates the width around $\theta^i_{N_{\mathrm{max}}}$, which results in the sudden lift. }
	\label{fig-degree-of-fine-tune}
\end{figure}

Based on these analytic formulation, we perform the following numerical analysis. We scan the parameter $\theta$ around each exact hill-climbing $\theta_N^i$, and for each value of $\theta$ we solve the background equations of motion Eqs.~\eqref{eq:eom1},~\eqref{eq:eom2}, and~\eqref{eq:eom3}. 
We use the results to evaluate the occupation number $n_k$ with Eq.~\eqref{eq-nk-non-exact} and the energy density of the Higgs fluctuation produced through the tachyonic instability with Eq.~\eqref{eq-tachyonic-production-estimate}. 
From the criterion~\eqref{eq-criteria-non-exact}, we determine the boundary values of $\theta$ for the successful preheating $\theta_{N+}^i$ and $\theta_{N-}^i$, and the required degree of fine-tuning $ \Delta \theta_{\mathrm{eff}, N}^i/\Delta\theta_N^i $. 
The result is shown in Fig.~\ref{fig-degree-of-fine-tune}, which is our main result of the present paper. 
We see that only $ \mathcal{O}(0.1) $ fine-tuning is required for smaller $N$ in Branch 1. 
This is because the typical Higgs mass squared is large for these cases and a relatively small amplification factor $\Omega_k$ is enough for the successful preheating as seen in Fig.~\ref{fig-omegak}.
As $ \theta $ gets closer to $ \pi/2 $ (for larger $N$ in Branch 1 and for smaller $N$ in Branch 2), the necessary fine-tuning becomes more severe. 
For the most severe case, $N=5$ in Branch 2, we need a fine-tuning of ${\cal O} (10^{-5})$. 
This can be understood intuitively that a larger $ \theta $ (smaller $ \xi $) corresponds to smaller tachyonic Higgs mass $ |m^2_h| $ during hill-climbing (see Fig.~\ref{fig-tachyonic-mass-comparison}), and hence the Higgs field needs to stay at the hilltop for a longer period in order to have stronger tachyonic effect. This naturally requires more fine-tuning, which is consistent with expectation from Fig.~\ref{fig-omegak}. 
Although our analysis here is based on a relatively simplified formulation, we expect that the actual degree of fine-tuning obtained by full numerical calculations with the full mode equations~\eqref{eq-exact-eom-delta-scalaron-redefine} and~\eqref{eq-exact-eom-delta-higgs-redefine}
is not significantly different from our results, 
because the exponential amplification of the Higgs particles takes place when the background Higgs field is climbing up the hill along $h \simeq 0$. 

Before concluding this section, let us note some issues in our analysis. 
Our numerical calculation starts from $ \sim 1 $ e-fold before the end of inflation while using the inflationary attractor Eq.~\eqref{eq-valley} corresponding to the large number of e-folds during inflation as the initial condition. If one chooses an earlier moment during inflation to start the computation using the same initial condition, the face values of $ \xi_N^i $ might appear slightly different from ours because numerical errors are accumulating with calculation time. Correspondingly, the numerical initial condition corresponding to the attractor should be formulated with much better accuracy. At the same time, it is more and more difficult to find the exact value of $ \xi_N^i $ when beginning the computation from a larger number of e-folds before the end of inflation. From this point of view, we argue that a small change of the face values of $ \xi_N^i $ for numerical calculations scanning different number of inflationary e-folds does not mean that our results depend on the choice of initial conditions.  Moreover, the estimation of degree of fine-tuning does not depend on the precision of $ \xi_N^i $, so our results on the degree of fine-tuning of model parameters are robust in this respect.

\section{Discussion on the phase direction and the longitudinal mode}
\label{Sec-5}

Thus far we have focused only on the physical, or radial, direction of the Higgs field and have not discussed the dynamics of the phase direction, or the Nambu-Goldstone mode, which determines the longitudinal mode of the gauge bosons~\cite{Ema:2016dny} in the gauged case. 
Before concluding, we give brief discussion on the tachyonic instability in the phase direction for the global U(1) case.
We can read off the implication to the SM SU(2)$_L \times$ U(1)$_Y$ case as done in Ref.~\cite{He:2018mgb}.
Note that the tachyonic instability for the longitudinal mode of the weak gauge boson are observed in Ref.~\cite{Bezrukov:2019ylq}.

As calculated in Ref.~\cite{He:2018mgb}, the mass for the Nambu-Goldstone mode is obtained by writing down the potential for the phase $\theta$ of the Higgs field defined as $\mathcal{H}=h e^{i\theta}/\sqrt{2}$ in Eq.~\eqref{jordanaction}, and moving to the Einstein frame,
\begin{align}
	m_{\theta_c}^2=
	-\frac{\alpha}{2} \frac{\partial U}{\partial\varphi} + \frac{e^{\alpha \varphi}}{h}\frac{\partial U}{\partial h} 
	-\frac{3}{4}\frac{U}{M^2_\mathrm{pl}} +\frac{5}{24}\frac{1}{M^2_\mathrm{pl}}\left(\dot{\varphi}^2+e^{-\alpha \varphi} \dot{h}^2\right),  \label{eq-meff-gauge-boson}
\end{align}
where the kinetic term of the phase direction is canonically normalized. 
During the hill-climbing epoch (when $\varphi < M_\mathrm{pl}, h\simeq 0$, and the kinetic terms are negligible), it is further simplified as
\begin{equation}
    m_{\theta_c}^2 \simeq -\frac{1}{\sqrt{6}} \frac{M^2}{M_{\rm pl}} (1+6\xi) \varphi .
\end{equation}
Thus, comparing with the physical Higgs mass (Eq.~\eqref{eq-scalaron-Higgs-mass-positive-phi-hill}), we find 
\begin{align}
	\frac{m^2_{\theta_c}}{m^2_h} =1+\frac{1}{6\xi}, 
\end{align}
which is order of unity unless $\xi \ll 1$. 
Since $\xi_N$ is no less than the order of unity even for $N=1$ in Branch 2 (see Eq.~\eqref{eq-exact-N-xi}), we conclude that the efficiency of the tachyonic instability for the phase direction, or the Nambu-Goldstone mode, is comparable to that for the physical Higgs fluctuations. 
We also expect that the same applies to the longitudinal mode of the gauge bosons, whose mass receives dominant contribution from the mass of the Nambu-Goldstone mode~\cite{Ema:2016dny}. 

While the mass for physical Higgs fluctuations around the potential valley is positive, the mass for the phase direction is given by
\begin{equation}
    m_{\theta_c}^2 \simeq - \frac{ \alpha}{2} \varphi {\tilde M}^2 + \frac{5}{24}\frac{1}{M^2_{\mathrm{pl}}} \left(\dot{\varphi}^2+\dot{h}^2\right)
\end{equation}
for $\varphi \ll M_\mathrm{pl}$, which can be tachyonic especially when the kinetic energy is small. 
Therefore, the Nambu-Goldstone mode and the longitudinal mode of the gauge bosons are more likely to receive a tachyonic contribution from field oscillations around the potential valley than from physical Higgs fluctuations, that is also seen in Ref.~\cite{Bezrukov:2019ylq}. 
However, the amplitude of the tachyonic mass is comparable to the time scale of the oscillations around the valley, 
and hence we expect that such tachyonic instability for the Nambu-Goldstone mode and for the longitudinal mode 
of the gauge bosons do not give a significant contribution compared to the particle production
during the hill-climbing. Tachyonic mass of the physical Higgs could also be realized during oscillations around the valleys with a large amplitude, but for the same reason we expect such effect to be relatively small. Detailed investigation of particle production during this epoch is beyond the scope of this work and left for future study. 

In summary, we conclude that the Nambu-Goldstone mode and the longitudinal mode of the gauge bosons also experience  tachyonic instability during hill-climbing with almost the same efficiency of amplification as the physical Higgs fluctuations. 
Therefore, taking into account their contributions, the total particle production will be enhanced accordingly.
Since particle production is an exponential effect, our basic results remain quantitatively unchanged even if we take them into account.

\section{Conclusions and outlook}
\label{Sec-6}

In this paper, we study one of the possible preheating mechanisms in the mixed Higgs-$R^2$ inflationary model pointed out in \cite{Bezrukov:2019ylq}, namely the tachyonic instability. 
Although some degree of fine-tuning is necessary for this mechanism to work, the resulting dynamics in the Universe can be interesting.
We give analytic conditions for this phenomenon to occur by investigating the Higgs field oscillations around the potential valley in the negative scalaron region $ \varphi < 0 $, and numerically find all the model parameters for this to happen.
We point out that tachyonic preheating can take place both in the Higgs-like regime $ \xi \lsim \xi_c $ (Branch 1) and the $R^2$-like regime $ M \gsim M_c$ (Branch 2).

Since tachyonic preheating in this model requires some degree of fine-tuning among the model parameters, we first find all the parameter values $ \theta = \theta^i_N $ realizing exact hill-climbing, which is the condition for the most efficient particle production
(with $N$ parameterizing the number of Higgs half-oscillations during the period when $\varphi < 0$ and $i$ labeling Branch 1 or 2: see Sec.~\ref{Sec-2-2} for definition).
We then analytically calculate particle production from the tachyonic instability. It is found that, for all these tuned parameter points, tachyonic particle production is strong enough to complete preheating except for $ N \leq 4 $ in Branch 2. However, even a slight deviation from $ \theta = \theta^i_N $ can significantly reduce the strength of the tachyonic effect. 
In order to estimate the necessary degree of fine-tuning, we scan the model parameters around each $ \theta^i_N $ and find the interval $ \Delta \theta^i_{\text{eff}, N} $ such that the preheating can be completed within $\theta \sim \theta^i_N \pm \Delta \theta^i_{\text{eff},N}/2 $. The result is given in Fig.~\ref{fig-degree-of-fine-tune}, which shows that the necessary fine-tuning becomes more severe as $ \xi $ gets smaller (closer to the $ R^2 $ limit). This is natural because we do not expect any tachyonic effect in $ R^2 $ inflation. 
While we mainly focus on the amplification of physical Higgs fluctuations instead of those in the phase direction or longitudinal gauge bosons, we find that the amplification of the latter is comparable to that of physical Higgs fluctuations by investigating their effective mass.
This suggests that the efficiency of the total particle production is enhanced by a factor of the order of unity, but our results remain basically unchanged and the required fine-tuning can be read off from Fig.~\ref{fig-degree-of-fine-tune}. 

Throughout this paper, we do not take into account the backreaction from produced particles on the homogeneous background. 
This is because we are interested in the growth of inhomogeneities until their energy becomes comparable to that of background inflaton oscillations.
As for the standard criterion for the end of reheating, it is defined as the onset of radiation-dominated epoch where the contributions from the homogeneous background fields, i.e. the correlated quantum
particles with zero spatial momentum in quantum language, become negligible. To determine whether the particles produced by the tachyonic instability are relativistic and all the energy in the homogeneous fields is transferred to inhomogeneities, one needs to take into account the backreaction which can only be done in numerical calculation and is beyond the purpose of this paper. 
We expect two possibilities. One is that, after a complete tachyonic period, the amplitude of the coherent oscillation of the background fields becomes very small (so do the masses of the produced particles) and the produced particles are relativistic with typical momentum $ k\sim |m_{h,\rm max}| $. In this case, the preheating is almost instantaneous and thermalization comes afterwards. The other possibility is that the backreaction terminates the tachyonic instability midway and rescattering between perturbations and background fields (turbulence) begins to take effect. In this case, however, the preheating takes longer time. Specifically, elastic (re)scattering that  conserves the number of particles is not sufficient for reheating and thermalization. Deeply inelastic scatterings with specially engineered initial conditions that increases the energy of particles at the cost of decrease of particle number is needed for this purpose. Without such special conditions, a natural hypothesis is that the final reheating temperature $ T_{\rm pre} $ cannot be larger than the initial momenta of particles coming from the background before rescattering, i.e. $k/a_e \geq T_{\rm pre} $. As a conservative estimation, if we take the momentum of the produced particles to be 
\begin{align}
    \frac{k^2}{a_e^2} = |m_h^2|_{\varphi=\varphi_2}= \left|3\xi M^2 C_1C_2 \frac{M_c}{M}\right| ~,
\end{align}
and the reheating temperature 
\begin{align}
    T_{\rm pre} =\left( \frac{90C_1^2C_2^2}{4g_{\rm pre}\pi^2} M_{\rm pl}^2 M_c^2 \right)^{1/4} ~,
\end{align}
one can obtain the condition 
\begin{align}
    M \gsim \sqrt{1+ 0.67 \left( \frac{0.01}{\lambda} \right)} M_c
\end{align}
which corresponds to $ \xi \gsim \xi^{i=2}_{N=26} $ for $\lambda =0.01$. Based on this hypothesis, the result means that for $R^2$-like regime, the instantaneous preheating by rescattering is not possible. Finally, 
in the case where preheating is not sufficient, late-time domination of scalar field oscillation is possible and perturbative decay may be needed to finish reheating. We leave these questions for our future work.

This paper focuses on the period right after the second zero-crossing of scalaron $ \varphi $ after the end of inflation, 
during hill-climbing of the scalaron along the potential hill at $h\simeq 0$. 
The possibility of having the tachyonic effect after a number of scalaron oscillations (instead of the second zero-crossing) is not addressed. This possibility is also pointed out in \cite{Bezrukov:2019ylq}. Such a question is beyond the scope of this paper, mainly because it involves consideration of the production of other particle species during field oscillations around the potential valley, which occurs before the possible tachyonic instability from late-time scalaron oscillations. 
We leave it for future study.

\vspace{5ex}
\noindent \textbf{Note added in proof}

\noindent At the same day when our paper was submitted to the hep-ph archive, the paper \cite{Bezrukov:2020txg} appeared there, too. In that paper, the authors used the lattice simulation to study the preheating process in this model. Our results are not in conflict with theirs. 

\section*{Acknowledgments}

We thank Fedor Bezrukov and Yohei Ema for useful discussions. 
MH was supported by the Global Science Graduate Course (GSGC) program of the University of Tokyo and the JSPS Research Fellowships for Young Scientists. 
RJ was supported by Grants-in-Aid for JSPS Overseas Research Fellow (No. 201960698). 
This work was partially supported by the Deutsche Forschungsgemeinschaft under Germany's Excellence Strategy -- EXC 2121 ``Quantum Universe'' -- 390833306.
This work was partially supported by JSPS KAKENHI, Grant-in-Aid for Scientific Research Nos.\ JP19K03842(KK), 15H02082(JY), 20H00151(JY)
and Grant-in-Aid  for Scientific Research on Innovative Areas Nos.\ 19H04610(KK), 20H05248(JY). 
AAS was partially supported by the Russian Foundation for Basic Research grant No. 20-02-00411.

\appendix 

\section{Analytical calculation of particle production through tachyonic instability}
\label{appendix-A}

In this appendix, we present analytic treatment of  particle production due to the tachyonic instability. 
In Sec.~\ref{Sec-3}, we give an intuitive formula Eq.~\eqref{eq-analytic-occupation-number}. 
Here we show its validity using an analytic method to solve the mode equation and calculate the Bogoliubov coefficients.

The mode function of the Higgs inhomogeneity at late time $t>t_{\mathrm{exit},0}$ when the adiabatic condition holds
is expressed with the WKB approximation as 
\begin{equation}
    \tilde{\delta h}_k (t) = \frac{\tilde{\alpha}_k}{\sqrt{2\omega_{h,k}(t)}} \exp \left(- i\int^{t}_{t_{\mathrm{exit},0}} \omega_{h,k} (t') dt' \right) +\frac{\tilde{\beta}_k}{\sqrt{2\omega_{h,k}(t)}} \exp \left(i\int^t_{t_{\mathrm{exit},0}} \omega_{h,k}(t') dt' \right),  \label{eq-WKB-after-exit}
\end{equation}
where ${\tilde \alpha}_k$ and ${\tilde \beta}_k$ are the Bogoliubov coefficients 
with the normalization condition $|\tilde{\alpha}_k|^2-|\tilde{\beta}_k|^2 = 1$. 
In the WKB limit, the Bogoliubov coefficients can be regarded as constants. The  occupation number of particles, $n_k$,
produced by the change of the mass (or $\omega_{h,k}$) is given by 
\begin{equation}
    n_k(t) = |{\tilde \beta}_k(t)|^2 ~.
\end{equation}
If the change of $\omega_{h,k}$ is adiabatic
\begin{align} \label{eq-adiabatic-condition}
    \left| \frac{\ddot{\omega}_{h,k}/\omega_{h,k}-3\dot{\omega}^2_{h,k}/(2\omega^2_{h,k})}{2\omega^2_{h,k}} \right| \ll 1 
\end{align}
at all the time of interest, 
one can simply solve the equation of motion for the Bogoliubov coefficients fully with the WKB approximation. 
However, now we are interested in the case where $\omega_{h,k}^2$ is negative when $t_{\mathrm{enter},0}<t<t_{\mathrm{exit},0}$, 
and hence the adiabatic condition is violated at $t=t_{\mathrm{enter},0}$ and $t_{\mathrm{exit},0}$ when $\omega_{h,k}$ crosses zero. As a result, the WKB approximation
is broken down at these moments. 
Fortunately, 
except for that, 
WKB approximation is valid. 
Therefore, the occupation number at the later time can be obtained 
by imposing appropriate matching conditions 
at $t=t_{\mathrm{enter},0}$ and $t_{\mathrm{exit},0}$ between the WKB solutions, 
for which we adopt the trick in Ref.~\cite{Dufaux:2006ee} (also in the Landau-Lifshitz's textbook~\cite{Landau:1991wop}).

Let us define the time when the adiabatic condition is violated as $t_\mathrm{enter}^-<t<t_\mathrm{enter}^+$ and $t_\mathrm{exit}^-<t<t_\mathrm{exit}^+$ around $t=t_{\mathrm{enter},0}$ and $t=t_{\mathrm{exit},0}$, respectively, and 
write the mode equation in the form of the WKB-type solutions 
$t<t_\mathrm{enter}^-$ and $t_\mathrm{enter}^+<t<t_\mathrm{exit}^-$ as
\footnotesize
\begin{align}
    \tilde{\delta h}_k (t) &= \frac{\alpha_k}{\sqrt{2\omega_{h,k}}} \exp \left(- i\int^{t}_{t_0} \omega_{h,k} (t') dt' \right) +\frac{\beta_k}{\sqrt{2\omega_{h,k}}} \exp \left(i\int^{t}_{t_0}  \omega_{h,k}(t') dt' \right)~~:~~t<t_\mathrm{enter}^-,  \label{eq-wkb-before-enter}\\
    \tilde{\delta h}_k (t) &= \frac{a_k}{\sqrt{2|\omega_{h,k}|}} \exp \left(-\int^t_{t_{\mathrm{enter},0}} |\omega_{h,k} (t')| dt' \right) +\frac{b_k}{\sqrt{2|\omega_{h,k}|}} \exp \left( \int^t_{t_{\mathrm{enter},0}} |\omega_{h,k}(t')| dt' \right)~~:~~t_\mathrm{enter}^+<t<t_\mathrm{exit}^-,  \label{eq-wkb-between-enter-and-exit}
\end{align}
\normalsize
where the Bogoliubov coefficients $\alpha_k, \beta_k, a_k$, and $b_k$ are approximated to be constant 
and $t_0$ is an initial time when  $\alpha_k$ and  $\beta_k$ are defined. 
Here we require the normalization condition $|{\alpha}_k|^2-|{\beta}_k|^2 = 1$ and $ a_kb^*_k-a^*_kb_k=i $, 
respectively. 

In the following, we solve the mode equations when the adiabatic condition is violated to determine the 
matching condition for the ``transfer matrices'' between the Bogoliubov coefficients. 
First, the matching condition at $t \simeq t_{\mathrm{enter},0}$ is examined. 
For $ t $ sufficiently close to the zero-crossing point, $t=t_{\mathrm{enter},0}$, 
one can Taylor expand $ \omega^2_k $ as 
\begin{align}
    \omega^2_{h,k} &\approx 0+ \left. \frac{d(\omega^2_{h,k})}{dt} \right|_{t=t_{\mathrm{enter},0}} (t-t_{\mathrm{enter},0}) 
    \equiv  A_k (t-t_{\mathrm{enter},0}), 
\end{align}
with $ A_k<0 $, 
so that the equation of motion for $ \tilde{\delta h}_k $ becomes 
\begin{align}
    \ddot{\tilde{\delta h}}_k+ A_k (t-t_{\mathrm{enter},0}) \tilde{\delta h}_k\approx 0 . 
\end{align}
This is just the same as the stationary Schr\"{o}dinger equation with a linear potential. The exact solution to it is known to be Airy functions, i.e.  
\begin{align}\label{eq-Airy-enter}
    \tilde{\delta h}_k (t)=B_{1k}\mathrm{Ai} \left(A^{1/3}_k(t_{\mathrm{enter},0}-t)\right) +B_{2k} \mathrm{Bi} \left(A^{1/3}_k(t_{\mathrm{enter},0}-t)\right),
\end{align}
where $ B_{1k} $ and $ B_{2k} $ are complex constants. For $ t<t_{\mathrm{enter},0} $, the argument in Eq.~\eqref{eq-Airy-enter} is negative, then the leading term of the asymptotic expansion of the Airy function at 
$ \left|A^{1/3}_k (t_{\mathrm{enter},0}-t)\right| \gg 1 $ reads
\begin{align} \label{eq-Airy-before-enter}
    \tilde{\delta h}_k (t) \rightarrow  \frac{1}{\sqrt{2\omega_{h,k}}} \frac{|A_k|^{1/6}}{\sqrt{\pi}} &\left[ (B_{1k}+B_{2k}) \cos \left( \int^{t_{\mathrm{enter},0}}_t \omega_{h,k}(t') dt' \right) \notag \right. \\ 
    &\left.+ (B_{1k}-B_{2k}) \sin \left( \int^{t_{\mathrm{enter},0}}_t \omega_{h,k}(t') dt' \right)  \right] ~\notag \\
    = \frac{1}{\sqrt{2\omega_{h,k}}} \frac{|A_k|^{1/6}}{\sqrt{\pi}} &\left[\left(\frac{1-i}{2} B_{1k} + \frac{1+i}{2}B_{2k}\right) \exp \left(-i \int^t_{t_{\mathrm{enter},0}} \omega_{h,k}(t') dt' \right)  \right. \notag 
    \\&\left. + \left(\frac{1+i}{2} B_{1k} + \frac{1-i}{2}B_{2k}\right) \exp \left(i \int^t_{t_{\mathrm{enter},0}} \omega_{h,k}(t') dt' \right)\right].
\end{align}
On the other hand, for $ t>t_{\mathrm{enter},0} $, the argument in Eq.~\eqref{eq-Airy-enter} is positive, 
then with the leading term of the 
asymptotic expansion of the Airy function at 
$ A^{1/3}_k (t_{\mathrm{enter},0}-t) \gg 1$, Eq.~\eqref{eq-Airy-enter} is approximated as 
\footnotesize
\begin{align}\label{eq-Airy-after-enter}
    \tilde{\delta h}_k (t) \rightarrow & \frac{1}{\sqrt{2|\omega_{h,k}|}} \frac{|A_k|^{1/6}}{\sqrt{\pi}} \left[ \frac{B_{1k}}{\sqrt{2}} \exp \left( -\int^t_{t_{\mathrm{enter},0}} |\omega_{h,k}(t')| dt' \right) +\sqrt{2} B_{2k} \exp \left( \int^t_{t_{\mathrm{enter},0}} |\omega_{h,k}(t')| dt' \right) \right] ~ .
\end{align}
\normalsize
One can use these results to connect the asymptotic solutions obtained with the WKB approximation, Eqs.~\eqref{eq-wkb-before-enter} and~\eqref{eq-wkb-between-enter-and-exit}. 
It should be noted that, to do so, we require there is a regime where both WKB and Taylor expansion are valid, which implies that the following condition should be satisfied 
\begin{align}\label{eq-WKB-Taylor} 
    |A_k^{1/3}(t_{\mathrm{enter},0}-t)| \gg 1, \quad |A_k| \gg \left. \frac{1}{2}\frac{d^2 \omega_{h,k}^2}{dt^2}\right|_{t=t_{\mathrm{enter},0}} |t-t_{\mathrm{enter},0}|
\end{align}
as well as Eq.~\eqref{eq-adiabatic-condition}, 
which we will confirm in the end of this appendix. 

The matching conditions for $t<t_{\mathrm{enter},0}$ and $t>t_{\mathrm{enter},0}$  give
\begin{align}
     &\begin{pmatrix} \alpha_k \\ \beta_k \end{pmatrix} = \frac{|A_k|^{1/6}}{2 \sqrt{\pi}} \begin{pmatrix} (1-i) e^{i\theta_k} & (1+i) e^{i\theta_k} \\ (1+i) e^{-i \theta_k} & (1-i) e^{-i\theta_k} \end{pmatrix} \begin{pmatrix} B_{1k} \\B_{2k} \end{pmatrix}, \quad \theta_k \equiv \int_{t_0}^{t_{\mathrm{enter},0}} \omega_{h,k}(t') dt'~,  \label{eq-trans-enter-before}\\[0.2cm]
     &\begin{pmatrix} a_k \\ b_k \end{pmatrix} = \frac{|A_k|^{1/6}}{ 2 \sqrt{\pi}} \begin{pmatrix} \sqrt{2} & 0 \\ 0 & 2\sqrt{2} \end{pmatrix} \begin{pmatrix} B_{1k} \\B_{2k} \end{pmatrix}, \label{eq-trans-enter-after}
\end{align}
respectively. 
Here $\theta_k$ is the phase accumulation from $t_0$ to the entry of the tachyonic 
regime. 
From Eqs.~\eqref{eq-trans-enter-before} and \eqref{eq-trans-enter-after}, 
we can obtain the transfer matrix as
\begin{equation}
    \begin{pmatrix} a_k \\ b_k \end{pmatrix}  = \frac{1}{2\sqrt{2}} \begin{pmatrix}(1+i) e^{-i \theta_k} & (1-i) e^{i \theta_k}  \\ 2 (1-i) e^{-i \theta_k}  & 2(1+i) e^{i \theta_k}  \end{pmatrix} \begin{pmatrix} \alpha_{k} \\ \beta_{k} \end{pmatrix}.\label{eq-trans-enter}
\end{equation}

Next we examine the matching condition at $t \simeq t_{\mathrm{exit},0}$. 
For $ t $ sufficiently close to the end of tachyonic regime, $t=t_{\mathrm{exit},0}$, 
we can Taylor expand $ \omega^2_k $ again as 
\begin{align}
    \omega^2_{h,k} &\approx 0 + \left. \frac{d(\omega^2_{h,k})}{dt} \right|_{t=t_{\mathrm{exit},0}} (t-t_{\mathrm{exit},0}) 
    \equiv  C_k (t-t_{\mathrm{exit},0}), 
\end{align}
with $ C_k>0 $, so that the equation of motion for $ \tilde{\delta h}_k $ becomes 
\begin{align}
    \ddot{\tilde{\delta h}}_k+ C_k (t-t_{\mathrm{exit},0}) \tilde{\delta h}_k\approx 0 . 
\end{align}
In the same way as done in the above, the exact solution is expressed in terms of the Airy function, 
\begin{align}\label{eq-Airy-exit}
    \tilde{\delta h}_k (t)=D_{1k}\mathrm{Ai} \left(C^{1/3}_k(t_{\mathrm{exit},0}-t)\right) +D_{2k} \mathrm{Bi} \left(C^{1/3}_k(t_{\mathrm{exit},0}-t)\right),
\end{align}
where $ D_{1k} $ and $ D_{2k} $ are complex constants. 
Once more, one can obtain the following asymptotic expressions of the mode functions. 
For $ \left|C^{1/3}_k (t_{\mathrm{exit},0}-t)\right| \gg 1 $, we have
\begin{align} \label{eq-Airy-after-exit}
    \tilde{\delta h}_k (t) \rightarrow  \frac{1}{\sqrt{2\omega_{h,k}}} \frac{C_k^{1/6}}{\sqrt{\pi}} &\left[ (D_{1k}+D_{2k}) \cos \left( \int^t_{t_{\mathrm{exit},0}} \omega_{h,k}(t') dt' \right) \notag \right. \\ 
    &\left.+ (D_{1k}-D_{2k}) \sin \left( \int^t_{t_{\mathrm{exit},0}} \omega_{h,k}(t') dt' \right)  \right] ~\notag \\
    = \frac{1}{\sqrt{2\omega_{h,k}}} \frac{C_k^{1/6}}{\sqrt{\pi}} &\left[\left(\frac{1+i}{2} D_{1k} + \frac{1-i}{2}D_{2k}\right) \exp \left(-i \int^t_{t_{\mathrm{exit},0}} \omega_{h,k}(t') dt' \right)  \right. \notag 
    \\&\left. + \left(\frac{1-i}{2} D_{1k} + \frac{1+i}{2}D_{2k}\right) \exp \left(i \int^t_{t_{\mathrm{exit},0}} \omega_{h,k}(t') dt' \right)\right].
\end{align}
For $ C^{1/3}_k (t_{\mathrm{exit},0}-t) \gg 1$, Eq.~\eqref{eq-Airy-exit} is approximated as 
\footnotesize
\begin{align}\label{eq-Airy-before-exit}
    \tilde{\delta h}_k (t) \rightarrow & \frac{1}{\sqrt{2|\omega_{h,k}|}} \frac{C_k^{1/6}}{\sqrt{\pi}} \left[ \frac{D_{1k}}{\sqrt{2}} \exp \left( \int^t_{t_{\mathrm{exit},0}} |\omega_{h,k}(t')| dt' \right) +\sqrt{2} D_{2k} \exp \left( -\int^t_{t_{\mathrm{exit},0}} |\omega_{h,k}(t')| dt' \right) \right] ~ .
\end{align}
\normalsize
By using these results, we connect the WKB solutions, Eqs.~\eqref{eq-WKB-after-exit} and
\eqref{eq-wkb-between-enter-and-exit}, with the matching conditions for $t>t_{\mathrm{exit},0}$ and $t<t_{\mathrm{exit},0}$ 
as
\begin{align}
     &\begin{pmatrix} {\tilde \alpha}_k \\ {\tilde \beta}_k \end{pmatrix} = \frac{C_k^{1/6}}{2 \sqrt{\pi}} \begin{pmatrix} 1+i & 1-i \\ 1-i & 1+i  \end{pmatrix} \begin{pmatrix} D_{1k} \\D_{2k} \end{pmatrix},   \label{eq-trans-exit-after}\\
     &\begin{pmatrix} a_k \\ b_k \end{pmatrix} = \frac{C_k^{1/6}}{ 2 \sqrt{\pi}} \begin{pmatrix}  0&2 \sqrt{2} e^{\Omega_k} \\ \sqrt{2} e^{-\Omega_k} & 0\end{pmatrix} \begin{pmatrix} D_{1k} \\D_{2k} \end{pmatrix}, \quad \Omega_k \equiv \int_{t_{\mathrm{enter},0}}^{t_{\mathrm{exit},0}} |\omega_{h,k}(t')| dt'\label{eq-trans-exit-before}
\end{align}
respectively. 
Here $\Omega_k$ is the ``phase'' accumulation during the tachyonic regime from $t_{\mathrm{enter},0}$ to $t_{\mathrm{exit},0}$.  
Consequently, the transfer matrix is given by 
\begin{equation}
    \begin{pmatrix} {\tilde \alpha}_k \\ {\tilde \beta}_k \end{pmatrix}  = \frac{1}{2\sqrt{2}} \begin{pmatrix}(1-i) e^{- \Omega_k} & 2(1+i) e^{\Omega_k}  \\  (1+i) e^{-\Omega_k}  & 2(1-i) e^{ \Omega_k}  \end{pmatrix} \begin{pmatrix} a_{k} \\ b_{k} \end{pmatrix}. \label{eq-trans-exit}
\end{equation}
From Eqs.~\eqref{eq-trans-enter} and \eqref{eq-trans-exit}, finally we obtain the transfer matrix
between before and after the tachyonic regime as 
\begin{equation}
    \begin{pmatrix} {\tilde \alpha}_k \\ {\tilde \beta}_k \end{pmatrix}  =  \begin{pmatrix} e^{-i\theta_k} (e^{\Omega_k}+e^{-\Omega_k}/4)& i e^{i\theta_k} (e^{\Omega_k}-e^{-\Omega_k}/4) \\  -i e^{-i\theta_k} (e^{\Omega_k}-e^{-\Omega_k}/4) & e^{i\theta_k} (e^{\Omega_k}+e^{-\Omega_k}/4) \end{pmatrix} \begin{pmatrix} \alpha_{k} \\  \beta_{k} \end{pmatrix}. \label{eq-trans-all}
\end{equation}
As a result, the occupation number of the particle production of a certain $ k $ mode is given by 
\begin{align}
    n_k=  \left|\tilde{\beta}_k\right|^2 =& \left| -ie^{-i\theta_k} \alpha_{k}  (e^{\Omega_k}-e^{-\Omega_k}/4) +e^{i\theta_k} \beta_{k}  (e^{\Omega_k}+e^{-\Omega_k}/4) \right|^2 ~.
\end{align}
By setting the vacuum initial condition, $ \alpha_{k}=1 $ and $ \beta_{k}=0 $, 
it is simply the expression we adopt in Sec.~\ref{Sec-3}, 
\begin{align}
    n_k = \left|\tilde{\beta}_k\right|^2 =&  \left| e^{\Omega_k}-e^{-\Omega_k}/4 \right|^2 
    \approx e^{2\Omega_k} \label{eq-tachyonic-nk-approx} 
\end{align}
where strong tachyonic instability is assumed in the last line.

Finally, we examine whether the connection between the WKB solutions and the Airy functions are valid, which means that one should check if there is a regime when both Eqs.~\eqref{eq-adiabatic-condition} and ~\eqref{eq-WKB-Taylor} are satisfied simultaneously.
In Sec.~\ref{Sec-3}, we have derived that 
\begin{align}\label{eq-omega-approx}
    \omega_{h,k}^2 =\frac{k^2}{a^2} +m^2_h \approx \frac{k^2}{a^2} -m^2_{h,\mathrm{max}} \sin \left[ M(t-t_{\mathrm{enter},0}) \right]
\end{align}
where $ m^2_{h,\mathrm{max}} \equiv 3\xi M M_c C_2 C_1 $. 
Note that $m_{h,\mathrm{max}} > M$ is satisfied for $\sin 2 \theta \gsim \xi_c^{-1}$, 
which is the case of our interest, (close to) the exact hill-climbing cases. 
The  dominant contribution to the energy density of the produced particles is from the modes with $ k^2/a^2 \lesssim m^2_{h,\mathrm{max}} $ and  $ m^2_{h,\mathrm{max}} - k^2/a^2  \simeq  m^2_{h,\mathrm{max}}$. 
The mode with $ m^2_{h,\mathrm{max}} - k^2/a^2  \ll  m^2_{h,\mathrm{max}}$ 
does not have a long time for the tachyonic period and $\Omega_k$ does not become so large. 
For the mode with $k^2/a^2 \ll m^2_{h,\mathrm{max}}$, the energy carried by each mode is not 
so large to give dominant contributions to the total energy density.

The left hand side of Eq.~\eqref{eq-adiabatic-condition} is expanded with respect to
$t-t_{\mathrm{enter},0}$ as
\begin{equation}\label{eq-adiabaticity}
    \left| \frac{\ddot{\omega}_{h,k}/\omega_{h,k}-3\dot{\omega}^2_{h,k}/(2\omega^2_{h,k})}{2\omega^2_{h,k}} \right| \simeq \frac{5}{16\sqrt{m_{h,\mathrm{max}}^4-(k/a)^4}M}|t-t_{\mathrm{enter},0}|^{-3},
\end{equation}
where we have omitted $\dot{a}$ since the cosmic expansion is smaller than the time scale of 
this dynamics. 
For the modes with $k/a\simeq m_{h,\mathrm{max}}$, the adiabatic condition reads
\begin{equation} \label{eq-adiabatic-condition-validity}
    |t-t_{\mathrm{enter},0}| \gg (m_{h,\mathrm{max}}^2 M)^{-1/3}. 
\end{equation}

On the other hand, from the first inequality of Eq.~\eqref{eq-WKB-Taylor}, 
one finds that the asymptotic expansion of the Airy function is valid for 
\begin{equation} \label{eq-asymptotic-validity}
    |t-t_{\mathrm{enter},0}| \gg |A_k|^{-1/3} = \left(\sqrt{1-\frac{k^4/a^4}{m_{h,\mathrm{max}}^4}}m_{h,\mathrm{max}}^2 M\right)^{-1/3} \simeq (m_{h,\mathrm{max}}^2 M)^{-1/3},  
\end{equation}
while from the second inequality of Eq.~\eqref{eq-WKB-Taylor} 
the linear approximation of the potential is found to be valid for 
\begin{equation} \label{eq-linear-validity}
    |t-t_{\mathrm{enter},0}| \ll 2 M^{-1} \sqrt{\frac{m_{h,\mathrm{max}}^4}{(k/a)^4}-1}  \simeq M^{-1}. 
\end{equation}
Therefore, from Eqs.~\eqref{eq-adiabatic-condition-validity},~\eqref{eq-asymptotic-validity}, 
and~\eqref{eq-linear-validity}, we can see that it is permissible to use the matching condition before and after the non-adiabatic period 
around $t\simeq t_{\mathrm{enter},0}$ 
for the case of our interest, i.e. $m_{h,\mathrm{max}}>M$ and $k/a \simeq m_{h,\mathrm{max}}$. 
One can easily show the validity of the matching condition 
around $t\simeq t_{\mathrm{exit},0}$ in the same way.

\section{The smallest \texorpdfstring{$ N $}{Lg} in Branch 2 to complete preheating}
\label{appendix-B}

In this appendix, we evaluate the energy density of the Higgs field fluctuations produced by the tachyonic instability in a more precise analytic way to
find the smallest $ \xi_N $ sufficient to complete preheating. 
This case is expected to be in the regime $ M\sim M_c $. 

\begin{figure}[h]
	\centering
	\includegraphics[width=.6\textwidth]{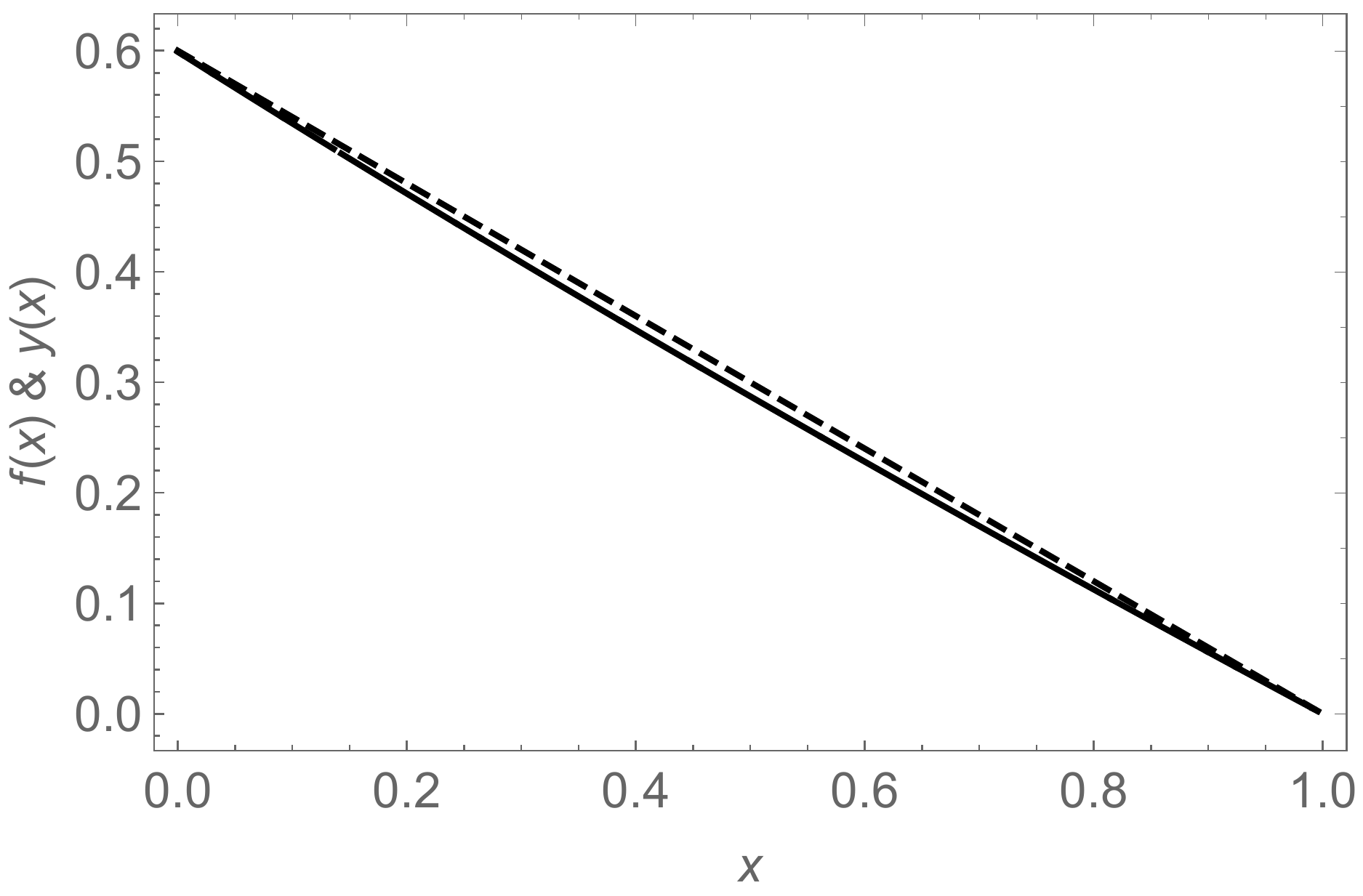}
	\caption{
	The comparison between the functions $f(x)$ (solid line) and $y(x)$ (dashed line).  
	It is clear that the simpler function $y(x)$ approximates the more precise function $f(x)$ very well at $0<x<1$.}
	\label{fig-Omegak-smallest-xi}
\end{figure}

\begin{figure}
	\centering
	\includegraphics[width=.6\textwidth]{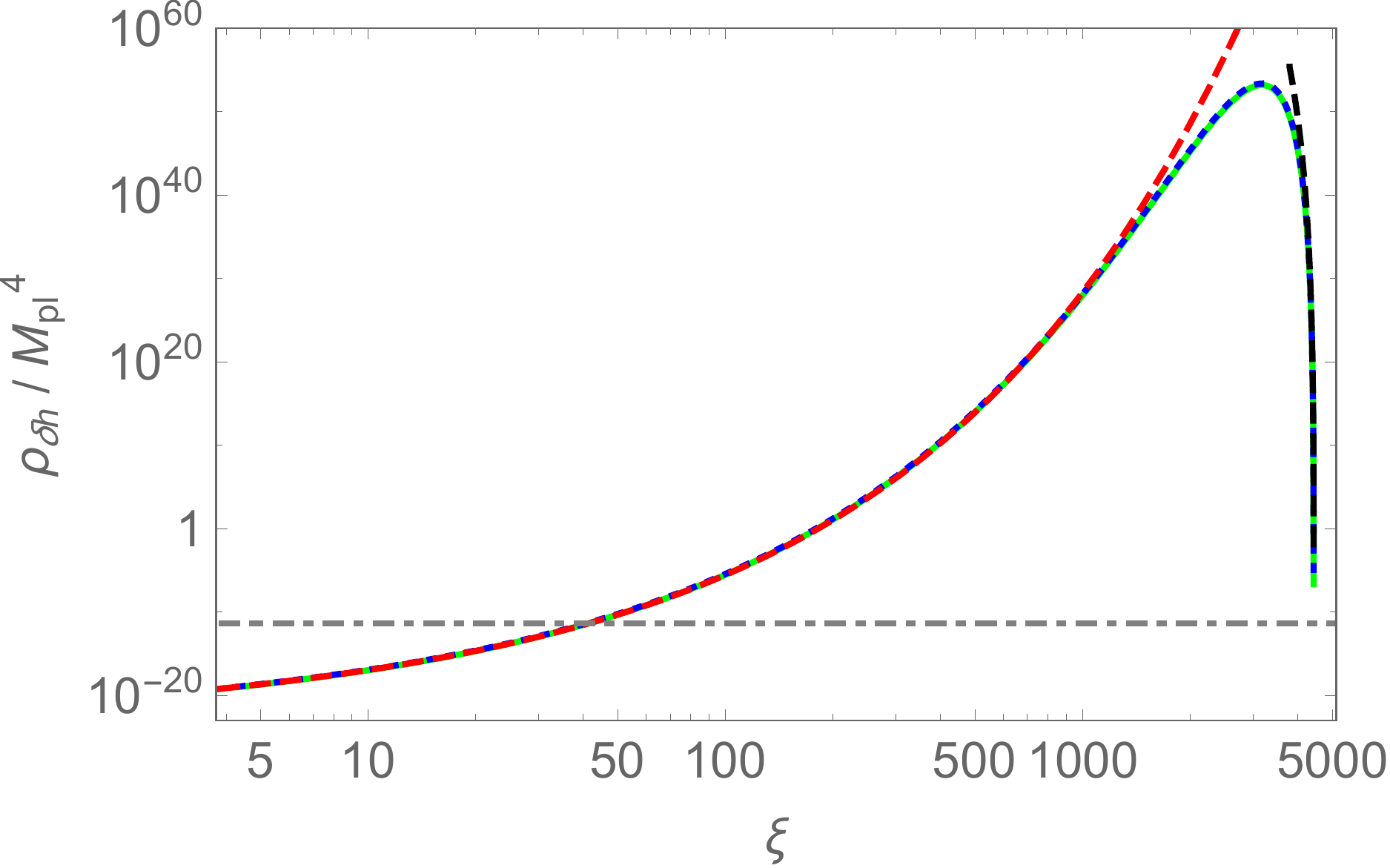}
	\caption{
	The energy density of produced Higgs fluctuations as a function of $\xi$ for $5<\xi<\xi_c$. 
	The light green line stands for the $\rho_{\delta h}$ calculated with the precise function $ f(x) $ in $ \Omega_k $ (Eq.~\eqref{eq-omegak-analytic-approx}). The blue dotted line is for the expression Eq.~\eqref{eq-rho-M-general}. The red dashed line and black dashed line represent the limits $ \xi\ll\xi_c $  and $ M\gg M_c $ (Eq.~\eqref{eq-rho-xi-limits}), respectively. The gray dotted-dashed line is $ C^2_1\rho_{\rm inf}/2 $.}
	\label{fig-rho-xi-general}
\end{figure}

\begin{figure}[h]
	\centering
	\includegraphics[width=.6\textwidth]{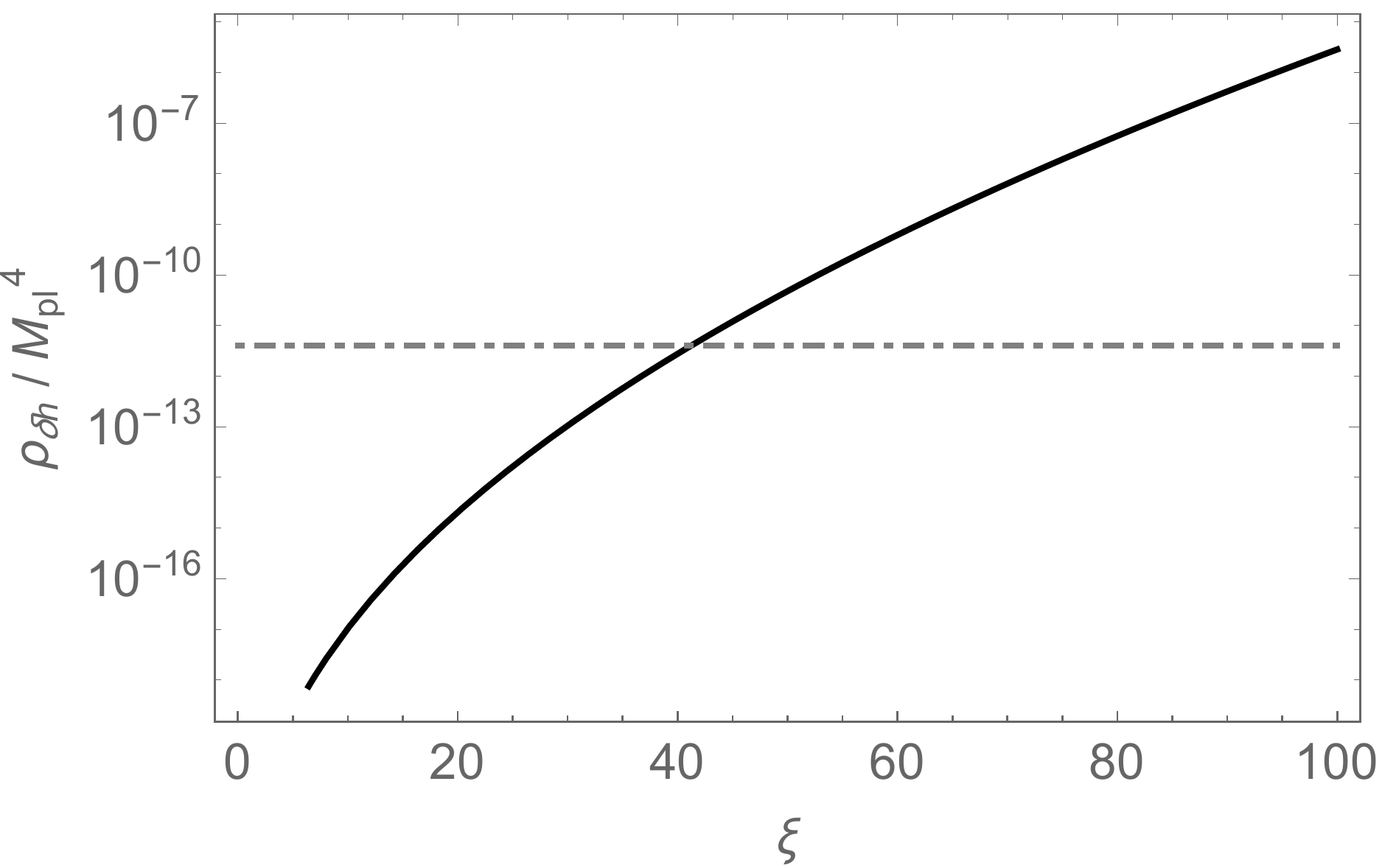}
	\caption{The same to Fig.~\ref{fig-rho-xi-general} but for $0<\xi<100$. Solid line is $ \rho(\xi) $ and the dashed line is $ C^2_1\rho_{\rm inf}/2 $.}
	\label{fig-rho-xi}
\end{figure}

In Sec.~\ref{Sec-3}, we give the analytical formula to estimate $ \Omega_k $ for a given $ \omega_{h,k}^2=k^2 -m^2_{h,\rm max} \sin [M(t-t_{\mathrm{enter},0})] $ where the cosmic expansion can be safely neglected. Now we estimate $ \Omega_k $ for a general $ k $ and $ \xi $ by assuming that the inflaton can always climb up the hill exactly 
\begin{align}
	\Omega_k &= \int^{t_{\rm{exit}}(k)}_{t_{\rm{enter}}(k)} | \omega_{h,k} (t') | dt' \nonumber \\
	&= \frac{m_{h,\rm max}}{M} \int^{\pi-\arcsin(k^2/m^2_{h,\rm max})}_{\arcsin(k^2/m^2_{h,\rm max})} \left( \sin t -\frac{k^2}{m^2_{h,\rm max}} \right)^{1/2} dt \nonumber \\
	&= 4\frac{m_{h,\rm max}}{M} \left( 1- \frac{k^2}{m^2_{h,\rm max}} \right)^{1/2} E\left[ \frac{1}{2} \arccos \left( \frac{k^2}{m^2_{h,\rm max}} \right), \frac{2}{1-k^2/m^2_{h,\rm max}} \right] \nonumber \\
	&\equiv 4 \sqrt{3C_2C_1} \left( \frac{\lambda}{3} \right)^{1/4} \sqrt{\frac{M_{\rm pl}}{M}} \left( 1-\frac{M_c^2}{M^2} \right)^{1/4} f\left( \frac{k^2}{m^2_{h,\rm max}} \right) \nonumber \\
	&\approx
	\begin{dcases}
	    4\sqrt{3C_2C_1} \sqrt{\xi} f\left( \frac{k^2}{m^2_{h,\rm max}} \right) &:~~\xi \ll \xi_c ~~\text{or}~~ M\simeq M_c, \\
	    4 \sqrt{3C_2C_1} \left( \frac{\lambda}{3} \right)^{1/4} \sqrt{\frac{M_{\rm pl}}{M}} f\left( \frac{k^2}{m^2_{h,\rm max}} \right) &:~~\xi \lsim \xi_c ~~\text{or}~~ M\gg M_c,
	\end{dcases} \label{eq-omegak-analytic-approx}
\end{align}
where $ k\leq m_{h,\rm max} $, $ E[\phi,x] =\int^{\phi}_0 (1-x \sin^2 t)^{1/2} dt $ is the elliptic integral of the second kind and $ f(x)\equiv (1-x)^{1/2} E[\arccos(x)/2, 2/(1-x)] $. Figure~\ref{fig-Omegak-smallest-xi} shows the function $ f(x) $ and a fitting function $ y(x) =0.6 (1-x) $.  One can see that 
$y(x)$ approximates $ f(x) $ at $0<x<1$ very well. 
Hereafter, we use $ y(x) =0.6 (1-x) $ to replace $ f(x) $ for simplicity. As a result, the number density of produced particles is given by 
\begin{align}
	n_k \simeq e^{2\Omega_k} = \exp \left[ 8\sqrt{3C_2C_1} \left( \frac{\lambda}{3} \right)^{1/4} \sqrt{\frac{M_{\rm pl}}{M}} \left( 1-\frac{M_c^2}{M^2} \right)^{1/4} \times 0.6 \left( 1-\frac{k^2}{m^2_{h,\rm max}} \right) \right]~. 
\end{align}
Putting the numerical values $ C_2=0.72 $ and $ C_1 =0.25 $ into the equation above and the typical value $ \lambda=0.01 $, we obtain 
\begin{align}
	n_k = \exp \left[ 0.85 \sqrt{\frac{M_{\rm pl}}{M}} \left( 1-\frac{M_c^2}{M^2} \right)^{1/4} \left( 1-\frac{k^2}{m^2_{h,\rm max}} \right) \right] ~.
\end{align}
With the help of this result, one can estimate the comoving energy density of produced particles $\rho_{\delta h}$ as a function of $M$ for Branch 1 and $ \xi $ for Branch 2, 
\begin{align}
	\rho_{\delta h} (\xi) &= \int \frac{d^3k}{(2\pi)^3}~\omega_{h,k} n_k \simeq \int^{m_{h,\rm max}}_0 \frac{k^3}{2\pi^2} \exp \left[ 0.85 \sqrt{\frac{M_{\rm pl}}{M}} \left( 1-\frac{M_c^2}{M^2} \right)^{1/4} \left( 1-\frac{k^2}{m^2_{h,\rm max}} \right) \right] dk \nonumber \\[0.15cm]
	&= \frac{m^4_{h,\rm max}}{2\pi^2} \int^1_0 k^3 \exp \left[ 0.85 \sqrt{\frac{M_{\rm pl}}{M}} \left( 1-\frac{M_c^2}{M^2} \right)^{1/4} \left( 1-k^2 \right) \right] dk \nonumber \\[0.15cm]
	&\simeq 3.4\times 10^{-5} M_{\rm pl} M^3 \sqrt{ 1- \frac{M^2_c}{M^2} } \nonumber \\[0.15cm]
	&~~~~\times \left\{\exp\left[0.85 \sqrt{\frac{M_{\rm pl}}{M}} \left(1-\frac{M^2_c}{M^2}\right)^{1/4}\right] -0.85\sqrt{\frac{M_{\rm pl}}{M}}\left(1-\frac{M^2_c}{M^2}\right)^{1/4} -1 \right\} \label{eq-rho-M-general}  \\[0.2cm]
	&\simeq
	\begin{dcases}\label{eq-rho-xi-limits}
	    5.9\times 10^{-4}M^4_c \xi \left( -1 +e^{3.5\sqrt{\xi}}- 3.5\sqrt{\xi} \right) &:~~\xi \ll \xi_c ~~\text{or}~~ M\simeq M_c ~, \\
	    3.4\times 10^{-5} M_{\rm pl} M^3 \left( -1+ e^{0.85\sqrt{\frac{M_{\rm pl}}{M}}} -0.85\sqrt{\frac{M_{\rm pl}}{M}} \right)  &:~~\xi \lsim \xi_c ~~\text{or}~~ M\gg M_c ~. 
	\end{dcases}
\end{align}
Here we evaluate $ \rho_{\delta h} $ by approximating $m_h^2=0$ at later times $t>t_\mathrm{exit}(k)$. 
Figure~\ref{fig-rho-xi-general} shows $ \rho (\xi) $ and its asymptotic forms. 
If we zoom in the region $ \xi \ll \xi_c $, we get Fig.~\ref{fig-rho-xi}. 
With these analytic investigation, we can see that $\rho_{\delta h}$ 
gets larger than the half of the background energy density $C_1^2 \rho_\mathrm{inf}/2$, 
that means the completion of preheating at $\xi\gtrsim 50$. 
It also means that the smallest $ \xi_N $ sufficient to complete preheating solely by tachyonic instability is around $ \xi_N \simeq 50 $ which corresponds to $ N=4 $ in Branch 2. Therefore, conservatively speaking, for $ \xi \leq \xi_{N=4, \rm{Branch2}} $, the tachyonic effect is not strong enough to complete preheating. 

\section{Survival rate}
\label{appendix-C}

In this appendix, we study the case when the inflaton cannot fully realize the tachyonic instability, namely the ending moment of the tachyonic effect $ t_{\rm drop} <t_{\mathrm{exit},0} $. We define a new quantity $ R \equiv (t_{\rm drop}-t_{\mathrm{enter},0})/ (t_{\mathrm{exit},0} -t_{\mathrm{enter},0}) = (t_{\rm drop}-t_{\mathrm{enter},0})M/\pi $ as the ``survival rate" of the inflaton on the potential hill which plays the same role as $t_{\rm drop}$. 
The drop-off time $t_\mathrm{drop}$ and the survival rate $R$ 
are determined by the model parameters, correctly speaking. 
In the following, we instead regard them as free phenomenological parameters and evaluate the particle production in terms of the survival
rate $R$. 
In this way, we can determine the required survival rate for the 
completion of preheating. 
The possible $k$ modes that experience tachyonic instability are different for $0\leq R<1/2$ and $1/2\leq R\leq 1$, as one can easily see that the maximal value of $|m_h^2|$ varies with $R$. So for convenience we separate the problem into two main parts, (1) $ 1/2 \leq R \leq 1 $ and (2) $ 0\leq R <1/2 $. 

\subsection{Case (1): \texorpdfstring{$ 1/2 \leq R \leq 1 $}{Lg}}

\begin{figure}[h]
	\centering
    \includegraphics[width=.6\textwidth]{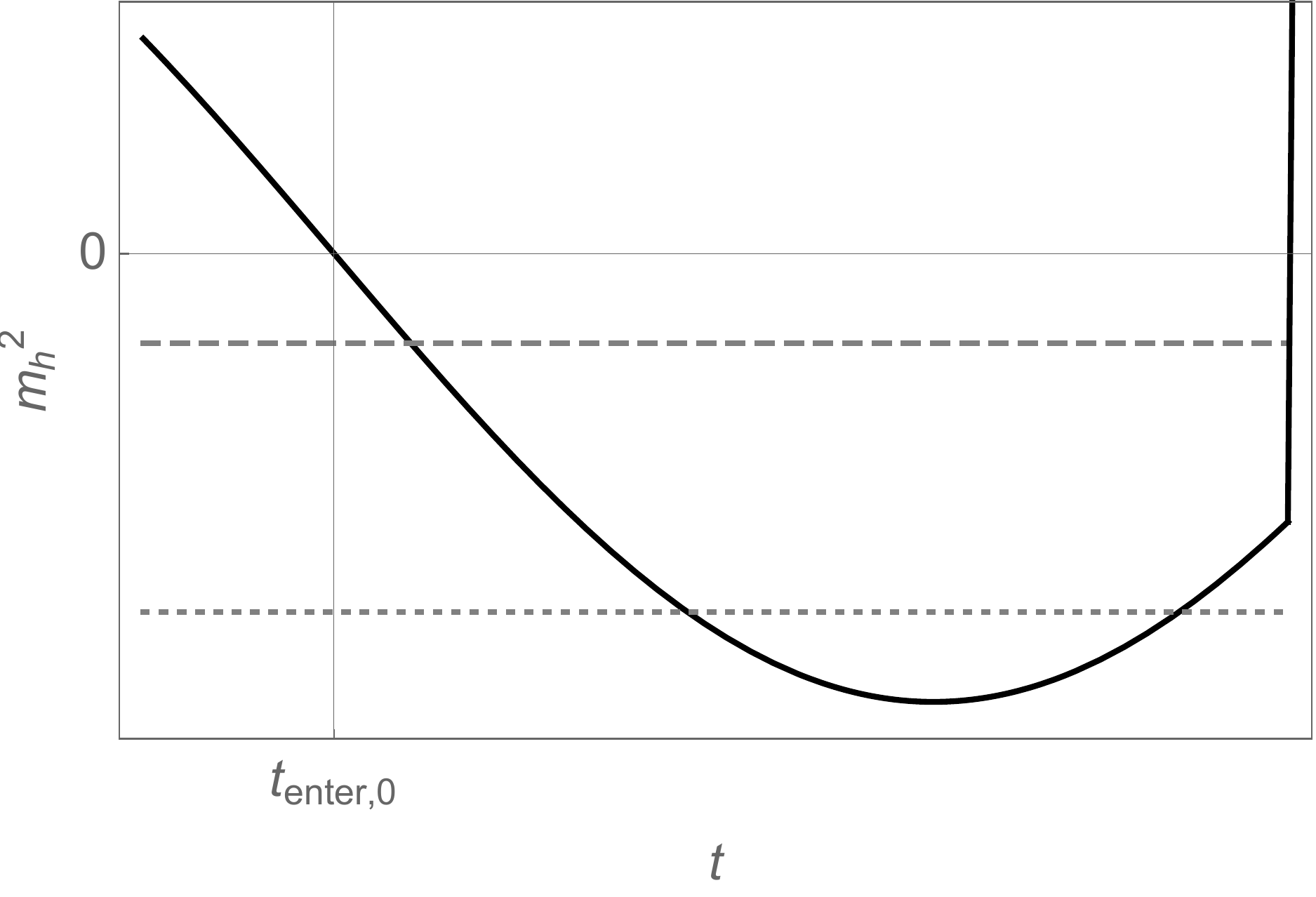}
	\caption{A schematic picture illustrating Case (1)-(a) and Case (1)-(b) is shown. The solid black line represents the effective mass squared of Higgs fluctuations $m_h^2(t)$. The gray dashed line corresponds to $k^2$ for the case (1)-(a) where $ 0\leq k^2/m^2_{h,\rm{max}} \leq \sin(\pi R) $ and the gray dotted line for the case (1)-(b) where $ \sin (\pi R) \leq k^2/m^2_{h,\rm{max}} \leq 1 $.}
	\label{fig-tdrop-case1}
\end{figure}

In this case, 
we need to further separate the problem into two parts in the $k$-domain, 
namely, (a) the case $\omega_{h,k}^2$ is negative at $t=t_\mathrm{drop}$, $ 0\leq k^2/m^2_{h,\rm max} \leq \sin (\pi R) $, 
and (b) the case $\omega_{h,k}^2$ is positive at $t=t_\mathrm{drop}$, $ \sin (\pi R) < k^2/m^2_{h,\rm max} \leq 1 $. 
See Fig.~\ref{fig-tdrop-case1} for the schematic picture that shows 
the representative $k$-modes for each case.
When we calculate $ \Omega_k $, 
in the case (a) the upper limit of the time integration is taken to be $ t_{\rm drop} $ which is $ k $-independent, while in the case (b) 
the upper limit integration is taken to be $t_\mathrm{drop}(k)$ (see Eq.~\eqref{eq-nk-non-exact}) 
which is $ k $-dependent. 

In a similar way to that used in Appendix~\ref{appendix-B}, in the case (a) one can calculate
\begin{align}
	\Omega_k &= \int ^{t_{\rm drop}}_{t_{\rm{enter}}(k)} |\omega_{h,k}(t')| dt'
	= \frac{m_{h,\rm max}}{M(\xi)} \int^{\pi R}_{\arcsin (k^2/m^2_{h,\rm max})} \left( \sin t -\frac{k^2}{m^2_{h,\rm max}} \right)^{1/2} dt \nonumber \\
	&= \frac{2m_{h,\rm max}}{M(\xi)} \sqrt{1-\frac{k^2}{m^2_{h,\rm max}}} \nonumber \\
	&~~~~\times \left( E\left[ \frac{1}{2}\arccos \left( \frac{k^2}{m^2_{h,\rm max}} \right), \frac{2}{1-k^2/m^2_{h,\rm max}} \right] +E\left[ \frac{\pi}{4} (2R-1), \frac{2}{1-k^2/m^2_{h,\rm max}} \right] \right) \label{eq-tdrop-case1-a} ~.
\end{align}
Here, we explicitly write $M(\xi)$ to show that $M$ is a function of 
$\xi$ through the equation Eq.~\eqref{eq-observation-constraint}. 
One might think $\Omega_k$ vanishes when $ k^2/m^2_{h,\rm max} =\sin (\pi R) $, 
but it is not the case. 
Since here we choose the argument of $ \arcsin (x) $ to be within $ 0 \leq x \leq \pi/2 $, 
$\arcsin (k^2/m^2_{h,\rm max}) \neq \pi R$ in this case, but instead we have 
\begin{align}
	\arccos \left( \frac{k^2}{m^2_{h,\rm max}} \right) &= \frac{\pi}{2} -\arcsin \left( \frac{k^2}{m^2_{h,\rm max}} \right) \\
	&= \frac{\pi}{2} -\pi (1-R) =\pi R -\frac{\pi}{2} ~.
\end{align}
Therefore, Eq.~\eqref{eq-tdrop-case1-a} is non-zero in the case $ k^2/m^2_{h,\rm max} =\sin (\pi R) $. 
In the case (b), we can calculate $\Omega_k$ in the same way 
in Appendix~\ref{appendix-B} as 
\begin{align}
	\Omega_k = \frac{4m_{h,\rm max}}{M(\xi)} f\left( \frac{k^2}{m^2_{h,\rm max}} \right) \label{eq-tdrop-case1-b}~.
\end{align}
As a result, the energy density of produced particles $ \rho_{\delta h} $ is obtained as a function of the model parameter $\xi$ and 
the survival rate $R$ 
by  integrating over the $ k $-space as 
\begin{align}
	\rho_{\delta h} (\xi, R) 
	&= \int \frac{d^3k}{(2\pi)^3} \omega_{h,k} n_k 
	\simeq \frac{1}{2\pi^2} \int k^3 e^{2\Omega_k} dk 
	= \frac{m^4_{h,\rm max}}{2\pi^2} \left( I_{11}(\xi,R) +I_{12}(\xi,R) \right)
\end{align}
where 
\footnotesize
\begin{align}
I_{11}(\xi, R) \equiv& \int^{\sqrt{\sin (\pi R)}}_0 k^3 \exp \left[ \frac{4m_{h,\rm max}}{M(\xi)} \sqrt{1-k^2} \left( E\left[ \frac{1}{2}\arccos \left( k^2 \right), \frac{2}{1-k^2} \right] +E\left[ \frac{\pi}{4} (2R-1), \frac{2}{1-k^2} \right] \right) \right] dk \\
	I_{12}(\xi, R) \equiv& \int^1_{\sqrt{\sin (\pi R)}} k^3 \exp \left[ 8\frac{m_{h,\rm max}}{M(\xi)} f(k^2/m_{h,\mathrm{max}}^2) \right] dk.  
\end{align}
\normalsize

\begin{figure}[h]
	\centering
	\includegraphics[width=.5\textwidth]{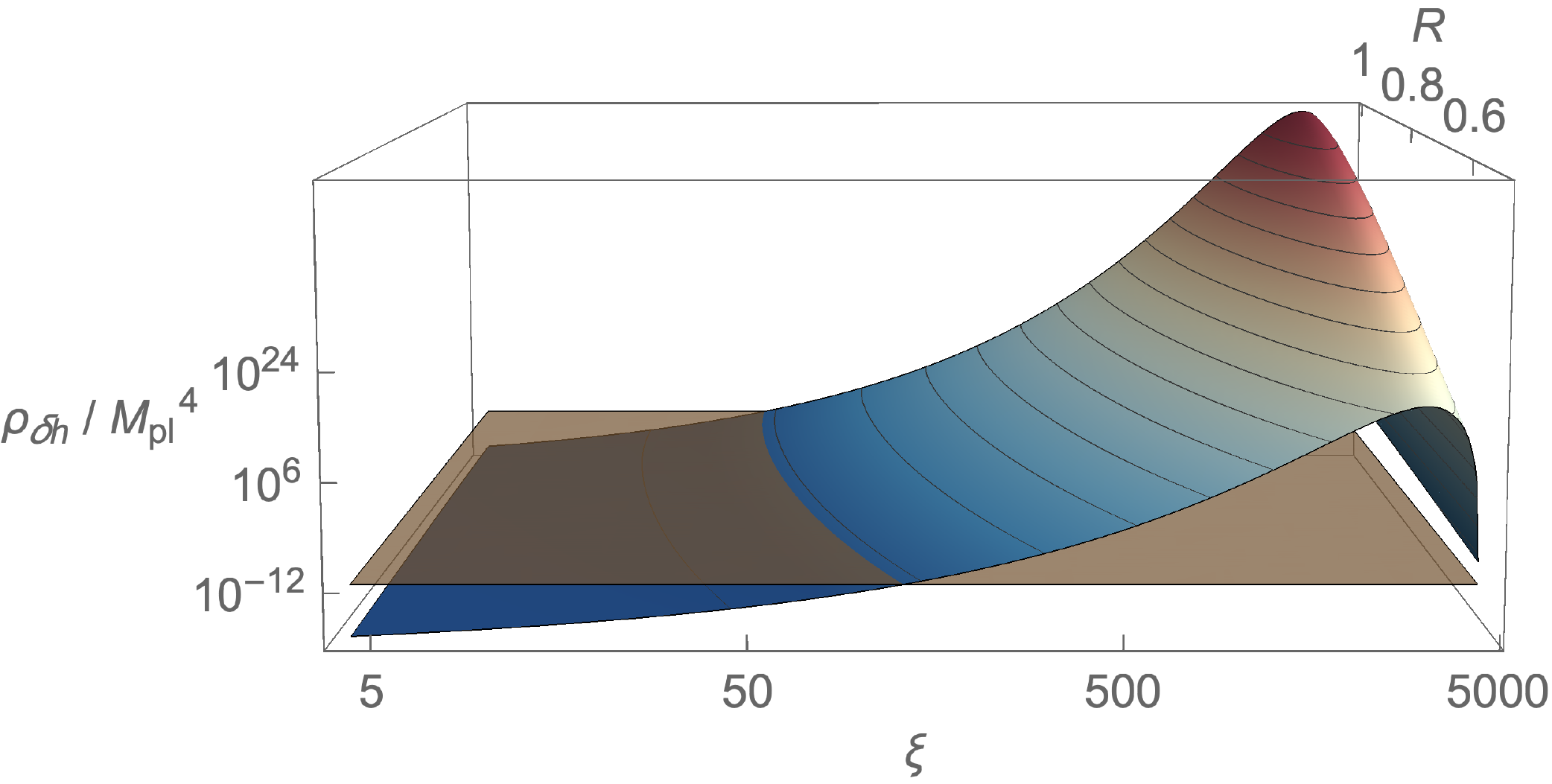}
	\caption{The energy density of Higgs fluctuations after the tachyonic instability as a function of $\xi$ and $R$ for $ 1/2 \leq R \leq 1 $. The gray transparent surface represents $ C^2_1\rho_{\rm inf}/2 $. The surface of section at $R=1$ corresponds to $\rho_{\delta h}$ evaluated in Appendix~\ref{appendix-B}.}
	\label{fig-rho-big-R}
\end{figure}

Fig.~\ref{fig-rho-big-R} shows the energy density of Higgs fluctuations after the tachyonic instability $\rho_{\delta h}$  as the function of $R$ and $\xi$ for  $ 1/2 \leq R \leq 1 $. 
We can see that the condition for the complete preheating is not 
sensitive to the survival rate $R$ for $R>1/2$. 
For  $ \xi \gsim 100 $, preheating is always completed for $R\geq 1/2$. 
On the contrary, for $\xi \lsim 50$, preheating cannot be completed solely by tachyonic effect even if $R=1$. 

In reality, the survival rate is very tiny in the most part of the parameter space and is order of the unity only around each exact hill-climbing parameter $ \xi_N $ (or $ M_N $). 
In other words, our investigation here is meaningful only around these parameters. 
Therefore, we shall understand that the survival rate $R$ is the one
around  $ \xi_N $ (or $ M_N $) and the calculated energy density
is the one around them, $ \rho_{\delta h}  (\xi_N, R) $. 
From that we can see how $ \rho_{\delta h} (\xi_N, R) $ depends on $R$
for each $\xi_N$. 

\subsection{Case (2): \texorpdfstring{$ 0\leq R <1/2 $}{Lg}}

In the case (2) with $ 0\leq R <1/2 $, 
$\omega_{h,k}^2$ is always negative at $t_\mathrm{drop}$ 
if the mode experiences the tachyonic instability at $t<t_\mathrm{drop}$. 
Then we can evaluate $\Omega_k$ as 
\begin{align}
	\Omega_k &= \int ^{t_{\rm drop}}_{t_{\rm{enter}}(k)} |\omega_{h,k}(t')| dt' 
	= \frac{m_{h,\rm max}}{M} \int^{\pi R}_{\arcsin (k^2/m^2_{h,\rm max})} \left( \sin t -\frac{k^2}{m^2_{h,\rm max}} \right)^{1/2} dt \nonumber \\
	&= \frac{2m_{h,\rm max}}{M} \sqrt{1-\frac{k^2}{m^2_{h,\rm max}}} \nonumber \\
	&~~~~\times \left( E\left[ \frac{1}{2}\arccos \left( \frac{k^2}{m^2_{h,\rm max}} \right), \frac{2}{1-k^2/m^2_{h,\rm max}} \right] -E\left[ \frac{\pi}{4} (1-2R), \frac{2}{1-k^2/m^2_{h,\rm max}} \right] \right), \label{eq-tdrop-case2} 
\end{align}
whose form is the same as Eq.~\eqref{eq-tdrop-case1-a}. However, one should notice the difference that, when $ k^2/m^2_{h,\rm max} =\sin (\pi R) $, 
\begin{align}
    \arccos \left( \frac{k^2}{m^2_{h,\rm max}} \right)
    &= \frac{\pi}{2} -\arcsin \left( \frac{k^2}{m^2_{h,\rm max}} \right) 
    = \frac{\pi}{2} -\pi R,
\end{align}
because $ \pi R <\pi/2 $. In other words, Eq.~\eqref{eq-tdrop-case2} vanishes if $ k^2/m^2_{h,\rm max} =\sin (\pi R) $. 
This is because if $ k^2/m^2_{h,\rm max}  \geq \sin (\pi R) $, 
the mode does not experience the tachyonic instability 
before the Higgs drop-off. 

\begin{figure}[h]
	\centering
	\begin{subfigure}
		\centering
		\includegraphics[width=.48\textwidth]{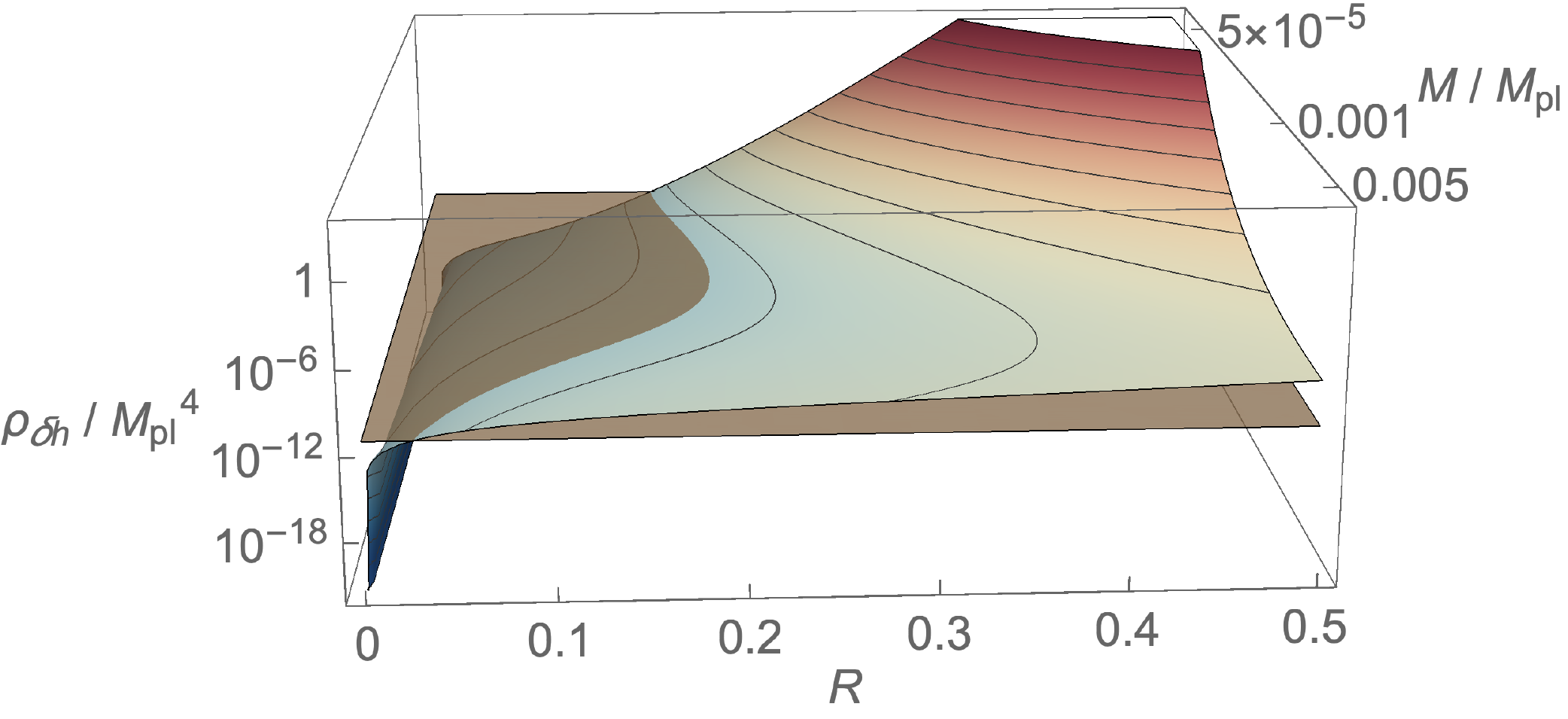}
	\end{subfigure}
	\begin{subfigure}
		\centering
		\includegraphics[width=.48\textwidth]{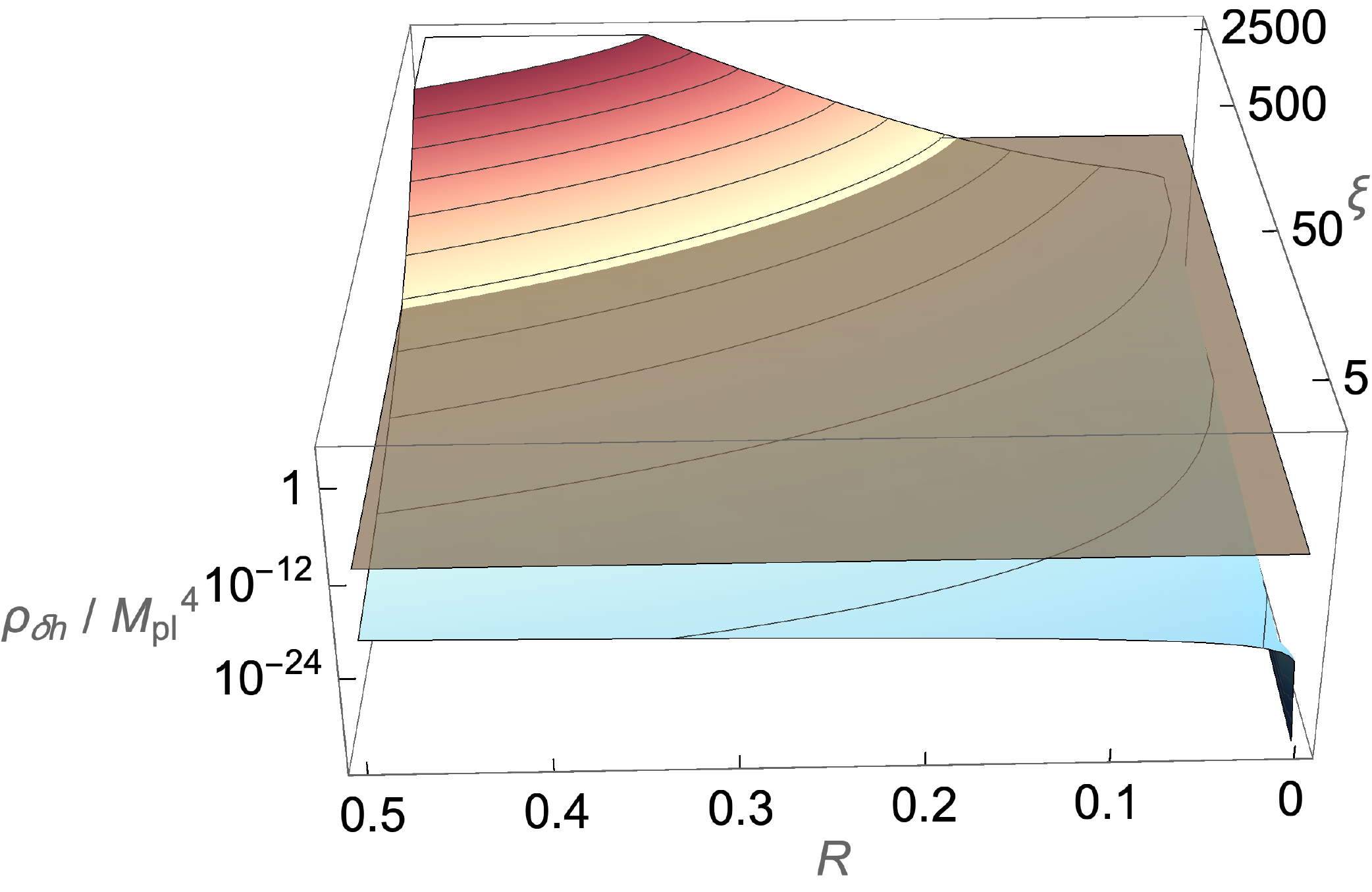}
	\end{subfigure}
	\caption{The energy density of Higgs fluctuations after the tachyonic instability as a function of $\xi$ (or $M$)  and $R$ for $ 0\leq R < 1/2 $ . The gray transparent surface is $ C^2_1\rho_{\rm inf}/2 $. Left: Branch 1 (figure shown in terms of $ M/M_{\rm pl} $). Right: Branch 2 (figure shown in terms of $ \xi $).}
	\label{fig-rho-small-R}
\end{figure}

\begin{figure}[h]
	\centering
	\begin{subfigure}
		\centering
		\includegraphics[width=.48\textwidth]{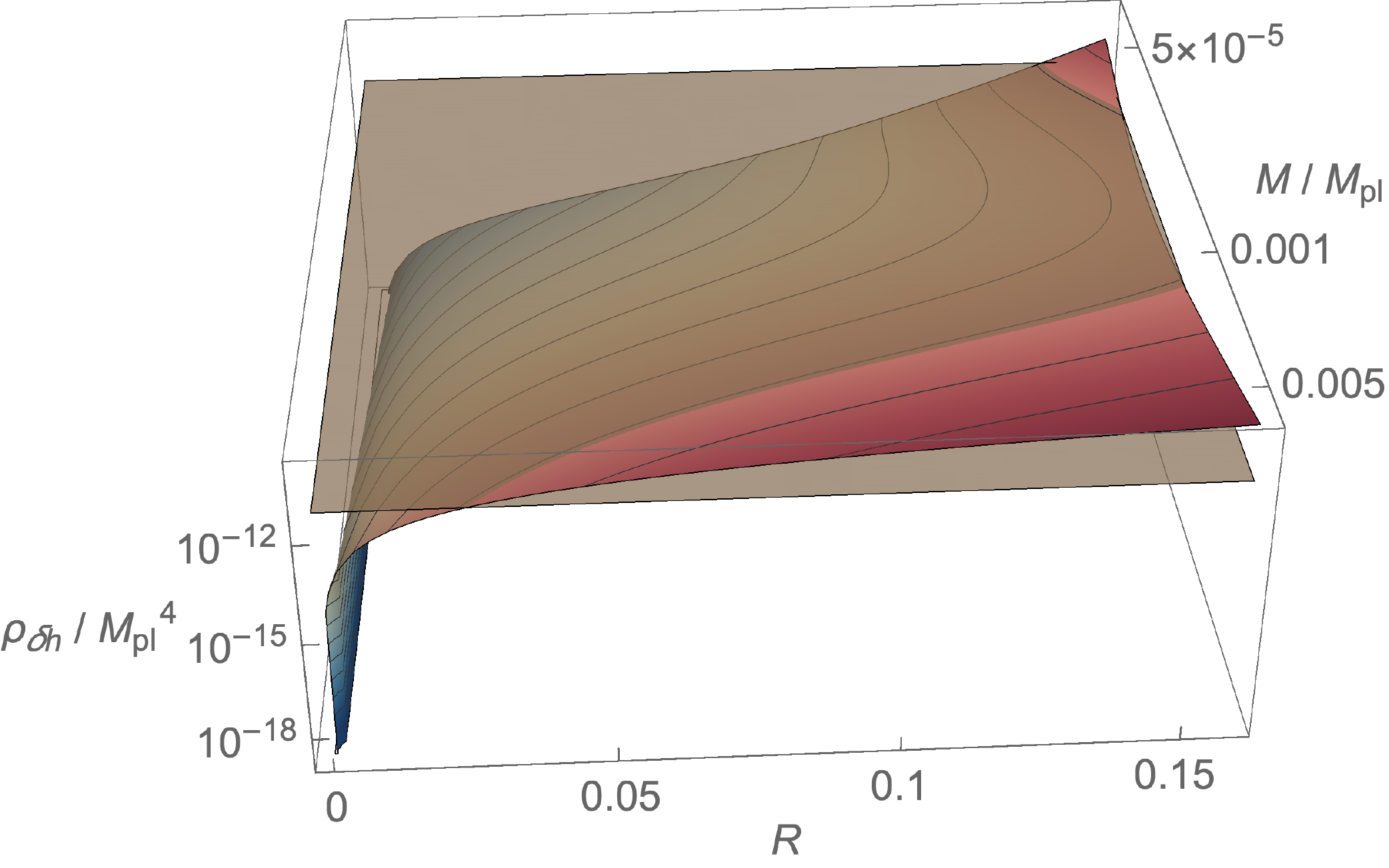}
	\end{subfigure}
	\begin{subfigure}
		\centering
		\includegraphics[width=.48\textwidth]{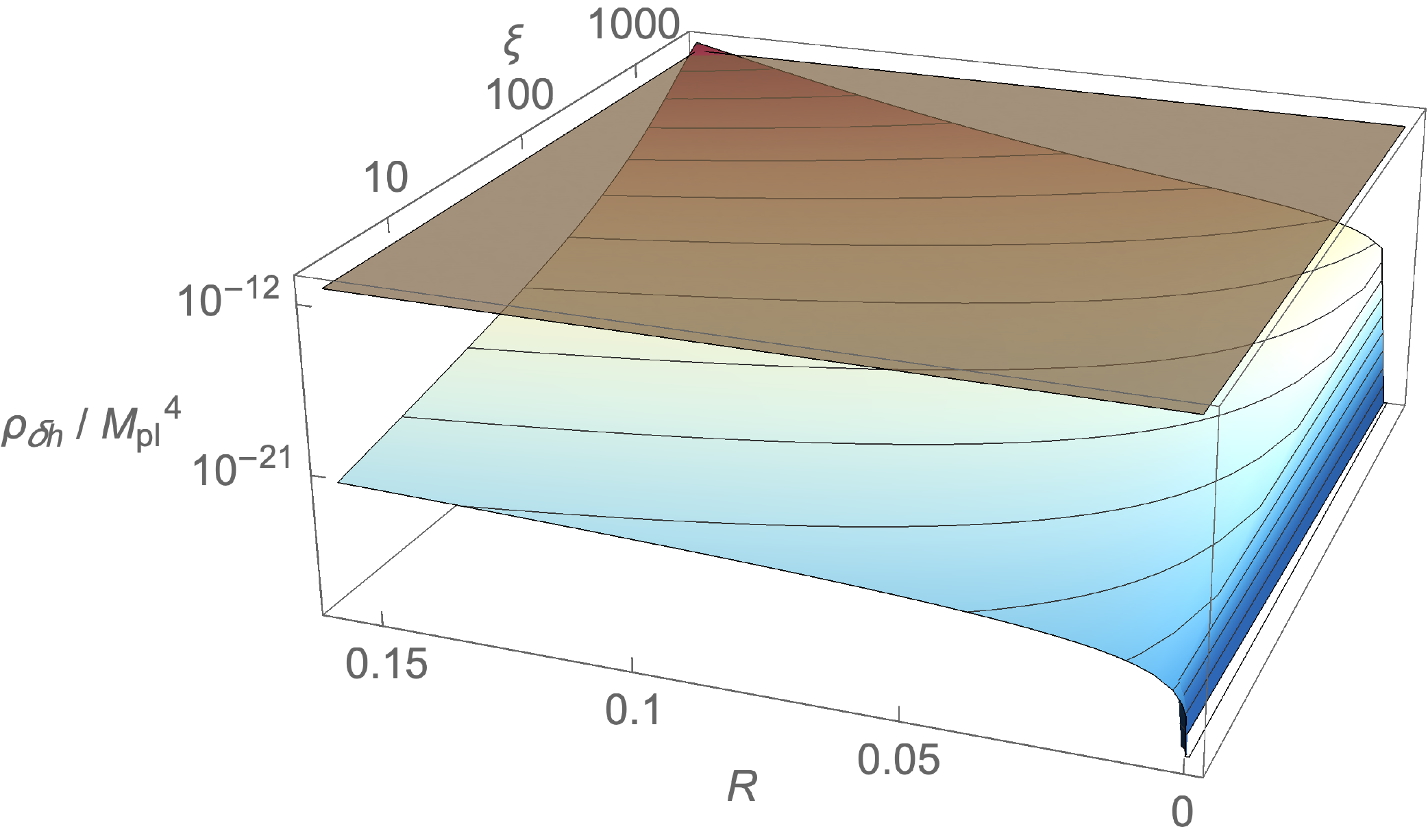}
	\end{subfigure}
	\caption{The same figure as Fig.~\ref{fig-rho-small-R} but for $0\leq R < 1/(2\pi)$. }
	\label{fig-rho-small-R-intersection}
\end{figure}

The energy density of produced particles is then 
\begin{align}
	\rho_{\delta h} (\xi, R) \simeq & \frac{1}{2\pi^2} \int k^3  e^{2\Omega_k} dk 
	= \frac{m^4_{h,\rm max}}{2\pi^2} I_2(\xi, R),
\end{align}
where we have defined
\footnotesize
\begin{align}
	I_2(\xi, R) &\equiv \int^{\sqrt{\sin (\pi R)}}_0 dk	\nonumber \\
	&~~~~\times k^3 \exp \left[ \frac{4m_{h,\rm max}}{M(\xi)} \sqrt{1-k^2} \left( E\left[ \frac{1}{2}\arccos \left( k^2 \right), \frac{2}{1-k^2} \right] -E\left[ \frac{\pi}{4} (1-2R), \frac{2}{1-k^2} \right] \right) \right] \nonumber \\
	&= \int^{\sqrt{\sin (\pi R)}}_0 dk \nonumber \\
	&~~~~\times k^3 \exp \left[ \frac{4m_{h,\rm max}}{M(\xi)} \sqrt{1-k^2} \left( E\left[ \frac{\pi}{4}-\frac{1}{2}\arcsin (k^2), \frac{2}{1-k^2} \right] -E\left[ \frac{\pi}{4} (1-2R), \frac{2}{1-k^2} \right] \right) \right],
\end{align}
\normalsize
which is different from $I_{11}$ due to the difference of the domain of $R$. 
The resultant $ \rho_{\delta h} (\xi, R) $ for Branch 2 and $ \rho_{\delta h} (M, R) $ for Branch 1 are shown in Fig.~\ref{fig-rho-small-R}. 
As one can see, for all $ \xi $ (or $ M $) within the unitary bound $M< 4.6\times 10^{-4}M_{\rm pl}$, at least $ R \gsim 1/(2\pi) $ is required to complete preheating, which is easier to see in Fig.~\ref{fig-rho-small-R-intersection}. 

As mentioned in the beginning of this appendix, $R$ in reality depends on the model parameters, i.e.  $R(\xi)$ or $R(M)$, and it plays an essential role to determine the degree of fine-tuning needed to complete preheating (see Sec.~\ref{Sec-4}). Thus, if one could relate $R$ and $\xi$ (or $M$), the required degree of fine-tuning can be predicted analytically. 
However, it is difficult to solve this relation analytically, especially due to the requirement $R \gsim 1/(2\pi)$. Generally speaking, the difficulty comes from the nonlinearity of the system. If the required survival rate is sufficiently small, i.e. tachyonic effect is strong enough even when the inflaton stays on the hill for very short time compared with the time scale of the scalaron oscillation, one can linearly approximate $m^2_h(t)$, so that the equations of motion for the background field dynamics for the scalaron and the Higgs field can be solved analytically. As a result, one can express the survival rate in terms of the model parameters. Unfortunately, as seen in Fig.~\ref{fig-rho-small-R} and \ref{fig-rho-small-R-intersection}, relatively large $R \gsim 1/(2\pi)$ is needed, which prevents one from using linear approximation to solve the equations of motion for the purpose of finding the necessary degree of fine-tuning to complete preheating.

\bibliography{ref}{}

\end{document}